\documentclass[]{jfm}

\usepackage{graphicx}
\usepackage{epstopdf,epsfig}
\usepackage{newtxtext}
\usepackage{newtxmath}
\usepackage{natbib}
\usepackage{hyperref}
\usepackage{amsmath}
\usepackage{amssymb}
\usepackage{bm}
\usepackage{color}
\usepackage{tikz}
\usepackage{soul}
\usepackage{multirow}
\DeclareRobustCommand\sampleline[1]{%
  \tikz\draw[#1] (0,0) (0,\the\dimexpr\fontdimen22\textfont2\relax)
  -- (2em,\the\dimexpr\fontdimen22\textfont2\relax);%
  }
\hypersetup{
    colorlinks = true,
    urlcolor   = blue,
    citecolor  = black,
}

\newcommand{\RomanNumeralCaps}[1]
\linenumbers

\title{On the vortex dynamics of synthetic jet -- turbulent boundary layer interactions}

\author{Joseph C. Straccia\aff{1},
 \and John A. N. Farnsworth\aff{1}
 \corresp{\email{john.farnsworth@colorado.edu}}}

\affiliation{\aff{1}Ann and H.J. Smead Department of Aerospace Engineering Sciences, \\University of Colorado Boulder,
Boulder, CO 80309, USA}

\begin{document}
\maketitle

\begin{abstract}
The vortex dynamics resulting from the interaction of synthetic jets with turbulent boundary layers was investigated experimentally using stereoscopic particle image velocimetry (SPIV). 
Three aspect ratio 18 rectangular orifice geometries were tested including spanwise and streamwise-oriented orifices issuing normal to the wall and a spanwise-oriented orifice pitched $45^\circ$ downstream. 
Additionally, three actuation conditions were tested to explore the impact of varying blowing ratio and Strouhal number.
SPIV data obtained on an array of measurement planes was used to reconstruct the three-dimensional mean and phase-locked velocity fields, enabling detailed analysis of the formation and development of the vortices.
The variations in the orifice geometry and the flow conditions produced a range of vortex structures including distorted vortex rings, hairpin vortices, and arch-shaped vortices. 
The jets from wall-normal orifices were dominated by vortices that rotated in the same direction as the boundary layer vorticity which enhanced the near-wall flow velocity by increasing mixing. 
Conversely, the dominant structures in the pitched jets rotated counter to the boundary layer vorticity, and these jets enhanced the flow velocity near the wall through direct momentum addition. 
Flow field analysis revealed that the synthetic jet-boundary layer interaction caused pitched jets operated at low blowing ratios to produce dominant vortices which were much weaker than those generated by wall-normal jets at the same conditions.
Finally, at the lowest Strouhal number tested the increased spacing between the vortices in the wall-normal jets resulted in much higher unsteady mixing than was observed at higher Strouhal numbers.
\end{abstract}


\newcommand{\greyline}{\raisebox{2pt}{\tikz{\draw[-,black!40!white,solid,line width = 0.5pt](0,0) -- (7mm,0);}}}

\section{Introduction}
\label{sec:Intro}

The interaction of synthetic jets with wall-bounded cross-flows is a fascinating dynamic phenomenon in which both the jet and the boundary layer (BL) may be substantially modified. 
The physical details of this interaction are an active area of research, especially with regard to the behaviour of the coherent structures. 
Such studies are valuable not only due to their scientific merit, but also because they provide the critical engineering knowledge needed to more effectively apply synthetic jet-based flow control to practical problems. 
In addition to exploring in depth the flow physics of synthetic jet–boundary layer interactions (SJBLI), this paper will also consider how changes to the jet may influence the effectiveness of the actuation in controlling smooth wall flow separation.

Synthetic jet actuators (SJA) are a type of active flow control device which adds momentum to a flow field without a net addition of mass. 
To generate this zero-net-mass-flux jet the driver draws external fluid into a cavity during the suction phase and then expels that fluid back out of the orifice during the blowing phase. 
The periodic expulsion of fluid by the actuator leads to the rapid formation of vortex rings which travel away from the orifice carrying the momentum introduced by the actuator outstroke~\citep{Xia_Mohseni_PoF_2015}. 
A vortex ring is formed with every cycle of the actuator, sometimes at a rate of thousands per second, resulting in a train of vortex rings which constitute the early jet~\citep{Smith_Glezer_PoF_1998}.

\subsection{Vortex dynamics in synthetic jets}
\label{sec:vortdynSJ}

Vortex dynamics plays an important role in establishing the characteristics of synthetic jets. 
Not only are the vortex rings the vehicles by which momentum is transported in the synthetic jet nearfield, but they also influence the jet's entrainment rate, shape, and momentum distribution. 
When the performance of equivalent steady and synthetic jets is compared, synthetic jets demonstrate a markedly higher entrainment rate, especially near the orifice~\citep{Smith_Glezer_PoF_1998, Cater_Soria_JFM_2002, Smith_Swift_2003}. 
This is attributed to the formation of the vortex rings which engulf surrounding fluid as the vortex sheet generated by the expelled fluid slug rolls up. 
Vortex ring-driven entrainment is further enhanced when the jet is non-circular.

Vortex rings induce a velocity in the fluid around them, as described by the Biot-Savart law. 
This effect extends to the fluid within the vortex core itself leading to what is known as a self-induced velocity. 
The speed of the local self-induced velocity varies with the curvature of the vortex ring axis; specifically, high curvature regions experience higher local self-induction than low or zero curvature zones~\citep{Arms_Hama_PoF_1965}. 
Thus, non-circular vortex rings, which by definition have variable curvature along their axes, deform as they propagate. 
In vortex rings with symmetric shapes these deformations lead to oscillatory changes in the ring's width known as axis-switching~\citep{Viets_Sforza_PoF_1972, Dhanak_Bernardinis_JFM_1981}. 
The oscillations in the vortex ring width are mirrored in the jet velocity field and can be quite significant in larger aspect ratio ($AR$: the ratio of orifice length to width) jets~\citep{Van_Buren_Whalen_JFM_2014, Straccia_Farnsworth_PRF_2021}. 
This growth and contraction of the vortex rings in the transverse direction is responsible for enhancing mixing between the jet and the surrounding fluid~\citep{Ho_Gutmark_JFM_1987, Wang_Feng_ExpFluids_2018}.

When a slug of fluid ejected from an orifice or nozzle issues into a cross-flow the resulting vortex structure may deviate significantly from the classic vortex ring shape. 
One effect of the cross-flow is to weaken the upstream side of the cylindrical vortex sheet formed by the ejection of the fluid slug~\citep{Lim_Lua_2008}. 
\citet{Sau_Mahesh_2008} observed this type of modification to the vortex rings formed by a circular nozzle in a laminar BL in their direct numerical simulations. 
At high blowing ratios (i.e., ratio of average nozzle exit velocity to the free-stream cross-flow velocity) roughly circular vortex rings formed, but as blowing ratio was reduced, no vortex rolled up on the upstream side of the nozzle, resulting instead in the formation of a hairpin vortex. 

To understand the shift from symmetric vortex rings in quiescent conditions to hairpin vortices in a cross-flow, it is helpful to think about the interaction from the perspective of the vorticity field. 
The vorticity generated on the upstream side of the jet is of the opposite sign to the vorticity in the viscous BL, and when fluids containing anti-parallel vorticity come into close proximity, segments of the counter rotating vortex lines are annihilated in the contact zone by viscous cross-diffusion~\citep{Melander_Hussain_CTR_1988}. 
Meanwhile, at the edge of the contact zone the remaining sections of the vortex lines crosslink to form a new vortex topology. 
The annihilation of vorticity in one part of a vortex causes the circulation in that segment to drop relative to the rest of the structure \citep{Oshima_Izutsu_FDR_1988, Straccia_Farnsworth_JFM_2020}. 
Simultaneously, the downstream vortex may be strengthened by the cross-flow interaction because the ambient fluid entrained during the vortex roll-up carries vorticity of the same sign~\citep{Cheng_Lou_PoF_2009}.

The difference in circulation strength between the upstream and downstream sides of a vortex ring formed in a cross-flow may also be explained from the perspective of the velocity field. 
When the cross-flow collides with the jet the pressure field at the orifice exit is modified such that fluid leaves the upstream edge at a slower speed, reducing vorticity production, and exits the downstream edge are a higher speed, increasing vorticity production \citep{Sau_Mahesh_2008}. 
These two explanations are both manifestations of the same effect, merely viewed from the perspective of the vorticity field versus the velocity and pressure fields.

This variability in the strength of the original vortex tube is made possible by branches in the tube that contain the `missing' circulation~\citep{Melander_Hussain_CTR_1988, Kida_Takaoka_JFM_1991}. 
For vortex rings formed in BLs, these branches look like `legs' which trail behind the main vortex and lead back towards the wall~\citep{Jabbal_Zhong_2010}. 
The legs are composed of BL vorticity which crosslinked to vorticity within the vortex during reconnection. 
As the circulation strength of the upstream side of the vortex is reduced by vorticity reconnection the strength of the vortex tube branches must increase by an equivalent amount \citep{Straccia_Farnsworth_JFM_2020}. 
For very low blowing ratios, where no detectable upstream side of the vortex ring forms, the vortex tube branches become the hairpin legs and the unannihilated downstream portion of the vortex ring becomes the hairpin head.

\citet{Jabbal_Zhong_2010} employed dye-visualization to study how the vortex structures formed by a circular orifice synthetic jet issuing into a laminar BL changed with blowing ratio. 
At low blowing ratios a train of hairpin vortices formed which remained relatively close to the wall. 
As blowing ratio increased the structures took the form of stretched vortex rings with trailing legs. 
Finally, at the highest blowing ratios tilted vortex rings formed which penetrated deeper into the cross-flow than in the other two scenarios. 
The vortex rings which formed at the higher blowing ratios rotated in the counter-clockwise (CCW) direction (cross-flow left-to-right) such that their direction of propagation was tilted into the oncoming flow. 
The vortex ring rotation was driven by the variable propagation speed around the ring resulting from uneven circulation strength in the upstream versus the downstream side of the ring~\citep{Lim_Lua_2008, Cheng_Lou_PoF_2009}.

The dynamics of vortex rings generated by rectangular orifice synthetic jets in a cross-flow differ in a few ways from those of circular jets. 
The cross-flow interaction still weakens the upstream side of the vortex ring; however, for moderate or high $AR$ orifices which are oriented with their long axis aligned with the spanwise direction (i.e., spanwise-oriented orifices) the vortex pair on centreline tilts clockwise (CW) instead of CCW (flow left-to-right)~\citep{Sahni_Wood_JFM_2011, Wang_Feng_JFM_2020, Elimelech_Vasile_2011}. 
The direction of rotation of the vortex pair on centreline is CW because motion is dominated by velocity induction from the adjacent counter-rotating side of the vortex ring instead of by local self-induction. 
Under the right conditions the upstream side of the vortex ring may even be rolled up and over the stronger downstream vortex segment and deposited near the wall~\citep{Sahni_Wood_JFM_2011}. 
If the upstream side of the vortex sheet is weakened sufficiently by the SJBLI, hairpin-like vortex structures may form in spanwise-oriented rectangular orifice synthetic jets~\citep{Belanger_Zingg_AIAA_2020}. 
Unlike the hairpin vortices from circular orifices, however, these form after a rapid spanwise contraction of the initial vortex structure due to self-induction and the influence of the wall (image vortex effect). 
Conversely, if conditions produce a sufficiently intact vortex ring a form of axis-switching occurs as the non-circular vortex rings propagate into the cross-flow~\citep{Wang_Feng_JFM_2020}. 

In both circular and rectangular orifice synthetic jets, increasing the blowing ratio increases the self-induced propagation speed of the vortex relative to the cross-flow speed and leads to an overall more complete vortex ring structure. 
Thus, the higher the blowing ratio, the deeper the synthetic jet tends to penetrate into the cross-flow~\citep{Berk_Hutchins_JFM_2018, Jabbal_Zhong_2010, Sahni_Wood_JFM_2011, Belanger_Zingg_AIAA_2020}. 
%However, this trend changes once the limiting vortex formation number is exceeded because increasing the actuator stroke length further does not result in a higher circulation strength of the primary vortex~\citep{Gharib_Rambod_JFM_1998,Sau_Mahesh_2008}.

\begin{figure}
\centering
\includegraphics[width=5.25in]{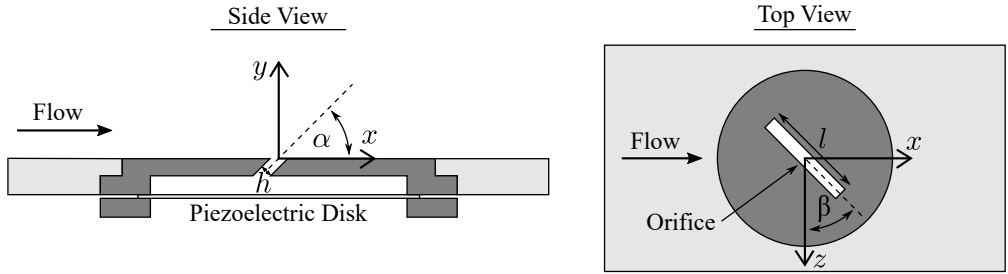}
\caption{Side view cross-section and top view of SJA installed in BL plate.}
\label{fig:sjascheme}
\end{figure}

Due to the lack of axisymmetry in rectangular orifices the orientation of the orifice major axis relative to the spanwise direction (i.e., the skew angle, $\beta$) also influences the resulting vortex dynamics (figure~\ref{fig:sjascheme}). 
%As a rectangular orifice is rotated about the wall-normal direction (i.e., $\beta$ is modified) the side of the vortex ring which is diminished as a result of the SJBLI changes, as does the location of the vortex tube legs~\citep{Van_Buren_Leong_PoF_2016, Wang_Feng_JFM_2020}. 
As a rectangular orifice's skew angle is modified the side of the vortex ring which is diminished as a result of the SJBLI changes, as does the location of the vortex legs~\citep{Van_Buren_Leong_PoF_2016, Wang_Feng_JFM_2020}. 
Jets from spanwise-oriented orifices are strongly influenced by the cross-flow because it is the long sides of the vortex ring which are modified. 
Conversely, jets from streamwise-oriented orifices (i.e., orifices with the major axis aligned with the cross-flow direction) present a much smaller frontal area to the cross-flow so, in this case, the vortex rings are minimally modified by the SJBLI. 
Thus vortex rings formed in the streamwise-orientation lift off more rapidly from the wall and penetrate deeper into the cross-flow than those formed in a spanwise orientation~\citep{Van_Buren_Leong_PoF_2016, Smith_2002, Wang_Feng_JFM_2020}. 

The interaction of a cross-flow with a synthetic jet issuing normal to the local surface produces a low momentum wake region in the lee of the jet. 
\citet{Berk_Ganapath_JFM_2019} investigated three potential contributors to the formation of a wake: (1) momentum lost by the cross-flow in accelerating the jet, (2) upwash of low momentum fluid off the wall by the induction from the vortex ring on the jet centreline, and (3) velocity induction from the vortex ring which may be oriented counter to the cross-flow direction. 
Based on their experiments they concluded that, to varying degrees, all three factors contributed to the formation of the wake in different regions of the interaction domain. 
When blowing ratio is increased, the wake moves further from the wall and has a larger peak velocity deficit~\citep{Berk_Hutchins_JFM_2018} which is due to the formation of stronger vortex rings that penetrated deeper into the cross-flow and induce higher velocities in the fluid around them.

Although the mixing introduced by wall-normal synthetic jets contributes to the formation of a low-momentum wake it also has a salutary effect on the near-wall BL. 
Specifically, the vortex rings or hairpin vortices induce a downward velocity around the outside of these structures, entraining high momentum fluid into the lower BL. 
The result is a thinned BL surrounding the footprint of the jet and accelerated flow near the wall~\citep{Smith_2002, Jabbal_Zhong_2010, Van_Buren_Leong_PoF_2016}. 
Therefore, wall-normal synthetic jets can be effective at enhancing the wall shear stress in a cross-flow despite reducing fluid momentum along the jet centreline further from the wall~\citep{Jabbal_Zhong_2010}.

The SJBLI is not the only factor that influences the structure of the synthetic jet vortex rings. 
The geometry of a nozzle or orifice can also lead to variable circulation and a branched topology of the vortices. 
For example, this occurs when the flow channel through which the fluid is ejected (i.e., the orifice or throat of the jet) is pitched or angled away from the wall-normal direction, i.e., pitch angle $\alpha \neq 90^\circ$ (figure~\ref{fig:sjascheme}). 
Pitched orifices produce vortex rings which have stronger circulation at the acute edge of the orifice than at the obtuse edge~\citep{Zhong_Garcillan_AIAA_2004, Yehoshua_Seifert_2006, Li_Sahni_2014}. 
Studies involving piston-driven vortex ring formation from inclined nozzles found that a difference in the magnitude of the pressure gradient on the short and long sides of the flow channel leads to non-uniform vorticity production at the nozzle exit~\citep{Webster_Longmire_1998,Le_Borazjani_2011}. 
%An alternate or potentially complementary explanation for the production of vortex rings with non-uniform circulation strength has been offered by several researchers for the case of a pitched orifice. 
%They hypothesized that the geometry of the inflow side of a pitched orifice may induce separation on one wall of the throat, resulting in a skewed velocity profile at the orifice exit~\citep[p.57-58]{Zhong_Garcillan_AIAA_2004, Bray_Thesis_1998}.
An alternate or potentially complementary explanation relates to the theory that orifices with pitched geometries tend to induce separation in the flow on one wall of the orifice channel resulting in a skewed velocity profile at the orifice exit~\citep[p.57-58]{Zhong_Garcillan_AIAA_2004, Bray_Thesis_1998}.

When an orifice is pitched in the downstream direction with respect to the cross-flow, the two modifying effects on the vortex circulation strength discussed above are in competition. 
On one hand, the SJBLI will cause vortex rings to form with lower circulation strength on the upstream side of the ring than on the downstream section, due to the rotational sign of the vorticity in the BL. 
On the other hand, pitching the orifice downstream has the opposite effect, which is the enhancement of the circulation strength of the upstream or acute edge of the orifice. 
Therefore, a balance exists which is governed by the orifice pitch angle and the relative characteristics of the cross-flow (i.e., blowing ratio and BL profile). 
\citet{Yehoshua_Seifert_2006} experimentally observed that the circulation asymmetry in the vortex pair produced by a pitched high $AR$ orifice was reduced as the cross-flow velocity was increased (i.e., blowing ratio was reduced). 
\citet{Li_Sahni_2014} found in their simulations of an infinite slot that under the right conditions the two effects essentially cancel each other out and a roughly symmetric vortex pair forms. 

The structure and behaviour of the vortices in a synthetic jet are also influenced by the frequency at which they are formed by the actuator.
As actuation frequency is increased the circulation strength per vortex decreases~\citep{Straccia_Farnsworth_JFM_2020}; therefore, the trajectories of synthetic jets formed at higher frequencies generally remain closer to the wall than those of jets formed at lower frequencies~\citep{Berk_Hutchins_JFM_2018, Berk_Ganapath_JFM_2019}. 
But while the circulation strength of an individual vortex decreases with increasing actuation frequency, the time-averaged (mean) circulation produced by a jet remains constant. 
\citet{Berk_Ganapath_JFM_2019} argued that this explained the minimal change in the jet wake momentum deficit when they varied the actuation frequency. 
%This insensitivity of the jet blockage to actuation frequency does not extend to frequencies above the synthetic jet formation limit, however, for which part or all of the vortex formed by the blowing cycle may be re-ingested during the suction phase~\citep{Holman_Utturkar_AIAAJ_2005}. 
%Specifically, above the formation limit vortex re-ingestion tends to reduce the influence of the jet on the cross-flow~\citep{Berk_Ganapath_JFM_2019}. 
Increasing actuation frequency also reduces the spacing between adjacent vortex structures within a BL. 
When the spacing is small enough, adjacent vortices will interact and may merge. 
Thus the flow field can shift from one characterized by a train of discrete vortex rings or hairpin vortices into one of persistent streamwise vortices in the domain downstream of the actuator~\citep{Sahni_Wood_JFM_2011, Van_Buren_Leong_PoF_2016, Belanger_Zingg_AIAA_2020}.

\subsection{Context for the present study}
\label{sec:sumpriorwork}

The combination of periodic suction and blowing, orifice skew, orifice pitch, and a wall-bounded cross-flow leads to complex vortex dynamics within the SJBLI zone and results in a range of vortex topologies. 
Investigating the details of these varied and multifaceted flow fields is one objective of the present study. 
To put the present work into context, table~\ref{tab:priorlit} provides a partial list of prior studies which sought to characterize the SJBLI on a flat plate.
The table includes information on (1) the BL type (laminar or turbulent), (2) the orifice shape (circular, slot or rectangular) and $AR$, (3) whether three-dimensional, three-component (3D3C) velocity field data are presented, (4) whether instantaneous or phase-locked flow field representations are presented (as opposed to the mean flow field), and (5) the type of measurements including flow visualization (FV), hot-wire anemometry (HW), particle image velocimetry (PIV) or computational fluid dynamics (CFD). 
A tilde symbol in the `yes' or `no' columns indicates that a particular category is only partially met or that the data relevant to that category are limited.

Although several dozen prior studies have sought better understanding of fundamental vortex dynamics in synthetic jets issuing into zero or near-zero pressure gradient cross-flows the configurations and conditions which may be most relevant to real-world flow control applications are less well studied (table~\ref{tab:priorlit}). 
For example, many of the prior vortex dynamics studies have focused on circular orifices. 
By contrast, applied research studies investigating the use of synthetic jets to solve flow control problems (e.g., separation control) frequently employ rectangular orifices with an $AR$ of 18 or larger. 
But the behaviour of vortices from circular versus finite span orifices can be quite different, as discussed earlier. 

\begin{table}
\centering
\begin{tabular}{c c c c c c c}
Article & BL Type & Orifice Shape; $AR$ & 3D3C & Instant. Data & Measurement \\ [0.5ex]
\hline
\citet{Zhong_Garcillan_AeroJ_2005} & Lam & Circ; 1 & No & Yes & FV \\
\citet{Shuster_Pink_2005} & Lam & Circ; 1 & No & Yes & PIV \\
\citet{Jabbal_Zhong_2008} & Lam & Circ; 1 & No & Yes & FV \\
\citet{Jabbal_Zhong_2010} & Lam & Circ; 1 & No & Yes & FV, PIV \\
\citet{Xia_Mohseni_AIAAJ_2017} & Lam & Circ; 1 & No & $\sim$ & HW, PIV \\
\citet{Zhou_Zhong_CompFluids_2009} & Lam & Circ; 1 & Yes & Yes & CFD \\
\citet{Mittal_Rampunggoon_AIAA_2001} & Lam & Slot; $\infty$ & No & Yes & CFD \\
\citet{Li_Sahni_2014} & Lam & Slot; $\infty$ & $\sim$ & Yes & CFD \\
\citet{Yehoshua_Seifert_2006} & Lam & Rect; 135 & No & Yes & FV, HW, PIV \\
\citet{Van_Buren_Beyar_PoF_2016} & Lam & Rect; 6, 12, 18 & Yes & $\sim$ & SPIV \\
\citet{Van_Buren_Leong_PoF_2016} & Lam & Rect; 18 & Yes & $\sim$ & SPIV \\
\citet{Wang_Feng_JFM_2020} & Lam & Rect; 3 & Yes & Yes & Tomo PIV \\
\citet{Milanovic_Zaman_AIAAJ_2005} & Turb & Circ; 1 & No & No & HW \\
\citet{Garcillan_Liddle_AIAA_2006} & Turb & Circ; 1 & No & Yes & PIV \\
\citet{Ramasamy_Wilson_JAircraft_2010} & Turb & Circ; 1 & No & No & HW, PIV \\
\citet{Cui_Agarwal_AIAA_2003} & Turb & Slot; $\infty$ & No & Yes & CFD \\
\citet{Smith_2002} & Turb & Rect; 45 & No & No & HW \\
\citet{Bridges_Smith_AIAAJ_2003} & Turb & Rect; 98 & No & No & HW \\
\citet{Milanovic_Zaman_2004} & Turb & Rect; 4, 6, 18 & No & No & HW \\
\citet{Berk_Ganapath_JFM_2019} & Turb & Rect; 13 & No & Yes & SPIV \\
\citet{Housley_PhDThesis_2020} & Turb & Rect; 18 & Yes & Yes & SPIV \\
\citet{Belanger_Zingg_AIAA_2020} & Turb & Rect; 13 & Yes & Yes & CFD \\ [1ex]
\end{tabular}
\caption{Summary of prior flat plate SJBLI studies.}
\label{tab:priorlit}
\end{table}

Many SJBLI studies are also conducted in laminar BLs. 
However, separation control problems in engineering flows often involve high Reynolds numbers and turbulent BLs. 
This distinction can be important because the distribution of vorticity in a BL may influence vorticity annihilation and the types of structures resulting from SJBLIs. 
Furthermore, vortices embedded in turbulent BLs distort the Reynolds stress field and, in turn, gradients in the Reynolds stress field accelerate diffusion of the embedded vortices~\citep{Perkins_JFM_1970, Pauley_Eaton_1988}. 
Therefore, one cannot necessarily expect a particular vortex structure in a laminar BL to evolve in the same manner as it would in a turbulent one. 
Another challenge with extending SJBLI results in laminar BLs to turbulent flow fields relates to flow transition. 
In fundamental flow control studies the BL shape or wall shear stress is often used as an indicator of the potential effectiveness for separation control. 
However, the tendency of a large perturbation to force transition in laminar BLs makes it difficult to determine whether a particular actuator induces superior mixing or is just more effective at tripping the flow, a task easily achieved by simpler passive devices.

Of the subset of experiments conducted to investigate the interaction between rectangular orifice synthetic jets and turbulent BLs many have only obtained time-averaged or mean data, typically using hot-wire anemometry. 
Although mean flow data may reveal the average effect of the actuation, it does not provide a definitive picture of the underlying vortex dynamics, i.e., the why. 
In a similar vein, data revealing the instantaneous vortex structure on just the centreline plane provides some information about the dynamics but can miss critical 3D effects like spanwise contraction of the jet or the presence of hairpins legs that straddle the symmetry plane. 
Therefore, it is helpful when trying to understand the detailed vortex dynamics in SJBLIs to obtain both time-averaged and the phase-locked measurements and to do so for the three-component velocity field throughout the interaction domain. 

\subsection{Outline of the present study}
\label{sec:outpresstudy}

The current study presents detailed experimental data revealing the vortex dynamics resulting from a SJBLI for an infrequently studied, but highly relevant, combination of actuator geometry and flow field conditions. 
For its relevance to common separation control problems, the baseline flow field, upon which the actuator acts, is that of a fully developed turbulent BL. 
Additionally, the actuator employs an $AR=18$ rectangular orifice which falls within the range of values often considered in applied studies. 
Three of the most common variations of the orifice orientation are explored including (1) a wall normal jet with the orifice major axis aligned with the spanwise direction, (2) a wall normal jet with the orifice major axis aligned with the streamwise direction, and (3) a jet pitched 45 degrees downstream with the orifice major axis aligned with the spanwise direction. 
Two blowing ratios of 0.5 and 1.0 are considered, values which fall within the practical range for typical engineering problems. 
Additionally, the influence of vortex spacing in the BL is also explored by varying the Strouhal number of the actuation. 
The data needed to reconstruct both the mean influence of the SJBLI along with the development of 3D instantaneous structures was captured using particle image velocimetry measurements on an array of planes throughout the domain. 
Finally, in addition to exploring the flow physics in these various cases the resulting flow fields are also analyzed to infer their potential effectiveness in delaying separation.

The experimental results of this study are presented as follows. 
Section~\ref{sec:expmethods} provides details on the actuator design, the conditions studied, and the experimental methods employed. 
Section~\ref{sec:results1} looks in detail at the dynamics in the three jets individually, at a single actuation condition. 
Section~\ref{sec:results2} compares the effects of the three jets on the flow field at that same condition. 
Next, the influences of different actuation conditions on the SJBLI are explored in \S~\ref{sec:results3}. 
Finally, the results are summarized and the conclusions are presented in \S~\ref{sec:conclusions}.
\section{Experimental methods}
\label{sec:expmethods}

\subsection{The facility}
\label{sec:facility}

The SJBLI experiments were conducted in the low-speed wind tunnel within the Experimental Aerodynamics Laboratory at the University of Colorado Boulder. 
This open-return facility is configured with a centrifugal blower at the wind tunnel inlet and is able to generate air speeds in the test section of up to 65~ms$^{-1}$. 
The test section has a square cross-sectional area of 0.76~m $\times$ 0.76~m and is 3.5~m long. 
%A boundary layer plate with a development length of 3.4~m was installed 0.31~m off of the test section floor and spanned the full width of the test section, introducing up to 4.75\% of cross-sectional geometric blockage. 
A boundary layer plate with a development length of 3.4~m was installed 0.31~m off of the test section floor and spanned the full width of the test section. 
%The BL plate leading edge had an 8:1 modified super elliptical shape designed to be tolerant to non-zero incidence angles while also free of discontinuities in curvature, ensuring smooth transition of the flow onto the plate surface~\citep{Lin_Reed_Saric_StabTranTurb_1992}. 
%At the trailing edge of the BL plate a flap was mounted which was adjusted to achieve a zero incidence at the leading edge, as determined from an array of embedded static pressure taps within the leading edge part. 
The BL plate leading edge had an 8:1 modified super elliptical shape while the trailing edge was mounted with a flap which was adjusted to achieve a zero incidence at the leading edge~\citep{Lin_Reed_Saric_StabTranTurb_1992}. 
%To trip the flow a 25.4~mm wide strip of 36 grit distributed sand grain roughness was installed across the full span of the BL plate at a distance of 0.19~m downstream from the leading edge. 
To trip the flow a 25.4~mm wide strip of 36 grit distributed sand grain roughness was installed across the BL plate at a distance of 0.19~m downstream from the leading edge. 
This provided over 1000 trip strip heights of development length for the flow before it reached the actuator which was installed on the plate centreline 1.62~m downstream from the leading edge.

\subsection{The actuator}
\label{sec:actuator}

To study the interaction of synthetic jets with turbulent BLs a modular actuator was designed. 
Interchangeable plates containing the actuator orifice were fabricated using a stereolithography (SLA) printer. 
The jet was driven by a 63.5~mm diameter piezoelectric disk actuator which was clamped against the orifice plate (figure~\ref{fig:sjascheme}). 
The resulting cylindrical cavity connected to the orifice was 58.4~mm in diameter and 3.18~mm tall. 
Under the piezoelectric disk was a second cavity that was sealed with a ring shaped gasket against a cast acrylic disk. 
%Clear acrylic was used for the floor of the actuator under-cavity to provide optical access for laser vibrometer measurements of the underside of the piezoelectric disk. 
Clear acrylic was used for the floor of the actuator under-cavity to provide optical access for laser vibrometer measurements of the piezoelectric disk. 
The actuator assembly was secured into a circular cut-out in the BL plate using a retaining ring, which, when loose, allowed the actuator to be rotated $360^{\circ}$. 
%The angular alignment of the actuator was determined using radial alignment markings on the actuator and BL plate. 

The present experiment involved three common rectangular orifice configurations. 
In all cases the orifice had a throat width, $h$, of 1~mm and an $AR$ of 18. 
What differentiates the three configurations are the orifice pitch angle ($\alpha$, the angle between the centreline of the orifice channel and the cross flow surface) and skew angle ($\beta$, the angle between the orifice major axis and the spanwise direction) as highlighted in figure~\ref{fig:sjascheme}. 
Using this nomenclature the three configurations are (1) $\alpha=90^\circ$, $\beta=0^\circ$, a spanwise-oriented orifice with the flow exiting normal to the cross-flow surface, (2) $\alpha=90^\circ$, $\beta=90^\circ$, a streamwise-oriented orifice with the flow exiting normal to the cross-flow surface, and (3) $\alpha=45^\circ$, $\beta=0^\circ$, a spanwise-oriented orifice with the flow pitched $45^\circ$ relative to the cross-flow surface. 
Orifice geometries (1) and (2) are referred to as ``wall-normal orifices'' while (3) is called the ``pitched orifice''.
The origin of the coordinate system used in this study sits at the centroid of the orifice exit while the $x$, $y$ and $z$ directions align with the cross-flow, wall-normal and spanwise directions, respectively (figure~\ref{fig:sjascheme}). 
%The $\alpha=45^\circ, \beta=0^\circ$ orifice was designed with the orifice inlet offset in the upstream direction such that the centroid of the orifice exit was at the same point in space as the other orifices, with respect to the wind tunnel. 
%In addition, the depth of the orifices in the $y$-direction was 2.63~mm in all three cases, therefore the $\alpha=45^\circ, \beta=0^\circ$ orifice had a length in the jet flow direction which was 41\% longer than the $\alpha=90^\circ$ orifice.

\subsection{Actuator performance}
\label{sec:actuatorperf}

The piezoelectric disk was driven with a sinusoidal voltage waveform with a constant amplitude and frequency. 
To establish the SJA performance at different driving conditions, benchtop experiments were conducted prior to cross-flow testing. 
An A.A. Lab Systems AN-1003 constant-temperature anemometer system was used to measure the synthetic jet velocity inside the orifice. 
These measurements were made on the $\alpha=90^\circ$ orifice centreline with a Dantec miniature single-component hot-wire probe inserted to half the depth of the orifice. 
Simultaneously, the velocity of the underside of the piezoelectric disk was measured using a Polytec PDV 100 laser vibrometer. 
Using these two instruments the performance of the actuator was mapped for a range of driving voltages and frequencies. 

\begin{equation} 
\label{eq:Uo}
%U_o = \frac{1}{T} \int_{0}^{T/2} u_o (t) dt = f L_o
U_o = \frac{1}{T} \int_{0}^{T/2} u_o (t) dt
\end{equation}

The average blowing velocity, $U_o$, of a synthetic jet is obtained by integrating the instantaneous velocity of the jet at the orifice throat $u_o(t)$ over the blowing phase of the cycle and averaged over the period of the full cycle, $T$ (\ref{eq:Uo}). 
The average outstroke velocity of the piezoelectric disk, $U_d$, can be computed by applying the same calculation to the instantaneous velocity of the disk centre point. 
A linear relationship was observed between the average blowing velocity and the average disk outstroke velocity and this relationship remained consistent even after disassembly and reassembly of the actuator. 
Therefore, during cross-flow testing the synthetic jet performance was set by matching the average disk outstroke velocity which corresponded to the desired average blowing velocity. 

Measuring the average blowing velocity of the pitched $\alpha=45^\circ$ orifice jet was not attempted due to the geometric challenges of inserting the hot-wire probe into the pitched orifice and the fact that the jet velocity profile no longer had a top-hat shape. 
Instead, the volumetric flux out of the wall-normal and pitched orifices was matched by selecting actuator driving voltages which achieved the same average disk outstroke velocities. 
Therefore, the average blowing velocities reported for the pitched orifice cases are actually the values measured in the wall-normal orifice for the same volume of ejected fluid and driving frequency.
Disassembly and reassembly of the actuator to install different orifice plates produced small variations in the actuator performance, as did the interaction of the jet with different speed cross-flows. 
To address these sources of variability the actuator was calibrated in situ before every day of testing using the laser vibrometer. 
To enable these measurements, the floor of the wind tunnel and underside of the BL plate were fabricated from clear cast acrylic. 
The average disk outstroke velocity was mapped for a range of driving voltages at all planned cross-flow speeds and the driving voltage necessary to achieve the desired test points was then interpolated from the calibration data.

\subsection{SPIV measurements}
\label{sec:SPIV}

\begin{figure}
\centering
\includegraphics[width=5.25in]{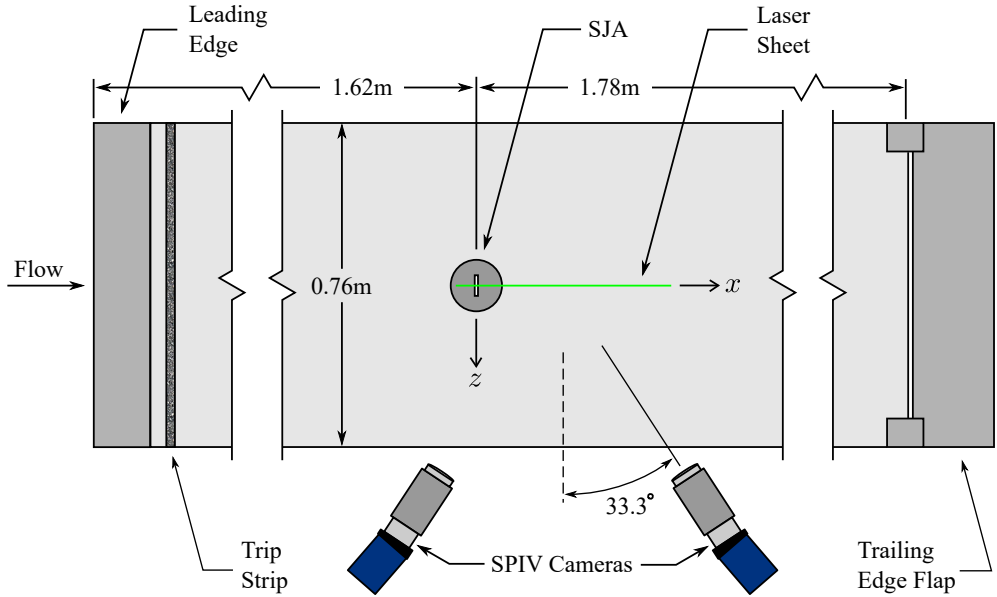}
\caption{Schematic of BL plate and SPIV measurement system.}
\label{fig:SPIVsetup}
\end{figure}

Stereoscopic particle image velocimetry (SPIV) was used to measure the three-component velocity field in the SJBLI experiments. 
The LaVision SPIV system was comprised of two 2560$\times$2160 pixel 16-bit dynamic range scientific CMOS cameras and a Quantel Evergreen 200 dual-pulsed 532~nm Nd:YAG laser. 
The laser was mounted above the test section on two Velmex linear traverses in a $y-z$ configuration. 
A series of optical elements were used to turn and focus the laser beam generating a thin sheet in the $x-y$ plane (figure~\ref{fig:SPIVsetup}).

The two SPIV cameras were mounted on one side of the test section with a three-axis ($x-y-z$) Velmex linear traverse system. 
One camera was positioned upstream of the actuator looking downstream towards the jet while the other was positioned downstream looking upstream, such that their orientations mirrored each other about the $x/h=40$ plane (figure~\ref{fig:SPIVsetup}). 
The $67^\circ$ of separation between the viewing angles of the two cameras enabled all three velocity components to be computed during processing of the raw images. 
Identical Nikon 105~mm 2.8D MicroNikkor lenses set to an aperture of f/8.0 were used on both cameras. 
The lenses were attached to the camera bodies by LaVision Scheimpflug mounts that allowed the full measurement plane to be brought into focus despite the off-axis arrangement of the cameras. 
The resulting image scale factor was 24.5 pixels~mm$^{-1}$.

Two types of SPIV measurements were obtained within the SJBLI domain: time-averaged and phase-locked data. 
For time-averaged measurements 500 instantaneous velocity fields were captured at a frequency which temporally aliased the actuation frequency. 
The result of averaging these 500 instantaneous measurements was the approximate mean jet flow field. 
Alternatively, in phase-locked measurements 250 instantaneous velocity fields were captured at a fixed phase angle of the actuation cycle. 
Averaging these measurements revealed the influence of the periodic actuation and the coherent structures at a particular instance in the actuation cycle. 
On the centreline of the jet eight evenly distributed actuator phase angles were obtained while off-centreline a subset of four actuator phases were measured.

Processing of the raw flow field images to obtain the 2D three-component vector field was conducted using the multi-pass stereoscopic cross-correlation algorithm in LaVision's DaVis 8.4.0 software. 
The first pass was run with 64$\times$64 pixel interrogation windows followed by two passes with 32$\times$32 pixel integration windows. 
A Gaussian weighting function was applied to the interrogation windows in the final pass and a 75\% overlap was used between the windows. 
The results after processing and masking was a 366$\times$253 vector velocity field with a vector resolution of 3.1 vectors per mm. 

\begin{figure}
\centering
\includegraphics[width=5.25in]{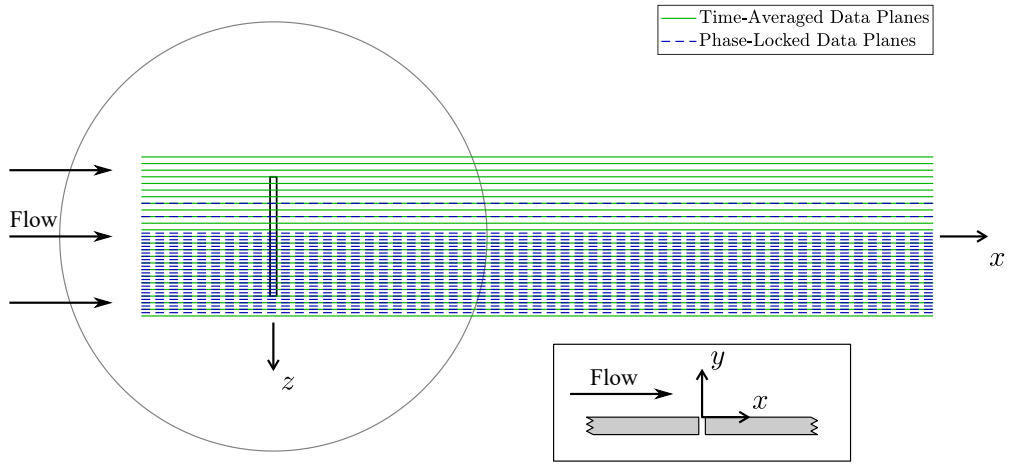}
\caption{SPIV measurement plane locations relative to the $\alpha=90^\circ$, $\beta=0^\circ$ orifice where solid green lines represent time-averaged data and dashed blue lines represent phase-locked data.}
\label{fig:SPIVplanes}
\end{figure}

The measured vector field after image alignment and cropping extended roughly $x=-19$ to 100~mm and $y=1$ to 83~mm, with the origin at the orifice centre. 
For off-centreline measurements the laser and cameras were traversed the same distance in the spanwise ($z$) direction such that the laser sheet and lens focal planes remained coplanar. 
Figure~\ref{fig:SPIVplanes} displays an example distribution of the time-average and phase-locked SPIV data planes in the jet interaction domain. 
The time-averaged data was obtained on planes spaced every 1~mm in the $z$-direction from $z=-$12~mm to 12~mm. 
Conversely, the phase-locked data was obtained every $\Delta z$ = 0.5~mm over a reduced span of the interaction domain to increase the data resolution within the coherent structures. 

In the $\alpha=90^\circ$, $\beta=0^\circ$ and $\alpha=45^\circ$, $\beta=0^\circ$ jets phase-locked data was obtained in one half of the interaction domain spanning from centreline out to a distance of $z=$11.5~mm. 
A subset of three more widely spaced planes were obtained on the opposite side of centreline from the high resolution data for the purposes of confirming the location of the jet plane of symmetry. 
To reconstruct the 3D flow field in these cases the data obtained in one half of the jet was mirrored onto the other side of the jet symmetry plane. 
In the $\alpha=90^\circ$, $\beta=90^\circ$ jet phase-locked data was obtained between $z=-$6~mm and 6~mm, therefore, no data mirroring was required for the 3D flow field reconstruction in this case.

%\begin{table}
%\centering
%\begin{tabular}{c c c}
%& Time-averaged & Phase-locked \\
%\hline
%$\sigma_u^{rms}/U_e$ & 0.26--0.45\% & 0.35--0.58\% \\ [1ex]
%$\sigma_v^{rms}/U_e$ & 0.14--0.48\% & 0.18--0.55\% \\ [1ex]
%$\sigma_w^{rms}/U_e$ & 0.18--0.39\% & 0.25--0.60\% \\ [1ex]
%\end{tabular}
%\caption{Uncertainty in SPIV measurements within interaction domain for all cases and conditions}
%\label{tab:uncert}
%\end{table}
\begin{table}
\centering
\begin{tabular}{c c c c}
& \multicolumn{1}{c}{Baseline} & \multicolumn{2}{c}{Actuation on} \\  [1.5ex]
\cline{2-4} \\  [-0.5ex]
& Time-averaged & Time-averaged & Phase-locked \\
\hline
$\sigma_u^{rms}/U_e$ & 0.19--0.28\% & 0.26--0.45\% & 0.35--0.58\% \\ [1ex]
$\sigma_v^{rms}/U_e$ & 0.09--0.15\% & 0.14--0.48\% & 0.18--0.55\% \\ [1ex]
$\sigma_w^{rms}/U_e$ & 0.13--0.21\% & 0.18--0.39\% & 0.25--0.60\% \\ [1ex]
\end{tabular}
\caption{Uncertainty in SPIV measurements within the interaction domain for all cases and conditions.}
\label{tab:uncert}
\end{table}

Uncertainty in the SPIV velocity measurements was quantified using the correlation statistics method implemented within DaVis~\citep{Wieneke_MeasSciTech_2015}. 
For reporting purposes the root-mean-square (RMS) of the uncertainty field within the SJBLI domain was calculated. 
The edge of this domain, $\delta_{99}(x)$, was defined as the height above the wall where the streamwise flow speed first dropped below $0.99U_e$ when moving from the upper edge of the measurement domain towards the wall, where $U_e$ is the velocity of the flow at the edge of the viscous BL. 
Restricting the RMS calculation to the interaction domain eliminated the outer flow velocity data where uncertainty was particularly low. 
The RMS values of the measurement uncertainty in the three-component directions are reported in table~\ref{tab:uncert} as percentages of $U_e$. 
The ranges represent the minimum and maximum RMS values observed in all of the averaged velocity fields obtained in the unactuated and actuated cases, inclusive of all measurement planes, phase angles, test conditions, and actuator configurations.

A number of steps were taken to mitigate laser reflections in the SPIV images. 
These included framing the SPIV images to cut-off the primary reflection on the BL plate and polishing build lines out of the SLA parts. 
An acrylic paint mixed with Rhodamine 6G powder, which absorbs 532~nm light and re-emits it at a lower wavelength, was also applied to the SLA parts. 
This coupled with band pass filters installed on the camera lenses helped reduce the intensity of laser reflections in the images. 
Two smaller reflections off of the inner edge of the ring retaining the SLA part were addressed in cases where they appeared in the 3D flow field reconstructions by either reflecting valid data from the opposite side of centreline or replacing bad vectors via interpolation. 

\subsection{Vortex identification and visualization}
\label{sec:vortexid}

Two methods were used to identify and visualize vortices in the SPIV data. 
To identify vortices in 2D planar data on the jet centreline the $\mathcal{Q}$–criterion was used, where $\mathcal{Q}$ is the second invariant of the velocity gradient tensor~\citep{Hunt_Wray_CTR_1988}. 
%$\mathcal{Q}$ was computed using (\ref{eq:Q}), where $||\bm{\Omega}||$ and $||\mathsfbi{S}||$ are the Frobenius norms of the rotation rate and strain rate tensors, respectively. 
Under this criterion a vortex is defined as a region of the flow where rotation dominates strain. 
%This allows the vorticity in the core of an established vortex to be distinguished from vorticity generated by shear stress in a BL, for example. 
%Although this criterion is technically satisfied when $\mathcal{Q}$ is greater than zero, a higher threshold was used to identify vortices in this study. 
Within a region of the $\mathcal{Q}$-field identified as containing a vortex, the vortex centre was defined as the location of the extremum in the vorticity field, obtained by linear interpolation of the discrete flow field data.
%
%\begin{equation} \label{eq:Q}
%\mathcal{Q} = \frac{1}{2}(u^2_{i,i}-u_{i,j}u_{j,i})=-\frac{1}{2}u_{i,j}u_{j,i}=\frac{1}{2}(||\bm{\Omega}||^2-||\mathsfbi{S}||^2)
%\end{equation}

The swirling strength criterion is used to visualize vortex structures in phase-locked 3D three-component velocity fields~\citep{Zhou_Adrian_JFM_1999}. 
This criterion examines the magnitude of the imaginary part of the velocity gradient tensor eigenvalues, denoted $\lambda_{ci}$ in (\ref{eq:lambdaci}). 
When $\lambda_{ci}$ is greater than zero swirling motion is present in the flow. 
Vortices may therefore be visualized by plotting isosurfaces at a chosen threshold of $\lambda_{ci}$. 
The particular thresholds used to generate the following figures were selected to highlight the structures of interest.

\begin{equation} \label{eq:lambdaci}
\mathrm{\bm{\nabla}} \bm{u} = [\bm{\nu_r} \; \bm{\nu_{cr}} \; \bm{\nu_{ci}}]\begin{bmatrix} \lambda_r&0&0\\
0&\lambda_{cr}&\lambda_{ci}\\
0&-\lambda_{ci}&\lambda_{cr}\end{bmatrix}[\bm{\nu_r} \; \bm{\nu_{cr}} \; \bm{\nu_{ci}}]^T
\end{equation}

Differentiation of the velocity field tends to amplify the noise within the SPIV data. 
To mitigate this effect the velocity gradient tensor used to compute $\lambda_{ci}$ was obtained using a non-local differentiation technique.
In this method the spatial derivate at a point is computed using the adjacent velocity field within a spherical stencil 1~mm in radius.
Additional details on this method are provided in \citet{Straccia_Farnsworth_JFM_2020}.

\subsection{Cross-flow conditions}
\label{sec:crossflowcond}

\begin{figure}
\centering
\includegraphics[width=2.5in]{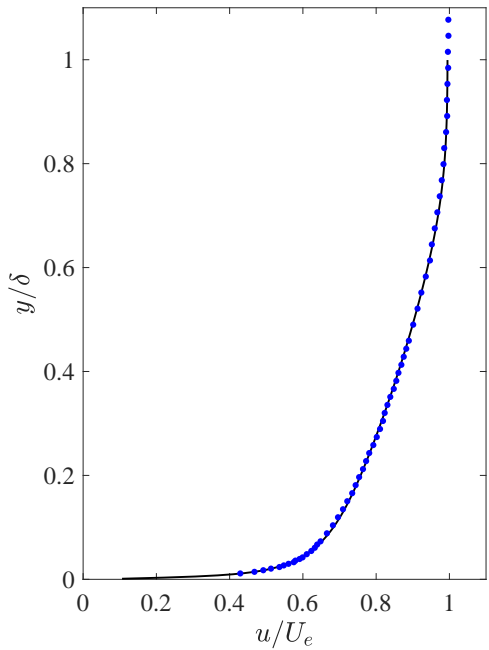}
\caption{Baseline turbulent BL profile at $x/h=0$, $z/h=0$ for $U_e$ = 12~ms$^{-1}$ (Hot-wire profile data, $\bullet$; Composite profile fit, \sampleline{}).}
\label{fig:BLprof}
\end{figure}

The baseline (no actuation) flow conditions and BL shape was characterized using a Dantec single element BL hot-wire probe which was traversed in the wall-normal direction. 
Data was obtained at three streamwise locations ($x=-$0.1, 0, 0.1~m) for all cross-flow speeds. 
To aid in assessing whether the measured BL profiles were consistent with a fully-developed zero pressure gradient turbulent BL the data was fit with the composite profile validated by \citet{Chauhan_Monkewitz_FluidDynRes_2009}. 
The fitting process also yielded estimates of the full BL thickness $\delta$ (i.e., the point on the edge of the BL where $u(y)=U_e$) and confirmed that wall-normal distances ($y$) were correctly zeroed at the wall~\citep{RodriguezLopez_Bruce_ExpFluids_2015}. 
Figure~\ref{fig:BLprof} presents an example fit of the $U_e$ = 12~ms$^{-1}$ BL at $x/h=0$. 
The remarkably good match between the data and the composite fit confirm that a canonical fully developed turbulent BL has been achieved. 
The fitting results were equally good at $U_e$ = 24~ms$^{-1}$. 

\begin{table}
\setlength{\tabcolsep}{6pt}
\caption{Conditions of the baseline cross-flow and boundary layer.}
\centering
\begin{tabular}{ccc}
\hline \hline
& $U_e$ = 12 ms$^{-1}$ & $U_e$ = 24 ms$^{-1}$  \\
\hline \\ [-2.0ex]
$\delta$ (mm) & 29.5 & 35.5 \\
$H$ & 1.400 & 1.358 \\
$Re_\tau$ & 837 & 1810 \\
$Re_\theta$ & 1780 & 4634 \\
$\beta_{PG}$ & -0.021 & -0.030 \\
$I_x$ & 0.27\% & 0.40\% \\
\hline \hline
\end{tabular}
\label{tab:crossflowcond}
\end{table}

The conditions of the baseline BL and outer flow above the orifice ($x$ = 0~mm) are summarized in table~\ref{tab:crossflowcond} for the two cross-flow speeds tested. 
The BL thickness values, $\delta$, were obtained as an output of the composite profile fitting process described earlier. 
Note that the BL at $U_e$ = 24~ms$^{-1}$ was slightly thicker than at $U_e$ = 12~ms$^{-1}$ because at $U_e$ = 24~ms$^{-1}$ the flow transitioned sooner and thus the BL grew at a faster rate for a longer streamwise distance. 
The shape of the BL is characterized using the shape factor, $H$, which is defined as the ratio between the displacement thickness $\delta^*$ and the momentum thickness $\theta$ (\ref{eq:H}).The shape factor values measured are consistent with a fully developed zero pressure gradient turbulent BL~\citep{Schlichting_2017}. 
The Reynolds number of the BL based on the friction velocity ($u_\tau$), defined in \ref{eq:Retau}, and the edge velocity ($U_e$), defined in \ref{eq:Retheta}, are also reported in table~\ref{tab:crossflowcond}. 
Finally, the freestream streamwise turbulence intensity ($I_x=\langle u'u' \rangle U_e^{-2}$), calculated from hot-wire anemometry data taken outside the BL ($y>2\delta$), is included.

\begin{equation} 
\label{eq:H}
H = \frac{\delta^*}{\theta} 
\end{equation}

\begin{equation} 
\label{eq:Retau}
Re_\tau = \frac{u_\tau \delta}{\nu} 
\end{equation}

\begin{equation} 
\label{eq:Retheta}
Re_\theta = \frac{U_e \theta}{\nu} 
\end{equation}

The pressure gradient in the wind tunnel test section was quantified using an array of twelve pressure taps running along the centreline of the BL plate. 
The Clauser pressure gradient parameter ($\beta_{PG}$) values computed using these data are listed in table~\ref{tab:crossflowcond}~\citep{Clauser_JAeroSci_1954}. 
The values indicate that the favorable pressure gradient that was present was very mild, which was expected given the excellent match between the BL profile measurements and the canonical zero pressure gradient turbulent BL profile from \citet{Chauhan_Monkewitz_FluidDynRes_2009}. 

\subsection{Synthetic jet-boundary layer interaction conditions}
\label{sec:SJBLIcond}

Several parameters are utilized in this paper to describe the interaction between the periodic jet and the cross-flow. 
The relative strength of the jet was quantified using blowing ratio, $C_b$, which relates the average blowing velocity of the jet to the velocity of the cross-flow (\ref{eq:Cb}). 
Low values of blowing ratio ($C_b<1$) describe cross-flow-dominated interactions while high values ($C_b>1$) correspond to jet-dominated flows. 
A common alternative to blowing ratio is momentum coefficient, $C_\mu$, which is the ratio of jet momentum to the cross-flow momentum (\ref{eq:Cmu}). 
Although the experiments were designed based on blowing ratio, momentum coefficient is also reported for convenience. 
To characterize the effect of actuation frequency the Strouhal number, $St$, which relates the actuation frequency to the freestream convective frequency, is also reported (\ref{eq:St}). 
As $St$ is increased coherent structures become more tightly spaced in the streamwise direction and the mutual interaction between coherent structures from different actuator cycles becomes stronger.

\begin{equation} 
\label{eq:Cb}
C_b = \frac{U_o}{U_e} 
\end{equation}

\begin{equation} 
\label{eq:Cmu}
C_{\mu} = \frac{U^2_o h}{U^2_e \delta} 
\end{equation}

\begin{equation} 
\label{eq:St}
St = \frac{f \delta}{U_o} 
\end{equation}

\begin{table}
\setlength{\tabcolsep}{6pt}
\caption{Synthetic jet-boundary layer interaction conditions tested.}
\centering
\begin{tabular}{ccccccc}
\hline \hline
Condition & $U_o$ (ms$^{-1})$ & $f$ (Hz) & $U_e$ (ms$^{-1}$) & $C_b$ & $C_\mu$ & $St$ \\
\hline \\ [-2.0ex]
1 & 12 & 540 & 12 & 1.0 & 0.0339 & 1.33\\
2 & 6 & 540 & 12 & 0.5 & 0.0085 & 1.33\\
3 & 12 & 540 & 24 & 0.5 & 0.0070 & 0.80\\
\hline \hline
\end{tabular}
\label{tab:caseconds}
\end{table}

The three SJBLI conditions investigated in this paper are summarized in table~\ref{tab:caseconds}. 
At the $C_b=1$ condition SPIV data was obtained on an array of parallel planes throughout the interaction domain such that the 3D flow field could be reconstructed. 
For the other two conditions where $C_b=0.5$ only centreline data are presented. 
The actuation frequency was held constant for all three conditions such that the Strouhal number at $U_e$ = 24~ms$^{-1}$ is roughly half that at $U_e$ = 12~ms$^{-1}$. 
%Therefore, the three conditions tested allow both the influence of blowing ratio (or momentum coefficient) to be compared with the effect of Strouhal number on the SJBLI.
Therefore, the three conditions tested allow both the influence of blowing ratio (or momentum coefficient) and Strouhal number to be investigated independently.
\section{SJBLI vortex dynamics at \texorpdfstring{$C_b=1$}{TEXT}}
\label{sec:results1}

The following section looks in detail at the 3D reconstructed flow field data for each of the three jet geometries at $C_b=1$ to establish a foundational understanding of the jet structure and behaviour. 

\subsection{The \texorpdfstring{$\alpha=90^\circ$}{TEXT}, \texorpdfstring{$\beta=0^\circ$}{TEXT} orifice SJA}
\label{sec:AR18Beta0}

\begin{figure}
\centering
\includegraphics[width=5.25in]{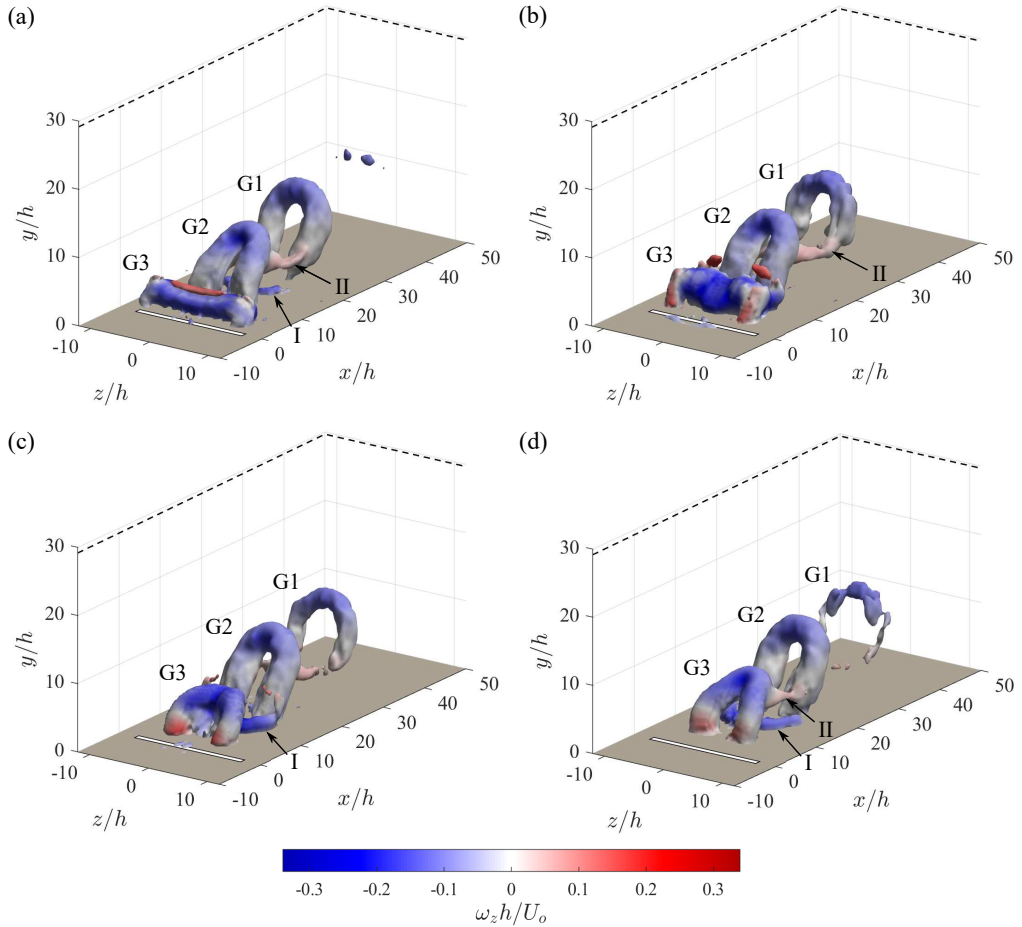}
\caption{Isosurfaces of phased-locked $\lambda_{ci}=0.5$ coloured by spanwise vorticity in the $\alpha=90^\circ$, $\beta=0^\circ$ jet at (a) $\phi=90^\circ$, (b) $\phi=180^\circ$, (c) $\phi=270^\circ$, and (d) $\phi=360^\circ$.}
\label{fig:AR18Beta0LambdaciQuad}
\end{figure}

A synthetic jet issuing from a $\alpha=90^\circ$, $\beta=0^\circ$ orifice interacts strongly with cross-flows due to the large spanwise extent of the interaction zone and the $90^\circ$ misalignment between the directions of the jet and the cross-flow. 
%The synthetic jet issuing from the $\alpha=90^\circ$, $\beta=0^\circ$ orifice interacts strongly with the cross-flow due to the large spanwise extent of the interaction zone and the $90^\circ$ misalignment between the directions of the jet and the cross-flow. 
This results in a large disturbance of the BL and a significant modification to the vortex structures.

\subsubsection{Vortex structure and development}
\label{sec:AR18Beta0vortex}

Figure~\ref{fig:AR18Beta0LambdaciQuad} presents phase-locked isosurfaces of $\lambda_{ci}$ for the purpose of visualizing the coherent structures produced by a $\alpha=90^\circ$, $\beta=0^\circ$ synthetic jet. 
The isosurfaces are coloured by spanwise vorticity ($\omega_z$) such that red represents positive or counterclockwise rotation (i.e., rotating counter to the BL vorticity), blue represents negative or clockwise rotation (i.e., rotating with the BL vorticity), and white represents regions with zero spanwise vorticity. 
For convenience, the different sections of a vortex ring will be distinguished by the direction of their rotation in the spanwise direction such that clockwise and counterclockwise rotating regions of the ring will be referred to as the CW and CCW side, respectively. 

The actuator phase angle, $\phi$, is quoted relative to the start of the blowing cycle ($\phi=0^\circ$), such that $90^\circ$ is roughly the point of peak blowing, $180^\circ$ is the start of the suction cycle, and $270^\circ$ is roughly the point of peak suction. 
The different generations of vortices are annotated at each phase angle in figure~\ref{fig:AR18Beta0LambdaciQuad}, with G1 being the eldest vortex visible and G3 being the youngest one.
Finally, the cross-flow direction is from bottom left to top right and the location of the orifice exit is indicated with a black outline.

\begin{figure}
\centering
\includegraphics[width=4in]{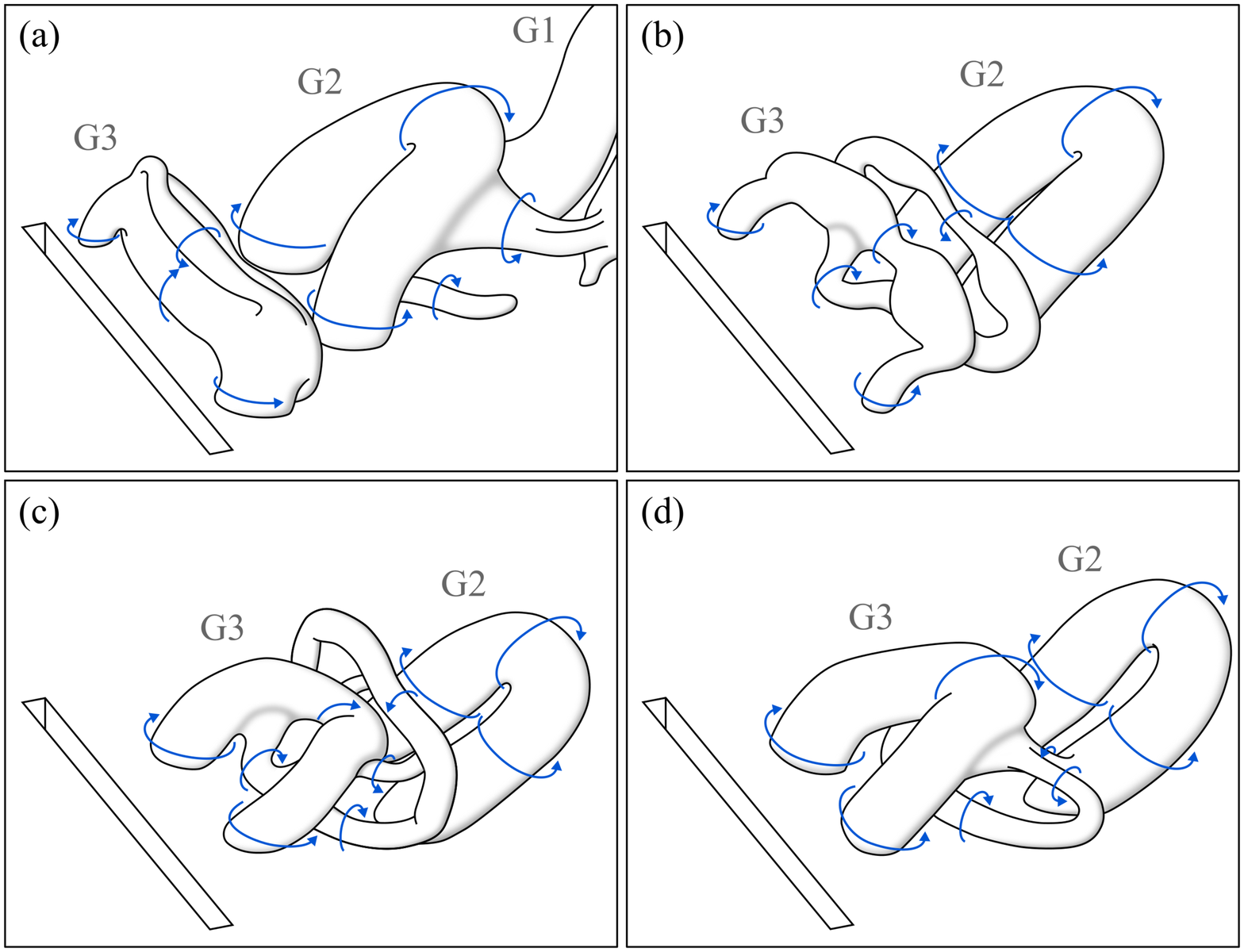}
\caption{Sketch of the vortices in the $\alpha=90^\circ$, $\beta=0^\circ$ jet with rotational direction indicated by blue arrows at (a) $\phi=180^\circ$, (b) $\phi=270^\circ$, (c) $\phi=360^\circ$, and (d) $\phi=90^\circ$.}
\label{fig:AR18Beta0Sketch}
\end{figure}

Obtaining a complete picture of the topology of the primary and secondary vortex structures and how they develop requires a detailed inspection of several parameters, from many viewpoints, and at numerous isosurface levels. 
Due to the infeasibility of presenting that quantity of data the authors have summarized their findings in a pictorial representation of the periodic structures (e.g., figure~\ref{fig:AR18Beta0Sketch}). 
These graphics are meant to conceptually relate our understanding of the flow field instead of faithfully reproducing a particular parameter at one specific contour level. 
%Therefore, nothing should be inferred from the relative width of the structures in these figures.

The periodic structure which forms during the blowing phase of the synthetic jet is a distorted vortex ring, visible just downstream from the orifice in figure~\ref{fig:AR18Beta0LambdaciQuad}a (G3). 
The upstream CCW rotating side of the vortex ring appears as a narrow discontinuous segment of the ring which has been diminished by vorticity cancelation with the counter-rotating BL vorticity. 
Conversely, the downstream CW rotating side of the vortex ring is a large robust structure, bolstered by the co-rotating BL vorticity. 
Note that the relatively stronger velocity induction from the dominant CW side of the vortex ring has already begun to rotate the weaker CCW segment off the wall and downstream. 
The balance of the circulation between the weak upstream side and the strong downstream side of the vortex ring is carried by short broad legs originating from a branch of the primary vortex tube at the spanwise ends of the vortex ring and terminating on their other end at the wall (figure~\ref{fig:AR18Beta0Sketch}a).

In the next phase of the cycle, the CW side of the vortex ring begins to deform due to self-induction (G3 in figure~\ref{fig:AR18Beta0LambdaciQuad}b). 
It is helpful to imagine the vortex legs together with the CW vortex as forming half of an elongated vortex ring, with the other half represented by the image vortex due to the effect of the wall. 
Thus the high curvature ends of this virtual vortex ring induce a velocity pointed in the upstream (negative $x$) direction which is stronger than the velocity induced locally on the roughly rectilinear mid-span portion of the ring. 
This differential causes the ends of the virtual ring to bend upstream and towards centreline in a manner similar to an axis-switching vortex ring propagating through quiescent fluid in the $-x$ direction \citep{Straccia_Farnsworth_PRF_2021}. 
Concurrently, the CCW side of the vortex ring continues to be rolled up and over the stronger CW vortex, and although only the curved ends of the structure are visible at this isosurface level, those segments are still connected in the spanwise direction by CCW vorticity (figure~\ref{fig:AR18Beta0Sketch}b).

\begin{figure}
\centering
\includegraphics[width=5.25in]{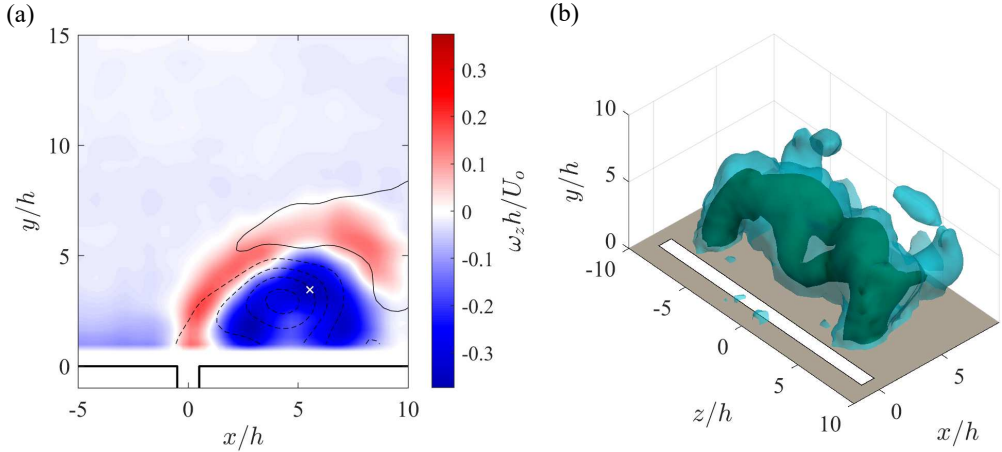}
\caption{Close up view of (a) coloured contours of spanwise vorticity overlaid with black contour curves of spanwise normal strain every 20\% of $|S_{zz}|_{max}$ from -90\% to 90\% (\sampleline{dashed}, $S_{zz}<0$; \sampleline{}, $S_{zz}>0$) and the location of the vortex centre ($\times$) on centreline and (b) isosurfaces of phased-locked $\lambda_{ci}$ at levels of 0.5 (aqua) and 1.75 (green) in the $\alpha=90^\circ$, $\beta=0^\circ$ jet at $\phi=180^\circ$.}
\label{fig:AR18Beta0Tight}
\end{figure}

%The structure of the G3 vortex in figure~\ref{fig:AR18Beta0LambdaciQuad}b is actually more complicated than it appears at this $\lambda_{ci}$ level. 
%The spanwise contraction of the virtual vortex ring induces compressive normal strain ($S_{zz}<0$) on centreline in the middle of the patch of CW vorticity associated with the vortex. 
%%This compressive strain causes a local drop in angular velocity and for the surrounding spanwise vorticity to spread out in the $x-y$ plane due to the requirements of continuity. 
%The resulting divergence in the $x-y$ plane causes a local drop in angular velocity and a spreading out of the surrounding spanwise vorticity due to the requirements of continuity.
%This phenomenon presents as a broadening of the CW vortex cross-section in the $y=x$ direction in figure~\ref{fig:AR18Beta0LambdaciQuad}b. 
%The net result is that a portion of the CW vorticity generated by the actuator outstroke splits off from the primary vortex and forms a secondary structure which remains close to the wall as it advects downstream (figure~\ref{fig:AR18Beta0Sketch}).
%The initial phase of this splitting process can be seen in the $\lambda_{ci}=1.75$ isosurface in figure~\ref{fig:AR18Beta0Tight}.
%
%\begin{figure}
%\centering
%\includegraphics[width=2.5in]{figures/AR18_beta0_540Hz_Uo12_Uinf12_3D3C_263deg_vortex_isosurface_closeup_thresh1_75_0_5_rfilter3D_r2w1_c.png}
%\caption{Close up view of isosurfaces of phased-locked $\lambda_{ci}$ at 0.5 (aqua) and 1.75 (green) in the $\alpha=90^\circ$, $\beta=0^\circ$ jet at $\phi=180^\circ$.}
%\label{fig:AR18Beta0Tight}
%\end{figure}
%
The structure of the G3 vortex in figure~\ref{fig:AR18Beta0LambdaciQuad}b is actually more complicated than it appears at this $\lambda_{ci}$ level. 
The spanwise contraction of the virtual vortex ring induces compressive normal strain ($S_{zz}<0$) on centreline in the middle of the patch of CW vorticity associated with the vortex (figure~\ref{fig:AR18Beta0Tight}a). 
The resulting divergence in the $x-y$ plane causes a local drop in angular velocity and a spreading out of the surrounding spanwise vorticity due to the requirements of continuity.
%This phenomenon presents as a broadening of the CW vortex cross-section in the $y=x$ direction in figure~\ref{fig:AR18Beta0LambdaciQuad}b. 
This phenomenon presents as a broadening of the CW vortex cross-section in figure~\ref{fig:AR18Beta0LambdaciQuad}b. 
The net result is that a portion of the CW vorticity generated by the actuator outstroke splits off from the primary vortex and forms a secondary structure which remains close to the wall as it advects downstream (figure~\ref{fig:AR18Beta0Sketch}).
The initial phase of this splitting process can be seen in the $\lambda_{ci}=1.75$ isosurface in figure~\ref{fig:AR18Beta0Tight}b.

In figure~\ref{fig:AR18Beta0LambdaciQuad}c the virtual vortex ring has continued to contract in the spanwise direction which introduces a bend in the vortex axis on centreline (see G3). 
The self-induced velocity of the CW rotating vortex is pointed upstream and away from the wall in the bend region and its magnitude increases as the bend becomes tighter (figure~\ref{fig:AR18Beta0Sketch}c).
%Thus a hairpin head forms in the vortex which lifts off of the wall. 
The progression of these deformations continue to follow those of the half-ring in the axis-switching analogy.

Meanwhile, the CW vorticity which split off from the primary vortex forms a secondary vortex close to the wall (see I in figure~\ref{fig:AR18Beta0LambdaciQuad}c). 
This upstream pointing chevron structure continues to be visible in the next two phases as it slowely advects downstream. 
At the ends of the chevron structure in figure~\ref{fig:AR18Beta0LambdaciQuad}c the vortex tube turns $90^\circ$ towards the wall-normal direction before turning again towards centreline to form the CCW branch of the original vortex ring (not visible at this $\lambda_{ci}$ level). 
In essence, this structure is like a secondary vortex ring which has partially separated from the upstream vortex and exists in the space near the wall between the different generations of the primary vortices (figure~\ref{fig:AR18Beta0Sketch}c). 

In the next few phases the primary structure continues to lift off the wall forming an arch-shaped vortex with a hairpin head and legs which terminate at the wall (see G2 in figure~\ref{fig:AR18Beta0LambdaciQuad}c). 
However, the topology of the legs begins to change by G2 in figure~\ref{fig:AR18Beta0LambdaciQuad}d. 
At this phase a new streamwise branch in the vortex tube appears in the $\lambda_{ci}$ isosurface which bridges the gap between the G2 and G3 vortices (see II in figure~\ref{fig:AR18Beta0LambdaciQuad}d). 
%The SPIV data implies that this branch appears due to complex distortion by tilting and stretching of the secondary vortex sitting in the interstitial space between the primary structures. 
%Where the secondary structure comes in proximity to the two primary structures vorticity reconnection occurs, forming a new interconnected topology. 
The SPIV data implies that this branch was created by a complex combination of tilting and stretching of the secondary vortex which sits in the interstitial space between the primary structures (figure~\ref{fig:AR18Beta0Sketch}c, d).
When the secondary structure comes close to the two primary structures, vorticity reconnection occurs, forming a new interconnected topology.

\begin{figure}
\centering
\includegraphics[width=5in]{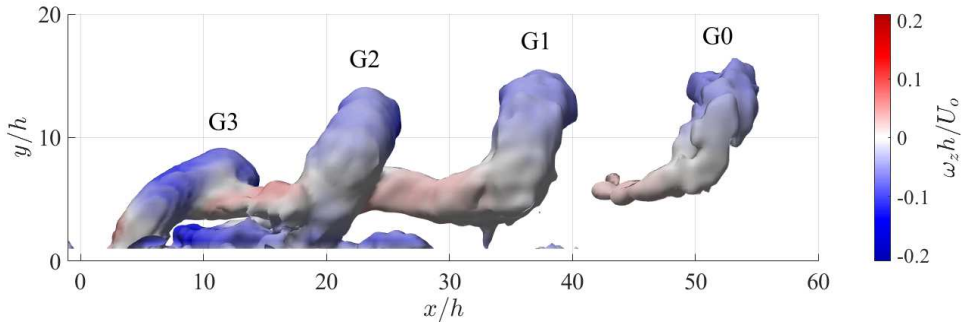}
\caption{Isosurfaces of phased-locked $\lambda_{ci}=0.25$ coloured by spanwise vorticity in the $\alpha=90^\circ$, $\beta=0^\circ$ jet at $\phi=360^\circ$ showing the train of hairpin vortices.}
\label{fig:AR18Beta0Train}
\end{figure}

Initially, the new streamwise branches are smaller and weaker than the legs of the arch-shaped vortex (see II in figure~\ref{fig:AR18Beta0LambdaciQuad}d), but as the structures travel downstream the streamwise branches grow at the expense of the original legs (see II in figure~\ref{fig:AR18Beta0LambdaciQuad}a) until they are the only set of legs visible (see II in figure~\ref{fig:AR18Beta0LambdaciQuad}b). 
Therefore, the primary vortex disconnects from the wall between $x/h=25$ and $x/h=35$ and its new streamwise-orientated legs terminate instead on the leeward side of the upstream vortex ring. 
In the subsequent phases this hairpin-like vortex continues to lift off the wall and eventually disappears from view at $\lambda_{ci}=0.5$ in figure~\ref{fig:AR18Beta0LambdaciQuad}.

In summation, the interaction of a $\alpha=90^\circ$, $\beta=0^\circ$ orifice jet with a cross-flow produces a train of vortices that look and behave much like hairpin vortex packets known to form naturally in turbulent BLs (figure~\ref{fig:AR18Beta0Train}). 
The role of natural hairpin vortices in driving sweeps and ejections that enhance wall-normal mixing makes these synthetic hairpin vortices a potentially desirable product of the SJBLI \citep{Adrian_PoF_2007}.

\subsubsection{Modification to velocity field}
\label{sec:AR18Beta0velocity}

\begin{figure}
\centering
\includegraphics[width=5.25in]{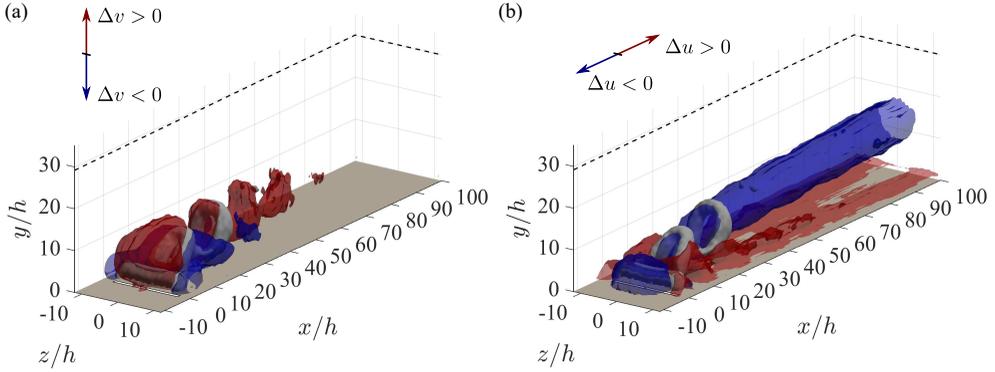}
\caption{Isosurfaces of phased-locked $\lambda_{ci}=0.5$ (grey) in the $\alpha=90^\circ$, $\beta=0^\circ$ jet at $\phi=90^\circ$ with isosurfaces of (a) $\Delta v/U_e=0.05$ (red), $\Delta v/U_e=-0.05$ (blue) and (b) $\Delta u/U_e=0.05$ (red), $\Delta u/U_e=-0.10$ (blue).}
\label{fig:AR18Beta0DeltaPavg}
\end{figure}

\begin{figure}
\centering
\includegraphics[width=5.25in]{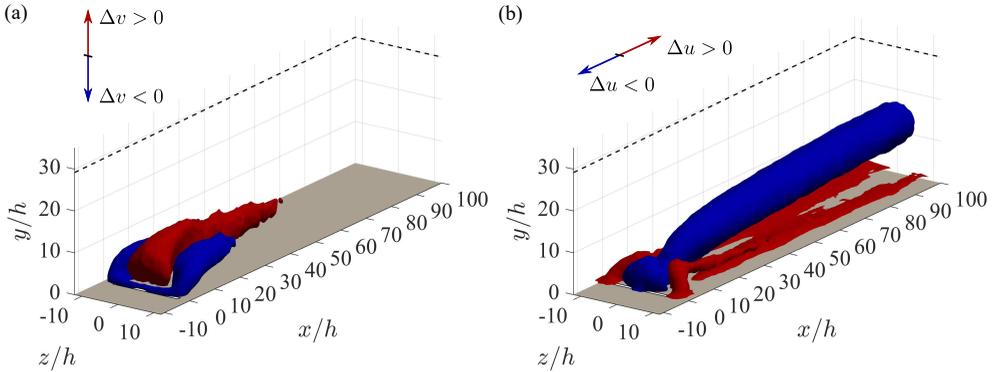}
\caption{Isosurfaces of time-averaged (a) $\Delta v/U_e=0.05$ (red), $\Delta v/U_e=-0.05$ (blue) and (b) $\Delta u/U_e=0.05$ (red), $\Delta u/U_e=-0.10$ (blue) in the $\alpha=90^\circ$, $\beta=0^\circ$ jet.}
\label{fig:AR18Beta0DeltaTavg}
\end{figure}

The influence of the flow control on the velocity field within the SJBLI domain is presented in figures \ref{fig:AR18Beta0DeltaPavg} and \ref{fig:AR18Beta0DeltaTavg}. 
Plotted are the normalized change in streamwise ($\Delta u/U_e$) and wall-normal velocities ($\Delta v/U_e$) which were obtained by subtracting the baseline velocity field (no-actuation) from the velocity field with actuation on. 
Accordingly, the red and blue isosurfaces enclose regions of the flow field where the flow velocity has increased or decreased, respectively, as a result of actuation. 
Figure~\ref{fig:AR18Beta0DeltaPavg} presents the phase-locked velocity field at a single representative actuator phase angle with gray isosurfaces of $\lambda_{ci}$ depicting the periodic coherent structures.
The $\Delta u/U_e$ and $\Delta v/U_e$ for the mean velocity field are also presented (figure~\ref{fig:AR18Beta0DeltaTavg}).
Additionally, the location of the edge of the BL, $\delta(x)$, in the baseline flow field is marked with a dashed line for reference. 

The hairpin vortices formed by the $\alpha=90^\circ$, $\beta=0^\circ$ SJA induce a velocity on centreline, below the head and between the legs, which is pointed upstream and away from the wall (figure~\ref{fig:AR18Beta0Sketch}).
The result is that every generation of vortex in the phase-locked flow field encloses a region of strong upwash (figure~\ref{fig:AR18Beta0DeltaPavg}a). 
This upwash along centreline is balanced by downward motion of the fluid surrounding the hairpin vortex, as indicated by the regions enclosed by the blue isosurfaces. 
In the mean flow field the average effect of the periodic structures advecting downstream is to lift low momentum fluid off the wall on centreline and entrain high momentum fluid towards the wall on both sides of the centreline (figure~\ref{fig:AR18Beta0DeltaTavg}a). 
In addition, a region of downward moving fluid which is generated during the suction cycle of the actuator appears on the upstream side of the orifice in the time-averaged flow field. 
Suction dominates blowing on average in this region of the mean flow field because the outward oriented momentum of the jet during the blowing cycle is turned downstream by the cross-flow.

The $\Delta u/U_e<0$ isosurfaces reveal that a low momentum wake with a roughly circular cross-section runs through the centre of the hairpin vortex heads (figure~\ref{fig:AR18Beta0DeltaTavg}b). 
An additional mound of low momentum fluid directly above the orifice forms during the blowing cycle of the actuator when fluid ejected from the orifice acts as an obstruction to the cross-flow. 
The effect of this viscous blockage is visible in the region of decelerated streamwise velocity directly above and upstream of the orifice in the phase-locked data (figure~\ref{fig:AR18Beta0DeltaPavg}b). 

Downstream of the orifice the velocity induced by the hairpin vortices on the surrounding fluid contributes to the development of a wake in two ways (figure~\ref{fig:AR18Beta0DeltaTavg}b). 
First, the upward velocity induced on centreline lifts low momentum fluid off the wall and deposits it deeper into the cross-flow. 
Second, the upstream pointing component of the induced velocity opposes the cross-flow at the centre of the hairpin heads, decelerating the fluid in the wake further (figure~\ref{fig:AR18Beta0Sketch}).
Therefore, the trajectory of the wake is inherently tied to the motion and changing shape of the hairpins vortices (figure~\ref{fig:AR18Beta0DeltaPavg}b).

On the sides of each hairpin vortex is an accompanying patch of high streamwise velocity introduced by the entrainment of high momentum fluid towards the wall by the vortices (figure~\ref{fig:AR18Beta0DeltaPavg}b). 
As the vortices travel downstream the regions of elevated streamwise velocity spread in the spanwise direction towards centreline due to induction from the hairpin vortex legs. 
Thus, the average effect of the actuation is to move low momentum fluid off the wall into a consolidated wake higher in the BL and to transport high momentum fluid down into a layer of accelerated flow close to the wall (figure~\ref{fig:AR18Beta0DeltaTavg}b).

\subsection{The \texorpdfstring{$\alpha=90^\circ$}{TEXT}, \texorpdfstring{$\beta=90^\circ$}{TEXT} orifice SJA}
\label{sec:AR18Beta90}

\begin{figure}
\centering
\includegraphics[width=5.25in]{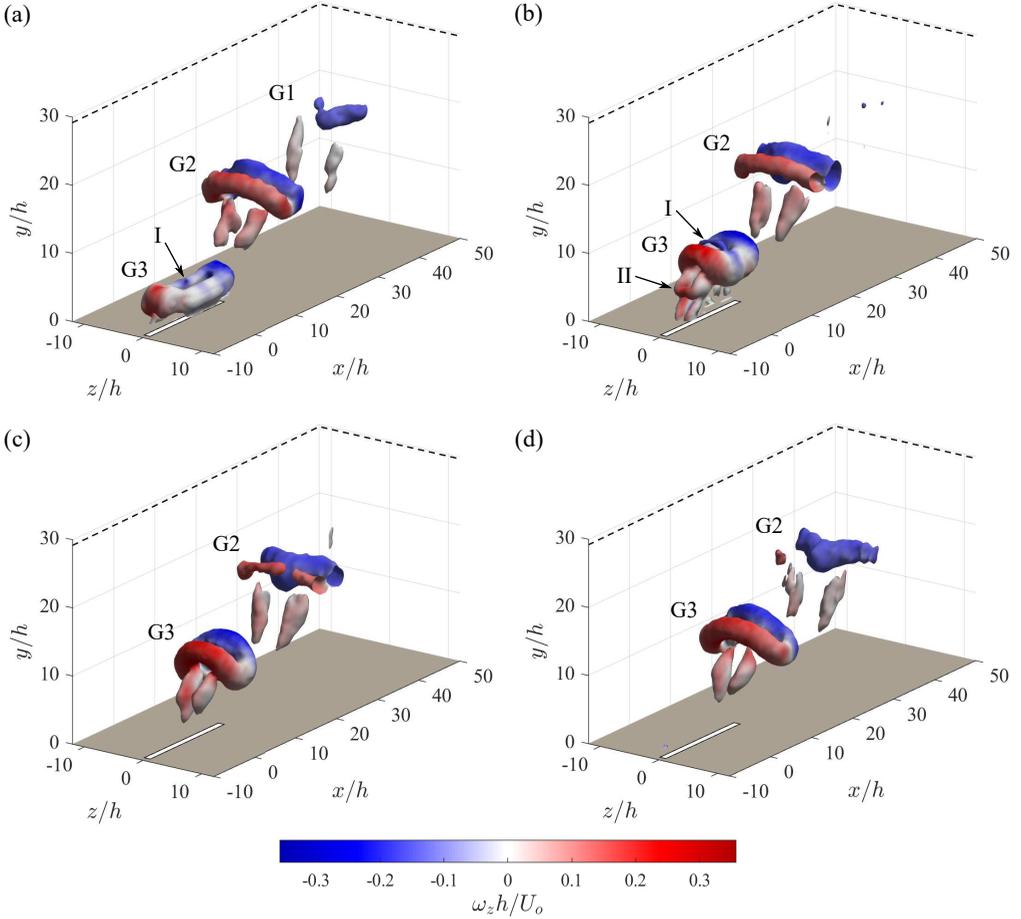}
\caption{Isosurfaces of phased-locked $\lambda_{ci}=0.75$ coloured by spanwise vorticity in the $\alpha=90^\circ$, $\beta=90^\circ$ jet at (a) $\phi=90^\circ$, (b) $\phi=180^\circ$, (c) $\phi=270^\circ$, and (d) $\phi=360^\circ$.}
\label{fig:AR18Beta90LambdaciQuad}
\end{figure}

With the actuator skewed $90^\circ$ such that the long axis of the orifice is aligned with the streamwise direction the initial jet presents a relatively small frontal area to the cross-flow, resulting in the formation of a more intact vortex ring than in the previous case. 

\subsubsection{Vortex structure and development}
\label{sec:AR18Beta90vortex}

\begin{figure}
\centering
\includegraphics[width=4in]{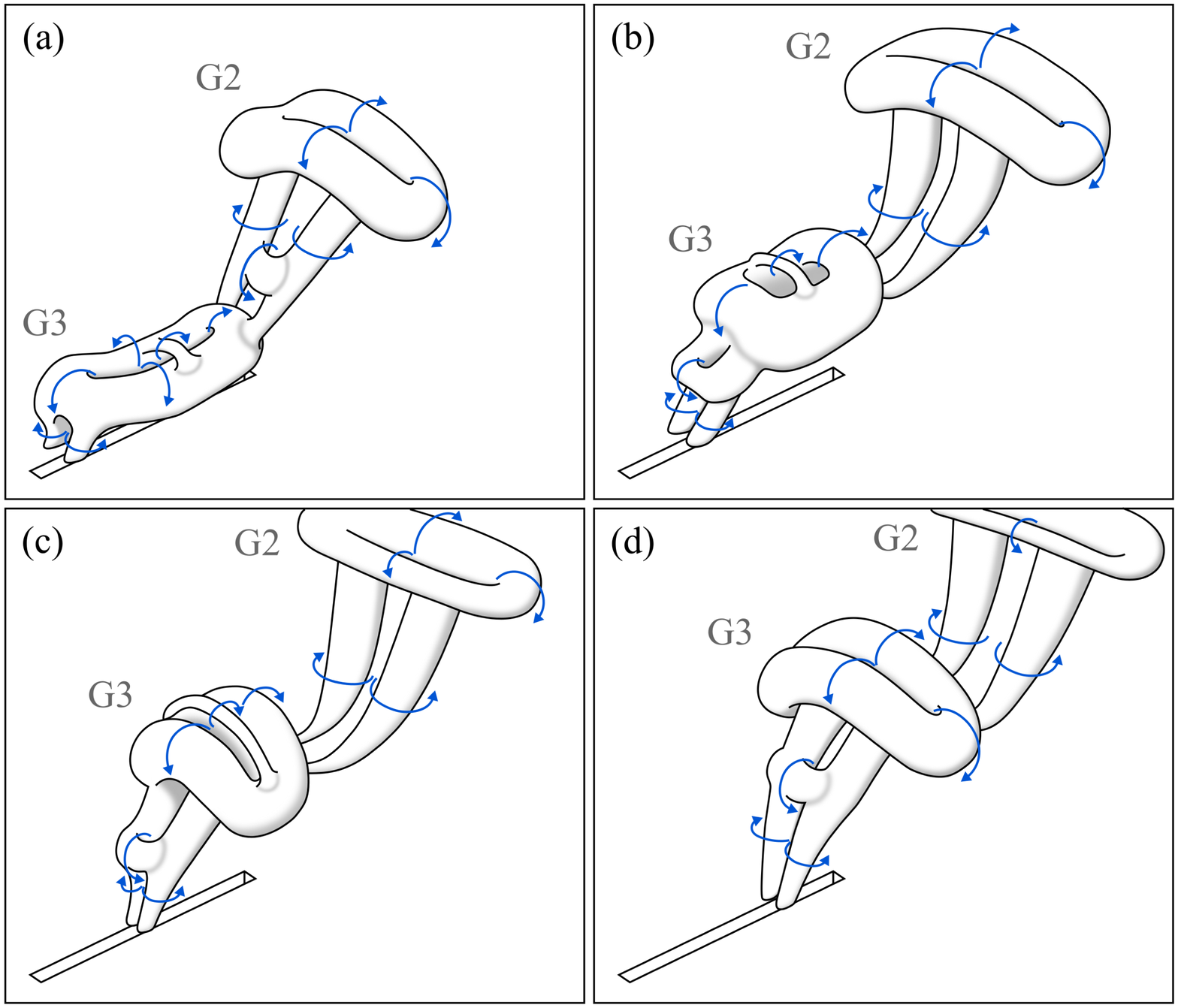}
\caption{Sketch of the vortices in the $\alpha=90^\circ$, $\beta=90^\circ$ jet with rotational direction indicated by blue arrows at (a) $\phi=90^\circ$, (b) $\phi=180^\circ$, (c) $\phi=270^\circ$, and (d) $\phi=360^\circ$.}
\label{fig:AR18Beta90Sketch}
\end{figure}

Figure~\ref{fig:AR18Beta90LambdaciQuad} depicts the coherent structures at four phases of the $\alpha=90^\circ$, $\beta=90^\circ$ jet. 
A pictorial representation of the vortex structures is presented in figure~\ref{fig:AR18Beta90Sketch}.
%The elongated vortex ring forming above the orifice in figure~\ref{fig:AR18Beta90LambdaciQuad}a (G3) looks much as it would have if there had been no cross-flow \citep{Straccia_Farnsworth_JFM_2020}. 
The elongated vortex ring forming above the orifice in figure~\ref{fig:AR18Beta90LambdaciQuad}a (G3) looks much as it would have if there had been no cross-flow.
The ends of the vortex ring have even started to lift away from the wall ahead of the ring centre due to the higher local self-induced velocity in those tightly curved segments of the ring. 
There are some subtle indications, however, that the vortex ring has been affected by the cross-flow. 
For one, the vortex ring has translated downstream a short distance. 
In addition, the velocity gradient in the BL has altered the vortex ring topology slightly. 
In particular, the upstream end of the vortex ring is slightly diminished in strength by vorticity cancelation with the BL, and from the upstream corners of the ring two short legs branch off carrying the balance of the circulation to the wall (figure~\ref{fig:AR18Beta90Sketch}a).

The legs become more evident at $\phi=180^\circ$ (G3 in figure~\ref{fig:AR18Beta90LambdaciQuad}b). 
It is also clear that the vortex ring is undergoing typical axis-switching deformation which has caused the ring to shrink in the $x$-direction and to grow slightly in the $z$-direction. 
In addition, two more structures are now visible on the vortex ring. 
The first is a CW rotating vortex running across the middle of the ring (see I in figure~\ref{fig:AR18Beta90LambdaciQuad}b). 
This structure is also present in figure~\ref{fig:AR18Beta90LambdaciQuad}a but is not visible at this isosurface level, except for a small stub annotated I (see figure~\ref{fig:AR18Beta90Sketch}a for the full structure).
Similar structures were detected in quiescent conditions by \citet{Straccia_Farnsworth_JFM_2020} which they called `formation bridges'. 
Like the bridge vortices formed by vorticity reconnection during the collision of a vortex ring with itself after axis-switching, formation bridges occur when the two sides of a wide-cored vortex ring come in contact during the vortex's formation. 
SPIV data obtained by \citet{Housley_PhDThesis_2020} in a similar experiment reveal that this initial vorticity reconnection event can be strong enough to cause the vortex ring to bifurcate into two smaller rings. 
What is curious on first look is the absence of a counter-rotating (CCW) formation bridge which should accompany the CW-rotating vortex. 
A closer look at the vorticity field on the jet centreline suggests that the reconnected CCW vorticity sits below the visible CW formation bridge in the central region of the vortex ring where the high induced strain prevents its detection by vortex identification schemes like $\mathcal{Q}$-criterion and $\lambda_{ci}$. 
This proposed topology appears to be supported by figure 3.44 in \citet{Housley_PhDThesis_2020}. 

The other structure visible on the vortex ring in figure~\ref{fig:AR18Beta90LambdaciQuad}b is a vortex running spanwise between the legs below the primary vortex ring (annotated II). 
Similar ladder-rung-like structures between the legs were detected by \citet{Wang_Feng_JFM_2020} on lower $AR$ vortex rings in a laminar BL. 
The interlinking structure rotates in the same direction as the upstream side of the vortex ring and appears to develop after the vortex ring.
It is likely that this vortex forms from trailing vorticity generated later in the actuator stroke that is not incorporated into the primary vortex ring, possibly due to a modified form of vortex pinch-off \citep{Gharib_Rambod_JFM_1998}. 
Such secondary vortices have been shown to form in synthetic jets issuing into quiescent fluid under the right conditions \citep{Straccia_Farnsworth_JFM_2020}. 
The CCW rotating interlinking vortex is detectable, although not visible in figure~\ref{fig:AR18Beta90LambdaciQuad}, for the next three phase-angles until it becomes too weak to distinguish from background noise by G2 at $\phi=180^\circ$. 
In addition to increasing unsteady mixing below the primary vortex ring, the interlinking vortex also induces a downstream pointing velocity component near the wall (figure~\ref{fig:AR18Beta90Sketch}b).
%
%\begin{figure}
%\centering
%\includegraphics[width=3in]{figures/AR18_beta90_540Hz_Uo12_Uinf12_3D3C_262deg_vortex_isosurface_closeup_thresh0_75_rfilter3D_r2w1_c.png}
%\caption{Close up view of isosurfaces of phased-locked $\lambda_{ci}=0.75$ coloured by spanwise vorticity in the $\alpha=90^\circ$, $\beta=90^\circ$ jet at $\phi=180^\circ$.}
%\label{fig:AR18Beta90Tight}
%\end{figure}

At $\phi=270^\circ$ the formation bridge is still stretched between the ever widening spanwise ends of the G3 vortex ring, although it is not visible in the figure~\ref{fig:AR18Beta90LambdaciQuad}c isosurface (see figure~\ref{fig:AR18Beta90Sketch}c for the full structure). 
The vortex ring continues to advect downstream and propagate away from the wall while widening in the spanwise direction due to the self-induced axis-switching deformations (figure~\ref{fig:AR18Beta90LambdaciQuad}d). 
When the subsequent vortex ring forms, the legs of the older G2 ring terminate slightly upstream of the end of the orifice. 
The result is that the legs from the prior vortex ring end up becoming tied to the downstream end of the new ring (figure~\ref{fig:AR18Beta90Sketch}a). 
Initially, this branch in the vortex tube is located towards the upper surface of the new vortex ring, but as the strong induced velocity around the ring acts, this connection point migrates to the underside of the upstream vortex ring.

The spanwise-oriented G2 vortex ring and its legs are visible downstream of the youngest ring in all four panels of figure~\ref{fig:AR18Beta90LambdaciQuad}. 
The vortex ring eventually extends past the edge of the measurement domain and then decays to a point where it is undetectable at the isosurface level set in figure~\ref{fig:AR18Beta90LambdaciQuad}, starting with the weaker CCW rotating side first. 
The vortex ring appears to roughly complete the axis-switching cycle by G2 in figure~\ref{fig:AR18Beta90LambdaciQuad}b. 
After that point the vortex rings deform very little and instead the older structures advect along as a train of spanwise rollers loosely tied together by their legs.

\begin{figure}
\centering
\includegraphics[width=5.25in]{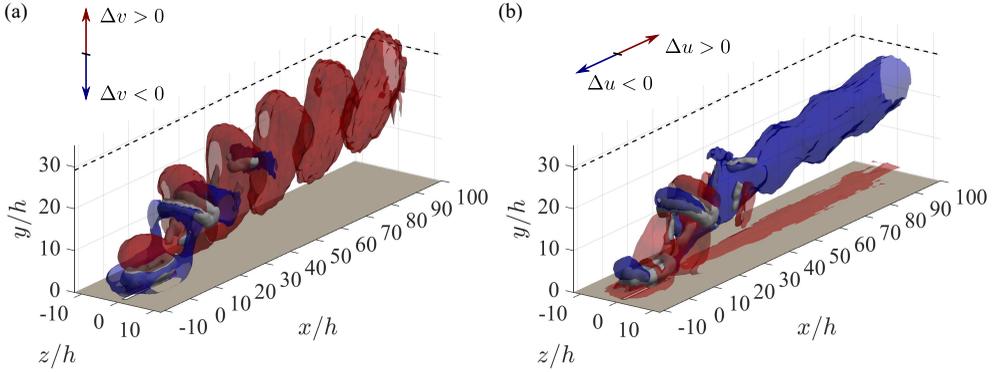}
\caption{Isosurfaces of phased-locked $\lambda_{ci}=0.75$ (grey) in the $\alpha=90^\circ$, $\beta=90^\circ$ jet at $\phi=90^\circ$ with isosurfaces of (a) $\Delta v/U_e=0.05$ (red), $\Delta v/U_e=-0.05$ (blue) and (b) $\Delta u/U_e=0.05$ (red), $\Delta u/U_e=-0.10$ (blue).}
\label{fig:AR18Beta90DeltaPavg}
\end{figure}

\subsubsection{Modification to velocity field}
\label{sec:AR18Beta90velocity}

\begin{figure}
\centering
\includegraphics[width=5.25in]{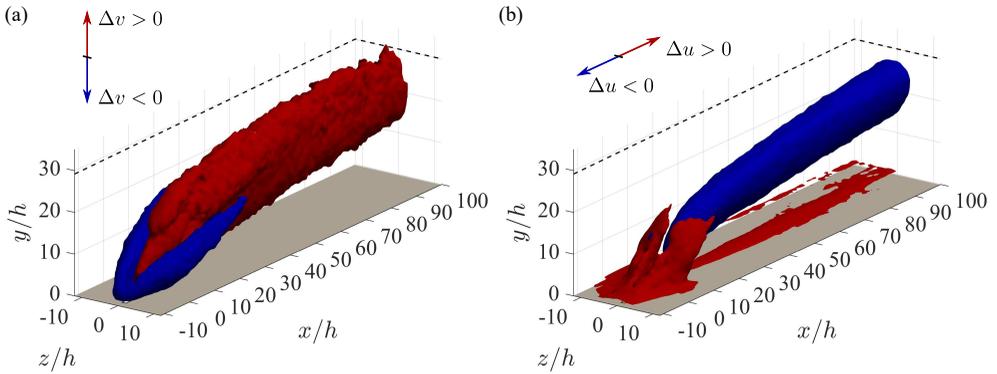}
\caption{Isosurfaces of time-averaged (a) $\Delta v/U_e=0.05$ (red), $\Delta v/U_e=-0.05$ (blue) and (b) $\Delta u/U_e=0.05$ (red), $\Delta u/U_e=-0.10$ (blue) in the $\alpha=90^\circ$, $\beta=90^\circ$ jet.}
\label{fig:AR18Beta90DeltaTavg}
\end{figure}

The effects that the $\alpha=90^\circ$, $\beta=90^\circ$ SJA has on the velocity field are similar to those of the $\alpha=90^\circ$, $\beta=0^\circ$ SJA in a general sense, but the strengths and locations of particular features are quite different. 
An especially strong upwash is induced at the centre of each vortex ring in this case because the collective action of the counter-rotating sides of the vortex ring results in an induced velocity vector that points predominantly in the wall-normal direction (figure~\ref{fig:AR18Beta90DeltaPavg}a). 
The vortex rotation also induces downward flow in the fluid around the circumference of the vortex ring (figure~\ref{fig:AR18Beta90Sketch}a).
In the time-averaged field the $\Delta v/U_e=0.05$ isosurface enclosing the upward moving fluid on the jet centreline extends to the end of the measurement domain thanks to the strength and longevity of the vortex rings (figure~\ref{fig:AR18Beta90DeltaTavg}a). 
 
A pronounced wake forms in the streamwise velocity field of the jet (figure~\ref{fig:AR18Beta90DeltaPavg}b). 
Upwash of low momentum fluid from lower in the BL is again a factor in producing the wake. 
In addition, the pair of legs induce a velocity that is pointed partially upstream, counter to the cross-flow direction, as does the underside of the dominant clockwise rotating side of the vortex ring (figure~\ref{fig:AR18Beta90Sketch}a).
The wake which results from the combined upwash and velocity induction has a mean cross-sectional shape at the end of the measurement domain which is broader at the top (due to the spanwise orientation of the vortex ring) and narrower at the bottom (due to the small gap between the legs) (figure~\ref{fig:AR18Beta90DeltaTavg}b). 
Notably absent from both the phase-locked and time-averaged flow fields is the region of low streamwise momentum fluid upstream of the orifice and near the wall which was seen in the $\alpha=90^\circ$, $\beta=0^\circ$ jet. 
This is because the streamwise-oriented jet only minimally obstructs the cross-flow and therefore there is less viscous blockage introduced into the flow than there is by the spanwise-oriented jet.

Streamwise velocity is enhanced above the CW rotating sides of the vortex ring, above the formation bridge, and on the outer sides of the legs (figures~\ref{fig:AR18Beta90Sketch}a and \ref{fig:AR18Beta90DeltaPavg}b). 
This is the reason for the sheath of elevated streamwise velocity upstream and to the sides of the wake near the orifice in the mean flow (figure~\ref{fig:AR18Beta90DeltaTavg}b). 
Further downstream a streak of elevated streamwise velocity forms near the wall and extends to the end of the measurement domain (figure~\ref{fig:AR18Beta90DeltaTavg}b). 
This high velocity stream is formed by the entrainment of fluid from higher in the BL along either side of the jet which rushes in to fill the space left by the upwash on centreline. 

\subsection{The \texorpdfstring{$\alpha=45^\circ$}{TEXT}, \texorpdfstring{$\beta=0^\circ$}{TEXT} orifice SJA}
\label{sec:AR18P45Beta0}

\begin{figure}
\centering
\includegraphics[width=5.25in]{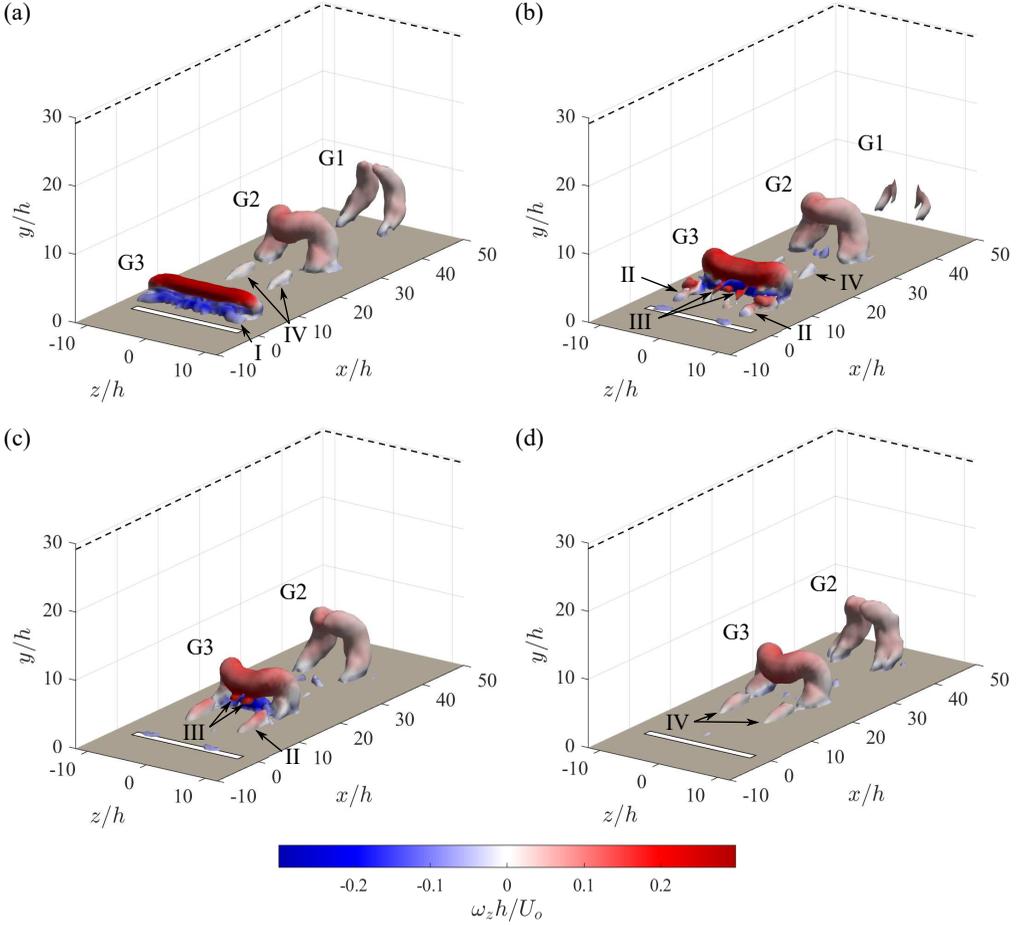}
\caption{Isosurfaces of phased-locked $\lambda_{ci}=0.5$ coloured by spanwise vorticity in the $\alpha=45^\circ$, $\beta=0^\circ$ jet at (a) $\phi=90^\circ$, (b) $\phi=180^\circ$, (c) $\phi=270^\circ$, and (d) $\phi=360^\circ$.}
\label{fig:AR18P45Beta0LambdaciQuad}
\end{figure}

Pitching a spanwise-oriented orifice $45^\circ$ downstream reduces the intensity of the SJBLI and gives the momentum introduced by the jet a streamwise component. 
The result is a dramatically different vortex structure and flow field, as compared with the wall-normal jets. 

\subsubsection{Vortex structure and development}
\label{sec:AR18P45Beta0vortex}

\begin{figure}
\centering
\includegraphics[width=4.25in]{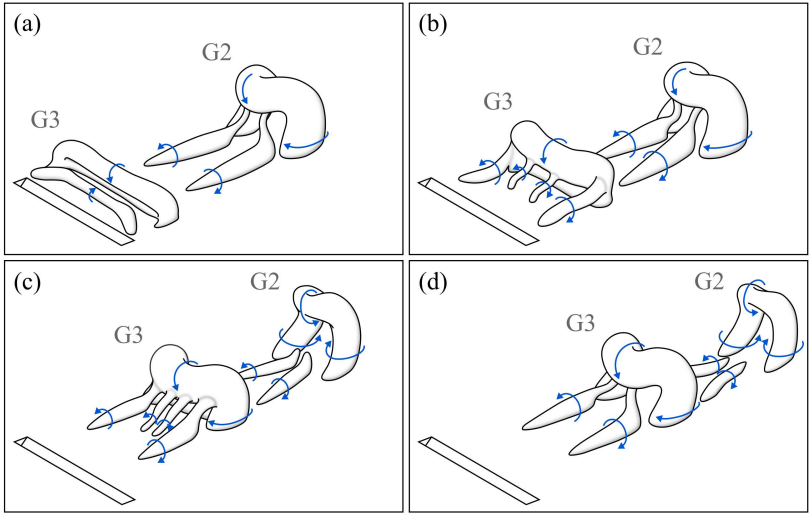}
\caption{Sketch of the vortices in the $\alpha=45^\circ$, $\beta=0^\circ$ jet with rotational direction indicated by blue arrows at (a) $\phi=90^\circ$, (b) $\phi=180^\circ$, (c) $\phi=270^\circ$, and (d) $\phi=360^\circ$.}
\label{fig:AR18P45Beta0Sketch}
\end{figure}

The coherent structures produced by the $\alpha=45^\circ$, $\beta=0^\circ$ jet are depicted in the $\lambda_{ci}$ isosurfaces of figure~\ref{fig:AR18P45Beta0LambdaciQuad} for four evenly spaced actuation phase angles. 
A pictorial representation of the vortex structures is presented in figure~\ref{fig:AR18P45Beta0Sketch}.

\begin{figure}
\centering
\includegraphics[width=3in]{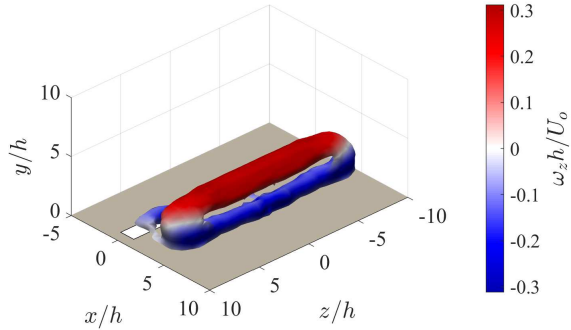}
\caption{Close up view of isosurfaces of phased-locked $\lambda_{ci}=0.75$ coloured by spanwise vorticity in the $\alpha=45^\circ$, $\beta=0^\circ$ jet at $\phi=90^\circ$.}
\label{fig:AR18P45Beta0Tight}
\end{figure}

The G3 vortex ring forming near the orifice in figure~\ref{fig:AR18P45Beta0LambdaciQuad}a is tilted downstream due to the pitching of the jet. 
Pitching the jet downstream enhances the circulation strength in the upstream side of the vortex ring, offsetting the vorticity cancelation between the CCW rotating vortex and the CW rotating BL vorticity. 
Simultaneously, the downstream CW rotating side of the vortex ring is weakened in the pitched jet. 
As a result, the flow field within the SJBLI domain is dominated by a CCW rotating vortex in this case. 
The CW side of the new vortex ring is partially obscured by the CCW rotating vortex in figure~\ref{fig:AR18P45Beta0LambdaciQuad}a but can be seen more clearly in figure~\ref{fig:AR18P45Beta0Tight}. 
%A secondary patch of $\lambda_{ci}$ upstream of the vortex ring in figure~\ref{fig:AR18P45Beta0LambdaciQuad}a is additional CW vorticity from the actuator outstroke that was not incorporated into the primary vortex. 
Note that there is a secondary patch of $\lambda_{ci}$ upstream of the vortex ring in figure~\ref{fig:AR18P45Beta0LambdaciQuad}a marking the location of additional CW vorticity from the actuator outstroke that was not incorporated into the primary vortex (see I in figure~\ref{fig:AR18P45Beta0LambdaciQuad}a). 

As the vortex ring propagates downstream the ends of the ring outrun the middle due to the higher local self-induction (G3 in figure~\ref{fig:AR18P45Beta0LambdaciQuad}b). 
Meanwhile, the trailing vorticity upstream of the vortex ring tilts into the streamwise direction and begins to coalesce into two pairs (four total) of counter-rotating streamwise vortices (see II and III in figure~\ref{fig:AR18P45Beta0LambdaciQuad}b and proposed topology in figure~\ref{fig:AR18P45Beta0Sketch}b). 
The inner vortex pair appears to form from tilted and consolidated vorticity in the trailing CCW rotating shear layer and is connected to the upstream side of the CCW rotating primary vortex. 
The outer vortex pair may form out of the CW vorticity identified in figure~\ref{fig:AR18P45Beta0LambdaciQuad}a, which can already be seen tilting into the streamwise direction at its ends. 
How the outer vortex pair connects to the primary ring is difficult to precisely resolve in the present data, but the connection points appear to initially be near the ends of the CW rotating side of the ring. 

By $\phi=270^\circ$ the CW-rotating side of the G3 vortex ring is greatly dissipated and is becoming difficult to detect in the $\lambda_{ci}$ field (figure~\ref{fig:AR18P45Beta0LambdaciQuad}c). 
The rest of the vortex ring has continued to progress through the typical axis-switching deformations, and a highly curved head has started to form along centreline. 
Upstream of the vortex ring the outer pair of secondary streamwise vortices are clearly visible (annotation II) while the inner pair show up only as small knobs that protrude upstream from the CCW-rotating side of the primary vortex near the centreline (annotation III). 

\begin{figure}
\centering
\includegraphics[width=5in]{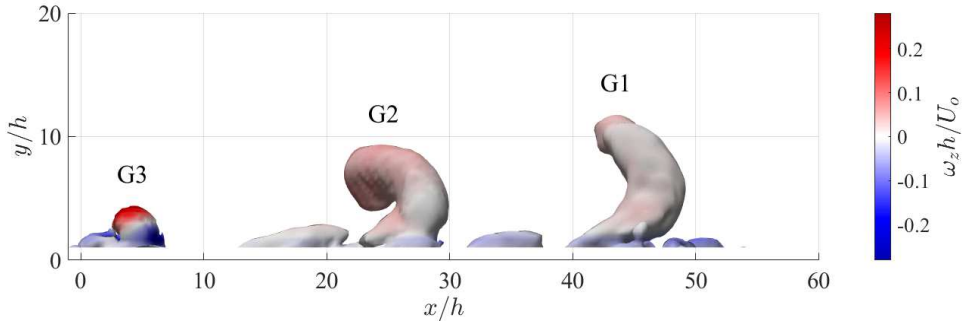}
\caption{Isosurfaces of phased-locked $\lambda_{ci}=0.3$ coloured by spanwise vorticity in the $\alpha=45^\circ$, $\beta=0^\circ$ jet at $\phi=90^\circ$ showing the train of arch-shaped vortices.}
\label{fig:AR18P45Beta0Train}
\end{figure}

In figure~\ref{fig:AR18P45Beta0LambdaciQuad}d the CW side of the G3 vortex is no longer detected and the primary structure now takes the form of an arch-shaped vortex with its apex leaned into the flow and its legs terminating at the wall. 
At this time only a single set of streamwise secondary vortices remain. 
The vortex tube topology at $\phi=360^\circ$ implies that the inner vortex pair have merged with their co-rotating partners from the outer vortex pair, at least at the surface of the primary vortex. 
Vorticity annihilation across centreline may also play a role in the disappearance of the inner vortex pair closer to the wall. 
The remaining streamwise secondary vortices (annotation IV) may appear to be extensions of the G3 primary vortex's legs in figure~\ref{fig:AR18P45Beta0LambdaciQuad}d but in fact the streamwise vorticity in the secondary structure is counter-rotating to that of the primary vortex. 
Instead, the secondary structures are branches off the primary structure and have a rotational direction which indicates that the head of the arch-shaped vortex has lower circulation strength than the legs do (figure~\ref{fig:AR18P45Beta0Sketch}b). 
This may explain why the apex of the arch-shaped vortex is the first to drop below the isosurface level for both vorticity magnitude and $\lambda_{ci}$ (see, for example, the broken isosurface of the G2 vortex in figure~\ref{fig:AR18P45Beta0LambdaciQuad}d and the G1 vortex in figure~\ref{fig:AR18P45Beta0LambdaciQuad}a).

At later phase angles the primary vortex continues to deform in such a way that the apex of the arch lifts up from the wall and rotates downstream relative to the legs due to its self-induced velocity (figure~\ref{fig:AR18P45Beta0Train}). 
The more upright arch-shaped vortex also becomes narrower in the spanwise direction. 
This geometric progression is similar to the deformations of an axis switching vortex ring traveling roughly in the $x$-direction. 
Compared to the other two orifice geometries, it is noteworthy that the vortices in the present case remain tied to the wall via their legs for as long as they are detectable, instead of detaching from the wall. 
Furthermore, while the $\alpha=90^\circ$, $\beta=0^\circ$ SJA produced a train of hairpin-shaped vortices, the somewhat similar train of arch-shaped vortices in the $\alpha=45^\circ$, $\beta=0^\circ$ jet rotate in the opposite direction to the hairpin structures (compare figures \ref{fig:AR18Beta0Train} and \ref{fig:AR18P45Beta0Train}). 
The secondary streamwise vortex pair advect downstream along the wall, located in the space between subsequent vortex generations (annotation IV). 
As it travels the streamwise pair also converges in the spanwise direction, a behaviour that is to be expected based on its direction of rotation and the wall-induced image vortex effect (figure~\ref{fig:AR18P45Beta0Sketch}a).
\citet{Housley_PhDThesis_2020} also detected a streamwise vortex pair in a SJBLI with a pitched orifice, and they speculated that this structure may have originated as the CW side of the vortex ring; however, they were unable to resolve that side of the vortex ring because of the gap between their lowest SPIV measurement and the wall.

\subsubsection{Modification to velocity field}
\label{sec:AR18P45Beta0velocity}

\begin{figure}
\centering
\includegraphics[width=5.25in]{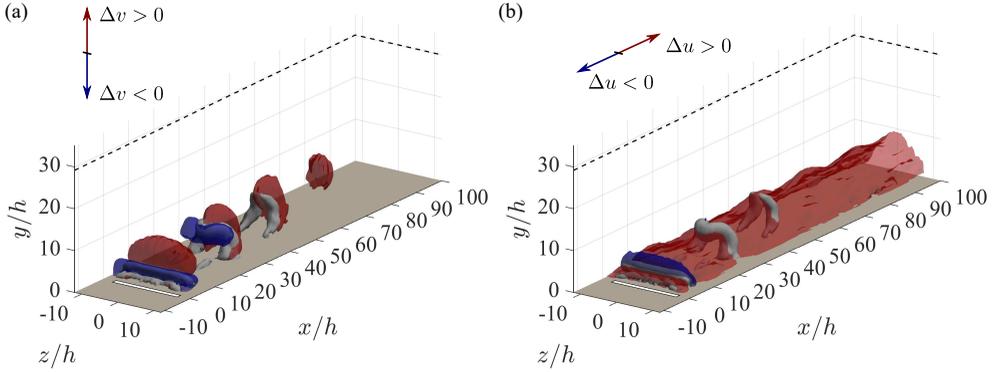}
\caption{Isosurfaces of phased-locked $\lambda_{ci}=0.5$ (grey) in the $\alpha=45^\circ$, $\beta=0^\circ$ jet at $\phi=90^\circ$ with isosurfaces of (a) $\Delta v/U_e=0.05$ (red), $\Delta v/U_e=-0.05$ (blue) and (b) $\Delta u/U_e=0.05$ (red), $\Delta u/U_e=-0.10$ (blue).}
\label{fig:AR18P45Beta0DeltaPavg}
\end{figure}

\begin{figure}
\centering
\includegraphics[width=5.25in]{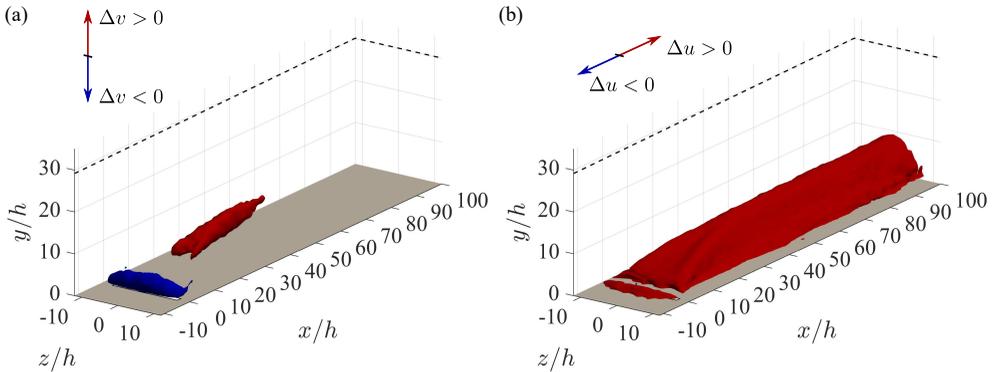}
\caption{Isosurfaces of time-averaged (a) $\Delta v/U_e=0.05$ (red), $\Delta v/U_e=-0.05$ (blue) and (b) $\Delta u/U_e=0.05$ (red), $\Delta u/U_e=-0.10$ (blue) in the $\alpha=45^\circ$, $\beta=0^\circ$ jet.}
\label{fig:AR18P45Beta0DeltaTavg}
\end{figure}

The pitched orifice has a markedly different effect on the velocity field within the BL than the other two orifice geometries have. 
The early vortex ring induces a strong upwash within the ring and downwash around the ring (figure~\ref{fig:AR18P45Beta0DeltaPavg}a), similar to the early $\alpha=90^\circ$, $\beta=90^\circ$ jet (figure~\ref{fig:AR18Beta90DeltaPavg}a). 
The difference is that by pitching the vortex ring downstream the induced velocity within the ring also has a significant streamwise component (figure~\ref{fig:AR18P45Beta0DeltaPavg}b). 
When the arch-shaped vortex first forms and its apex leans upstream, upwash is induced downstream of the crook formed by the vortex while fluid is induced towards the wall on the upstream side of the structure (figures~\ref{fig:AR18P45Beta0Sketch}a and \ref{fig:AR18P45Beta0DeltaPavg}a). 
As the arch-shaped vortex straightens out, velocity induction in the wall-normal direction steadily drops. 
Instead, the upright arch-shaped vortex induces a strong streamwise velocity under the arch (figure~\ref{fig:AR18P45Beta0DeltaPavg}b). 

The time-averaged isosurfaces of wall-normal velocity change are relatively small for this case (figure~\ref{fig:AR18P45Beta0DeltaTavg}a). 
A patch of downward motion upstream of the orifice is the result of the actuator suction cycle while an additional region of upward moving fluid exists within the pitched jet itself. 
Generally speaking, the velocity field does not indicate strong wall-normal mixing in a time-averaged sense. 
Rather, the large region of enhanced streamwise momentum which exists downstream of the actuator is the result of direct momentum addition from the pitched orifice (figure~\ref{fig:AR18P45Beta0DeltaTavg}b). 
From a vortex dynamics perspective, the arch-shaped vortices are the vehicles which carry the high streamwise momentum in the pitched synthetic jet. 
In addition to producing a much thicker and faster moving layer of accelerated fluid near the wall than the other two orifices, the pitched orifice also does not generate a detectable wake in figure~\ref{fig:AR18P45Beta0DeltaTavg}b. 
It will be shown that a layer of fluid of reduced streamwise velocity does exist further from the wall, due to the upstream pointing velocity induction above the arch-shaped vortex (figure~\ref{fig:AR18P45Beta0Sketch}a), but that the velocity reduction is small and the slower moving fluid is not consolidated into a well-defined wake.

\section{Comparison of the three SJAs at \texorpdfstring{$C_b=1$}{TEXT}}
\label{sec:results2}

Having established the types of periodic structures formed by each of the jets and how they alter the flow field in \S~\ref{sec:results1}, this section focuses on comparing the flow fields of the three jets and assessing their relative potential to control separation.

\subsection{Trajectory and propagation speed of periodic structures}
\label{sec:vorttraj}

\begin{figure}
\centering
\includegraphics[width=5in]{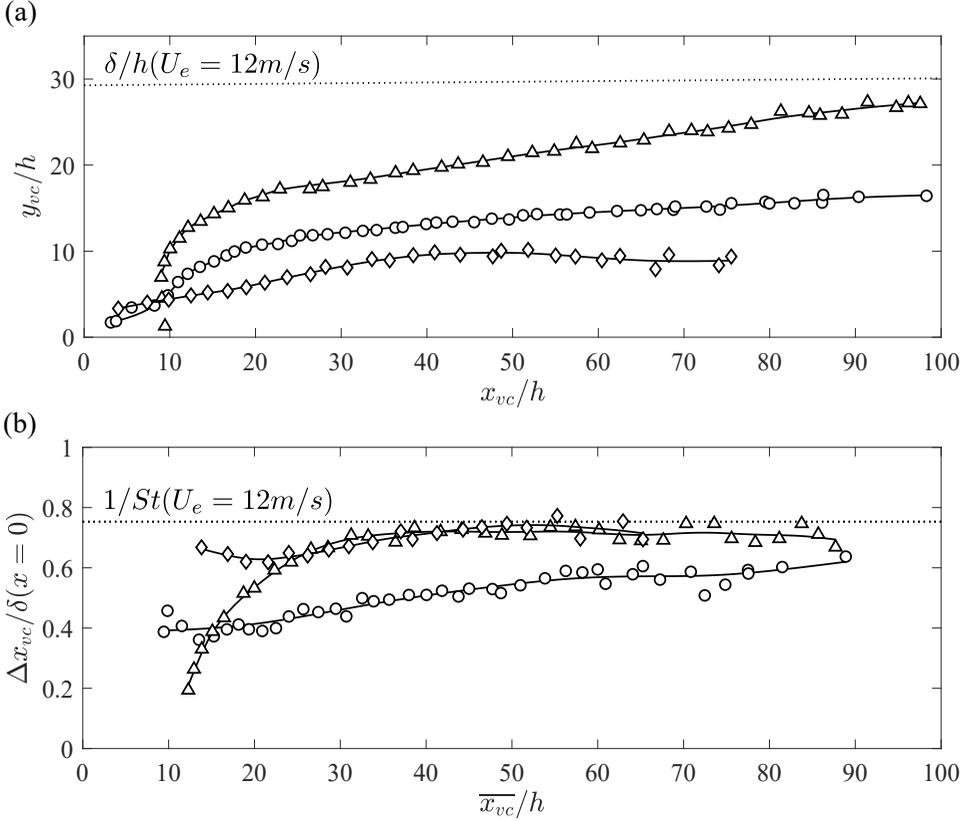}
%\caption{Centerline (a) vortex centre locations and (b) vortex centre streamwise spacing from phase-locked data ($\alpha=90^\circ$, $\beta=0^\circ$, $\bigcirc$; $\alpha=90^\circ$, $\beta=90^\circ$, $\vartriangle$; $\alpha=45^\circ$, $\beta=0^\circ$ $\lozenge$)}
\caption{Centerline vortex centre (a) locations and (b) streamwise spacing from phase-locked data for the $\alpha=90^\circ$, $\beta=0^\circ$ ($\circ$), $\alpha=90^\circ$, $\beta=90^\circ$ ($\vartriangle$), and $\alpha=45^\circ$, $\beta=0^\circ$ ($\lozenge$) jets.}
\label{fig:AllOrfVortTraj}
\end{figure}

The distinctive periodic structures produced by the three actuator geometries travel through the BL at different rates and on different paths. 
The spatial trajectories of the vortices for these cases are depicted in figure~\ref{fig:AllOrfVortTraj}a. 
Indicated are the centres of the dominant vortex on the jet centreline, extracted from the eight phased-locked data sets. 
Note that the CW rotating vortex is the dominant structure in the $\alpha=90^\circ$, $\beta=0^\circ$ and $\alpha=90^\circ$, $\beta=90^\circ$ jet cases while the CCW rotating vortex is dominant in the $\alpha=45^\circ$, $\beta=0^\circ$ jet. 
The discrete vortex location data was fitted with spline curves which are plotted as solid lines in figure~\ref{fig:AllOrfVortTraj}a while a dotted line near the top of the figure indicates the height of the BL edge for the baseline case. 

The initial trajectory of the $\alpha=90^\circ$, $\beta=0^\circ$ jet vortex follows an S-curve path emblematic of axis-switching in moderate $AR$ vortex rings (figure~\ref{fig:AllOrfVortTraj}a). 
This motion is brought about by the self-induced deformations of the virtual vortex ring which is formed by the primary structure together with its wall image. 
The vortex trajectory up to $x/h=25$ closely matches that presented in figure 9i of \citet{Straccia_Farnsworth_PRF_2021} for an $AR=19$ rectangular orifice synthetic jet issuing into quiescent fluid, albeit rotated by $90^\circ$. 
At $x/h \approx 27.5$ the vortex detaches from the wall and propagates downstream as a hairpin vortex on a roughly linear trajectory. 
The arrangement of the hairpin vortex train produced by the SJBLI is reminiscent of the linear ramp structure observed in naturally occurring turbulent BL hairpin vortex packets \citep{ Adrian_Meinhart_JFM_2000}. 
By the end of the measurement domain the head of the hairpin vortices have penetrated to a height that is roughly half the thickness of the baseline BL. 

Axis-switching of the largely intact vortex ring in the $\alpha=90^\circ$, $\beta=90^\circ$ jet case also modified the initial trajectory of the vortices on the jet centreline (figure~\ref{fig:AllOrfVortTraj}a). 
The CW vortex moves almost straight up from the wall initially, because the rapid contraction in width of the vortex ring offsets the downstream transport of the ring by the cross-flow. 
Once the self-induced deformations cease and the vortex ring assumes a spanwise orientation, the vortex follows an essentially linear path downstream of $x/h \approx 20$. 
The upward motion of the vortex rings in the streamwise-oriented jet is more rapid than the motion of the vortices of the $\alpha=90^\circ$, $\beta=0^\circ$ jet case, leading to their deeper penetration into the cross-flow. 
In fact, by $x/h=100$ the upper surface of the vortex rings have passed through the original edge of the BL as measured in the baseline case.

The motion of the vortices in the $\alpha=45^\circ$, $\beta=0^\circ$ jet is initially driven by velocity induction between the counter-rotating sides of the vortex ring, which is pointed downstream in the pitched jet (figure~\ref{fig:AllOrfVortTraj}a). 
By $x/h=15$ the CW vortex is no longer detectable, and the motion of the CCW vortex is dictated from there on by the self-induced motion of the resulting arch-shaped vortex. 
As the apex of the arch-shaped vortex rotates downstream relative to the legs, the arch straightens out and the CCW vortex travels away from the wall. 
The CCW vortex thus reaches its furthest distance from the wall at $x/h \approx 45$, roughly where the vortex had deformed into an upright arch shape. 
Downstream of this point the CCW vortex penetrated no further into the cross-flow. 
Of the three cases the jet from the pitched orifice remains the closest to the wall, with the vortex penetrating to a height no deeper than a third of the thickness of the baseline BL.

The vortex centre location data plotted in figure~\ref{fig:AllOrfVortTraj}a was also used to compute the spacing of the periodic structures in the BL. 
%The results are presented in figure~\ref{fig:AllOrfVortTraj}b where $\overline{x_{vc}}$ is the streamwise mid-point and $\Delta x_{vc}$ is the streamwise space between subsequent generations of the dominant vortex. 
The results are presented in figure~\ref{fig:AllOrfVortTraj}b where $\overline{x_{vc}}$ is the streamwise mid-point between two adjacent vortices of subsequent generations and $\Delta x_{vc}$ is the streamwise distance between those vortices.
The parameter $\Delta x_{vc}$ has been normalized by the thickness of the baseline BL. 
Strouhal number, as defined in \ref{eq:St}, may be physically interpreted as the spatial frequency of the periodic structures within a cross-flow. 
Therefore, the ordinate of figure~\ref{fig:AllOrfVortTraj}b is the reciprocal of $St$ and accordingly the dotted line in the figure indicates the value of $1/St$, i.e., the vortex spacing that would be achieved if the periodic structures advected downstream at a speed of $U_e$. 

The vortices in the $\alpha=90^\circ$, $\beta=0^\circ$ jet are spaced more closely than is implied by the $St$ value (figure~\ref{fig:AllOrfVortTraj}b). 
On the other hand, after initial self-induced deformations subside, the vortices in both the $\alpha=90^\circ$, $\beta=90^\circ$ and $\alpha=45^\circ$, $\beta=0^\circ$ jets converge to a much larger vortex spacing, one that is only slightly smaller than the theoretical $St$ value. 
The $St$ equation typically used in cross-flow studies makes the simplifying assumption that the periodic structures advect at the BL edge velocity, $U_e$, which is not physically accurate while the structures are embedded in the slower moving BL fluid. 
But there are more factors that influence the vortex spacing than just the distance of the structures from the wall.
This is clear from the fact that the vortices in the $\alpha=90^\circ$, $\beta=0^\circ$ and $\alpha=45^\circ$, $\beta=0^\circ$ jets have very similar streamwise spacing despite being at very different heights in the BL (figure~\ref{fig:AllOrfVortTraj}). 
At this point it is helpful to look at the advective speed of the periodic structures in these jets.

\begin{figure}
\centering
\includegraphics[width=5in]{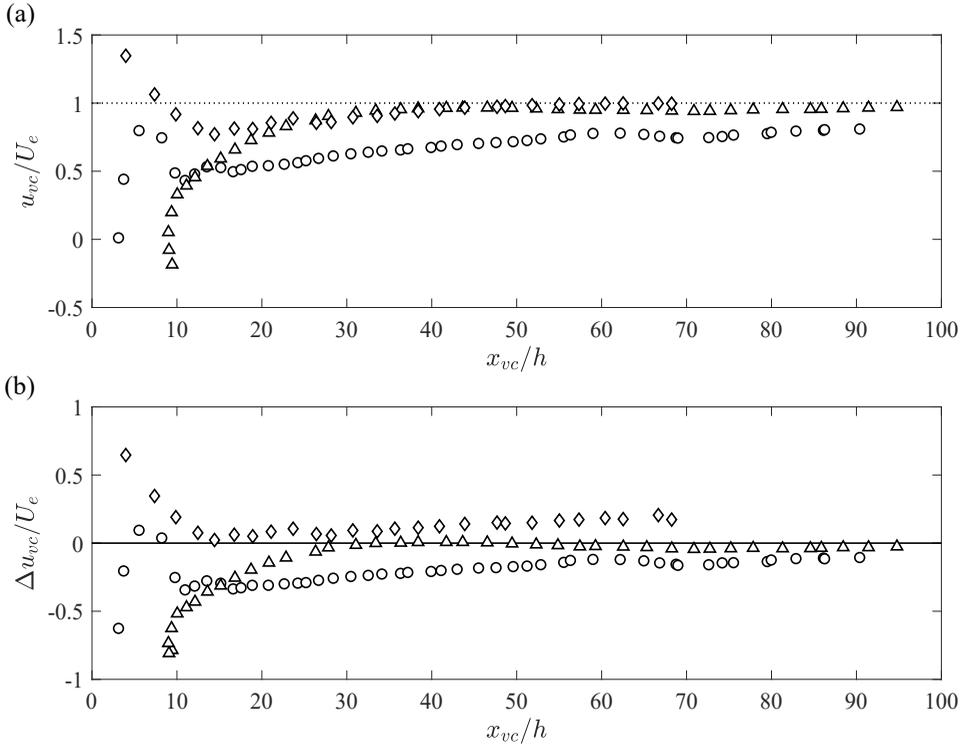}
\caption{Centerline vortex (a) absolute streamwise velocity and (b) change in streamwise velocity for the $\alpha=90^\circ$, $\beta=0^\circ$ ($\circ$), $\alpha=90^\circ$, $\beta=90^\circ$ ($\vartriangle$), and $\alpha=45^\circ$, $\beta=0^\circ$ ($\lozenge$) jets.}
\label{fig:AllOrfVortVel}
\end{figure}

Figure~\ref{fig:AllOrfVortVel}a is a plot of the streamwise advective velocity of the dominant vortices for the three cases. 
These velocities were computed using the spline curves fit to the vortex centre locations (solid lines in figure~\ref{fig:AllOrfVortTraj}a) and the known time increments separating the eight measured phases of the actuator. 
Additionally, figure~\ref{fig:AllOrfVortVel}b compares the vortex propagation speed to the speed of the unforced flow. 
The parameter $\Delta u_{vc}$ is the difference between the vortex advective velocity and the baseline BL flow speed at the height of the vortex centre. 
Therefore, when $\Delta u_{vc}>0$ the vortex travels faster than a passive scalar would, on average, at that location in the baseline BL, and when $\Delta u_{vc}<0$ the vortex travels slower than the average baseline flow.

The CW rotating vortex in the $\alpha=90^\circ$, $\beta=0^\circ$ jet initially has low streamwise velocity because the jet issues normal to the wall (figure~\ref{fig:AllOrfVortVel}a). 
After formation, the vortex quickly accelerates in the downstream direction due to momentum transfer from the cross-flow. 
The velocity of the coherent structure is also influenced by vortex induction. 
For example, between $x/h=5$ and $x/h=8.5$ the streamwise velocity of the vortex actually exceeds the average baseline flow speed at that height in the BL (figure~\ref{fig:AllOrfVortVel}b). 
During this period in the jet development the CCW vortex is rolled up and over the CW vortex such that the velocity induction between the counter rotating sides of the vortex ring is pointed downstream. 
The streamwise velocity of the CW vortex then declines as the CCW vortex rolls back towards the wall and dissipates. 
This deceleration continues as axis-switching deformations in the virtual vortex ring cause the CW rotating head of the structure to rotate off the wall and move upstream, relative to the legs.

While the streamwise velocity of the CW vortex increases slowly between $x/h=11$ and $x/h=90$ it remains lower than the baseline flow speed for the duration of the measurement (figure~\ref{fig:AllOrfVortVel}). 
The advective speed of the hairpin vortices is slow because the hairpin head induces an upstream pointing velocity on itself. 
The gradual convergence of the vortex streamwise velocity with the baseline flow speed comes about as the strength of self-induction drops due to dissipation of the vortex (figure~\ref{fig:AllOrfVortVel}b). 
By the end of the measurement domain the hairpin head travels at $0.8U_e$ which is similar to the speed at which naturally occurring hairpin vortex packets travel in turbulent BLs \citep{Adrian_Meinhart_JFM_2000}.

The combination of a wall-normal pitch angle in the $\alpha=90^\circ$, $\beta=90^\circ$ jet, along with the self-induced deformations depicted in figure~\ref{fig:AllOrfVortTraj}a, lead to low initial streamwise velocity in the CW rotating side of the vortex ring. 
Once the axis switching deformations reorient the vortex ring major axis in the spanwise direction the streamwise component of the self-induced velocity drops significantly and the CW vortex is transported downstream by the cross-flow. 
This motion in the later vortex ring is evident from the essentially zero value of $\Delta u_{vc}/U_e$ starting around $x/h=28$. 
The streamwise velocity of the vortex in the furthest downstream measurement is $0.97U_e$ due both to the propagation of the vortex ring to the edge of the BL (figure~\ref{fig:AllOrfVortTraj}a), where the cross-flow speed is high, and to the passive nature of the ring's motion after the axis switching deformations are complete (figure~\ref{fig:AllOrfVortVel}b).

Unsurprisingly, the initial streamwise velocity of the CCW vortex is very high in the $\alpha=45^\circ$, $\beta=0^\circ$ jet because the vortex ring formed by the pitched orifice is oriented downstream and therefore induces a high streamwise velocity on itself (figure~\ref{fig:AllOrfVortVel}b). 
The induced velocity drops however as the core of the CW rotating vortex loses coherence. 
The streamwise velocity of the CCW vortex reaches a minimum between $x/h=15$ and $x/h=20$, roughly where the CW vortex becomes undetectable. 
Streamwise velocity rises again starting around $x/h=20$ as the CCW vortex deforms into an arch-shaped structure. 
This arch-shaped vortex induces a downstream velocity on itself, and so these structures travel faster than the fluid in the baseline case (figure~\ref{fig:AllOrfVortVel}b). 
As the apex of the arch lifts off the wall, and rotates downstream relative to the legs the streamwise velocity continues to rise. 
By $x/h=70$ the CCW vortex is traveling at over $0.99U_e$ (figure~\ref{fig:AllOrfVortVel}b) despite remaining relatively close to the wall (figure~\ref{fig:AllOrfVortTraj}a). 

Interestingly, the dominant vortices in the $\alpha=90^\circ$, $\beta=90^\circ$ and $\alpha=45^\circ$, $\beta=0^\circ$ jets end up with roughly the same velocities downstream despite the periodic structures being very different from each other (figure~\ref{fig:AllOrfVortVel}a). 
In the $\alpha=90^\circ$, $\beta=90^\circ$ jet the high streamwise velocity is due to rapid penetration of the vortex ring into the faster moving fluid near the edge of the BL while the arch-shaped vortices in the $\alpha=90^\circ$, $\beta=90^\circ$ jet achieved a high streamwise velocity due to a self-induced velocity that is oriented downstream. 
The relatively slow streamwise velocity of the vortices in the $\alpha=90^\circ$, $\beta=0^\circ$ jet is in turn caused by only moderate penetration of the periodic structures into the BL and a self-induced velocity that was pointed upstream. 
These factors explain the significant differences in streamwise spacing of the dominant vortices in the three jets observed in figure~\ref{fig:AllOrfVortTraj}b.
%These factors explain the significant differences in streamwise spacing of the dominant vortices in the three jets observed in figure~\ref{fig:AllOrfVortTraj}b (compare also figure \ref{fig:AR18Beta0Train} to \ref{fig:AR18P45Beta0Train}).

\subsection{Velocity field near the wall}
\label{sec:wallU}

\begin{figure}
\centering
\includegraphics[width=5in]{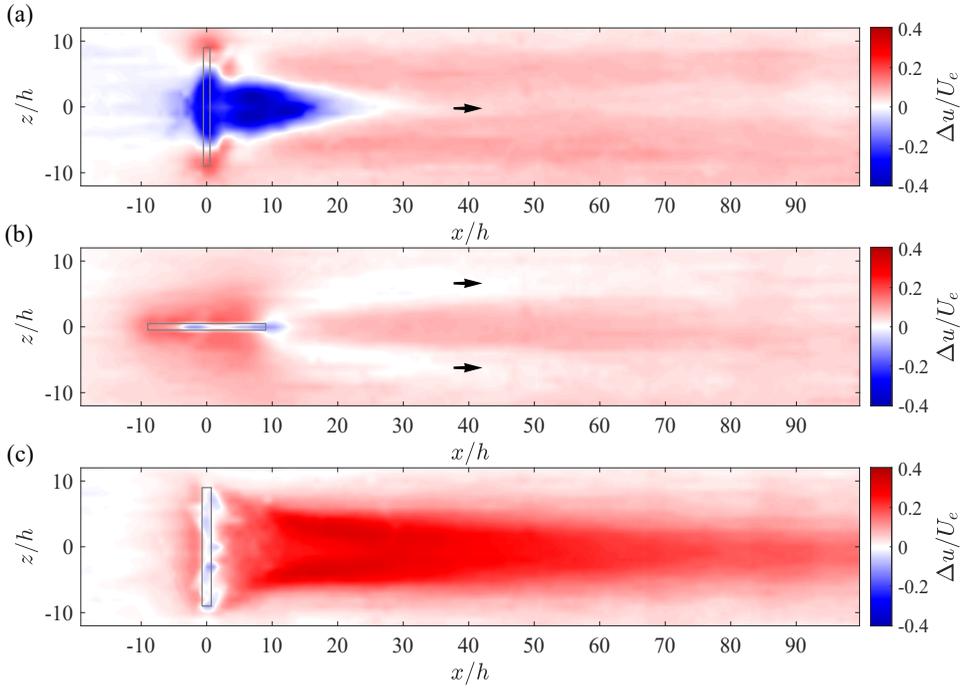}
\caption{Contours of time-averaged streamwise velocity change near the wall ($y/h=1$) for the (a) $\alpha=90^\circ$, $\beta=0^\circ$, (b) $\alpha=90^\circ$, $\beta=90^\circ$, and (c) $\alpha=45^\circ$, $\beta=0^\circ$ jets.}
\label{fig:AllOrfWallVel}
\end{figure}

The influence that the SJA has on the flow velocity near the wall is of particular interest because of the implication it has for wall shear stress and the susceptibility of the BL to separation. 
Figure~\ref{fig:AllOrfWallVel} displays the normalized change in streamwise velocity for the three orifice geometries on a measurement plane at a height of $y/h=1$ ($y/\delta=0.035$ or $y^+=29$). 
The contour plots are from the time-averaged data sets, and the position of the orifice is indicated by a rectangular outline with flow traveling from left to right. 

A prominent feature of the $\alpha=90^\circ$, $\beta=0^\circ$ jet in the near-wall plane is the region of reduced flow velocity downstream of the orifice (figure~\ref{fig:AllOrfWallVel}a). 
Initially, the primary vortex is connected to the wall by the legs which induce an upstream pointing velocity that decelerates the fluid between them. 
As the vortex contracts in the spanwise direction and the legs detach from the wall the region of decelerated flow shrinks in width. 
%By $x/h=27.5$ the hairpin vortex lifts off the wall, and the legs, which have reconnected in the streamwise direction, now induce high momentum fluid down around their sides and then inwards towards the jet centreline, increasing the velocity there.
Between $x/h=25$ and $x/h=35$ the hairpin vortex lifts off the wall, and the legs, which have reconnected in the streamwise direction, now induce high momentum fluid down around their sides and then inwards towards the jet centreline, increasing the velocity there. 
However, that redistribution of fluid is weakly opposed by the chevron-shaped secondary structures which rotate in the opposite direction as the hairpin legs (figure~\ref{fig:AR18Beta0Sketch}a). 
The net result is a persistent streak of slightly lower momentum fluid on centreline as compared with the fluid located to the sides of centreline (see arrow in figure~\ref{fig:AllOrfWallVel}a). 
Despite the spanwise variability in the fluid momentum near the wall all of the fluid in that layer still travels faster than it did before actuation was applied (downstream of $x/h \approx 30$).

In the $\alpha=90^\circ$, $\beta=90^\circ$ jet the streamwise velocity is higher than it was in the baseline flow almost everywhere in the near-wall plane (figure~\ref{fig:AllOrfWallVel}b). 
%This is owing to the rapid lift off vortex ring and the fact that the relatively weaker leg vortices do not remain tied to the wall past the orifice. 
A region of low velocity fluid does not exist downstream of the orifice in this case because the relatively weaker legs emanating from the vortex rings do not remain connected to the wall past the end of the orifice.
Downstream of the orifice the region of highest streamwise velocity is found on centreline where the upwash from the vortex ring is strongest. 
Regions of slightly lower velocity on either side of centreline (see arrows in figure~\ref{fig:AllOrfWallVel}b) sit just outboard of two streaks of secondary streamwise vorticity found very close to the wall. 
That streamwise vorticity, which rotates counter to the streamwise vorticity in the vortex ring, may be generated by the viscous interaction of spanwise velocity induction from the primary vortex and the no-slip condition at the wall. 

The effectiveness of the $\alpha=45^\circ$, $\beta=0^\circ$ SJA in enhancing streamwise velocity near the wall is remarkable when compared with the performance of the wall-normal jets (figure~\ref{fig:AllOrfWallVel}c). 
The highest streamwise velocity occurs in the region enclosed by the arch-shaped vortices. 
This region narrows downstream of the orifice as the arch-shaped vortices contract in width. 
However, even outside of the footprint of the primary vortices streamwise velocity is elevated as a result of the actuation, presumably due to mixing and diffusion of momentum.

\subsection{Mean mixing in a midfield plane}
\label{sec:meanmix}

\begin{figure}
\centering
\includegraphics[width=5.25in]{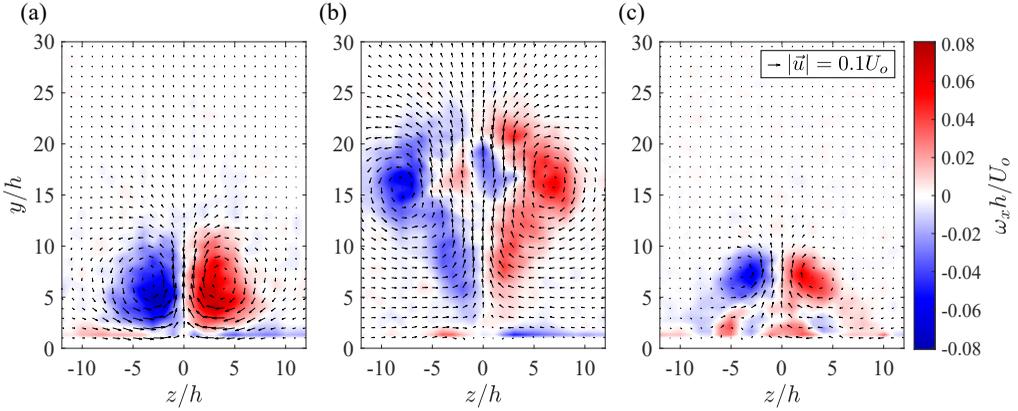}
\caption{Contours of time-averaged streamwise vorticity overlaid with velocity vectors on the $x/h=25$ plane for the (a) $\alpha=90^\circ$, $\beta=0^\circ$, (b) $\alpha=90^\circ$, $\beta=90^\circ$, and (c) $\alpha=45^\circ$, $\beta=0^\circ$ jets.}
\label{fig:AllOrfMidfield}
\end{figure}

The advecting train of vortices downstream of the actuator lead to a rotational mean flow within the SJBLI domain, as revealed in the time-averaged streamwise vorticity contours of figure~\ref{fig:AllOrfMidfield} on the $x/h=25$ plane. 
Overlain on the contour plot are the in-plane time-average velocity vectors, which have a consistent scale factor applied across the three panels. 

In all of the jets the advecting periodic structures establish counter-rotating patches of fluid in the mean field such that the common flow on centreline is pointed upward off the wall. 
It is tempting to interpret the time-average vorticity field as representing ``vortices'' but in an unsteady flow field this interpretation is inaccurate and can easily be misleading. 
To be rigorous these fields should be viewed as the result of the mean influence of the evolving and advecting vortices within the unsteady field. 
In the $\alpha=90^\circ$, $\beta=0^\circ$ jet the mean rotation is induced by the sides and legs of the hairpin vortices (figure~\ref{fig:AllOrfMidfield}a). 
In this jet the rotation is strong and proximate to the wall which makes the actuation particularly effective at entraining fast moving fluid into the lower BL. 

In the $\alpha=90^\circ$, $\beta=90^\circ$ jet the centres of the mean flow rotation, seen at a height of $y/h \approx 16$ in figure~\ref{fig:AllOrfMidfield}b, align with the ends of the vortex ring, which by $x/h=25$ has completed its axis switch and is now widest in the spanwise direction. 
The influence of the legs and interlinking vortex structures is apparent closer to the wall between $y/h=5$ and 10. 
In the early vortex ring, while the long sections of the vortex are aligned in the streamwise direction and are relatively close together in the spanwise direction, the wall-normal velocity of the mean flow is very high. 
This region of rapid mean upwash occurs directly above the orifice. 
But as the vortex ring axis-switches and the ends of the ring --- which carry the streamwise vorticity --- move apart, mean upwash on centreline drops. 
Thus by $x/h=25$ the average wall-normal flow speed on centreline is higher in the $\alpha=90^\circ$, $\beta=0^\circ$ jet than in the $\alpha=90^\circ$, $\beta=90^\circ$ jet. 
The rapid upward propagation of the vortex rings in the $\alpha=90^\circ$, $\beta=90^\circ$ jet also causes the induced flow redistribution to act further from the wall. 
This lessening influence of the primary vortex on the fluid at lower $y/h$ is mitigated somewhat by the trailing legs which remain closer to the wall.

\begin{figure}
\centering
\includegraphics[width=5in]{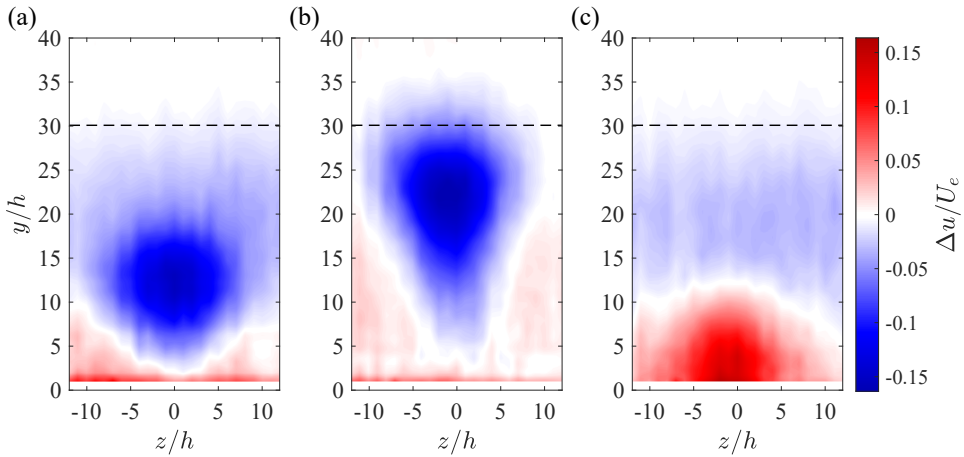}
\caption{Contours of time-averaged streamwise velocity change at $x/h \approx 100$ for the (a) $\alpha=90^\circ$, $\beta=0^\circ$, (b) $\alpha=90^\circ$, $\beta=90^\circ$, and (c) $\alpha=45^\circ$, $\beta=0^\circ$ jets.}
\label{fig:AllOrfExitPlane}
\end{figure}

The mean redistribution of flow is generally weakest in the $\alpha=45^\circ$, $\beta=0^\circ$ jet (figure~\ref{fig:AllOrfMidfield}c). 
This close to the orifice the head of the arch-shaped vortex is still leaned upstream, introducing a streamwise component to rotation in the mean flow field around $y/h=7$. 
Unlike the wall-normal orifices, however, the mean rotation does not persist downstream because the straightening out of the arch-shaped vortex tilts the streamwise component of the vorticity into the wall-normal direction. 

\subsection{Velocity field in the \texorpdfstring{$x/h=100$}{TEXT} plane}
\label{sec:exitplaneu}

Figure~\ref{fig:AllOrfExitPlane} displays the normalized change in streamwise velocity at the end plane of the measurement domain ($x/h \approx 100$) for all three orifice geometries tested. 
Actuation enhanced the near-wall flow velocity in all cases, but the flow fields further from the wall differ significantly from each other. 
The $\alpha=90^\circ$, $\beta=0^\circ$ SJA produced a roughly circular wake which sat above a region of accelerated flow near the wall (figure~\ref{fig:AllOrfExitPlane}a). 
The $x/h \approx 100$ plane of the $\alpha=90^\circ$, $\beta=90^\circ$ jet looks similar to the spanwise-oriented jet, albeit stretched in the $y$-direction (figure~\ref{fig:AllOrfExitPlane}b). 
The streamwise-oriented jet penetrates almost twice as deeply into the cross-flow as the spanwise-oriented one, and therefore the wake sits much further from the wall. 
A larger area of the BL has elevated streamwise velocity in the $\alpha=90^\circ$, $\beta=90^\circ$ jet case, but the magnitude of the increase is generally lower than it is in the $\alpha=90^\circ$, $\beta=0^\circ$ jet. 
The flow field of the $\alpha=45^\circ$, $\beta=0^\circ$ jet is distinct from the other two jets in that a localized wake did not form (figure~\ref{fig:AllOrfExitPlane}c). 
Instead, a diffuse region with a relatively small reduction in streamwise velocity exists above the jet. 
Unsurprisingly, the core region of accelerated fluid near the wall takes the shape of a semi-ellipse that mirrors the projection of the arch-shaped vortices onto the $x$-plane. 

\begin{figure}
\centering
\includegraphics[width=4in]{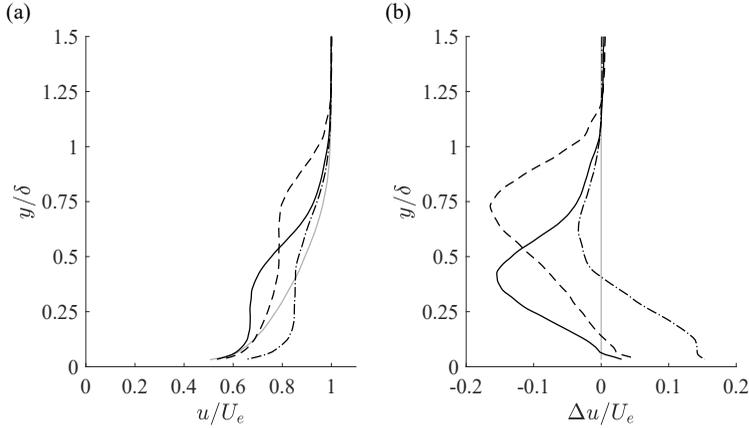}
\caption{Centerline time-averaged (a) streamwise velocity and (b) change in streamwise velocity profiles at $x/h \approx 100$ for the baseline flow (\protect\greyline), $\alpha=90^\circ$, $\beta=0^\circ$ jet (\sampleline{}), $\alpha=90^\circ$, $\beta=90^\circ$ jet (\sampleline{dash pattern=on .475em off .3em on .475em off .3em}), and $\alpha=45^\circ$, $\beta=0^\circ$ jet (\sampleline{dash pattern=on .55em off .2em on .05em off .2em}).}
\label{fig:AllOrfBLProf}
\end{figure}

\begin{figure}
\centering
\includegraphics[width=4in]{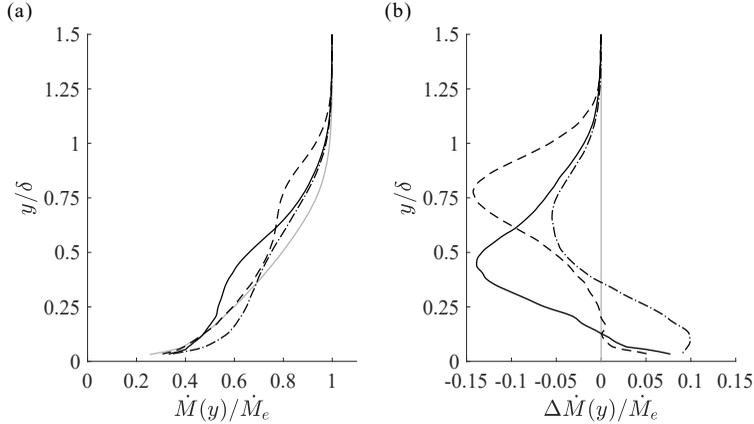}
\caption{Time-averaged (a) streamwise momentum and (b) change in streamwise momentum profiles at $x/h \approx 100$ for the baseline flow (\protect\greyline), $\alpha=90^\circ$, $\beta=0^\circ$ jet (\sampleline{}), $\alpha=90^\circ$, $\beta=90^\circ$ jet (\sampleline{dash pattern=on .475em off .3em on .475em off .3em}), and $\alpha=45^\circ$, $\beta=0^\circ$ jet (\sampleline{dash pattern=on .55em off .2em on .05em off .2em}).}
\label{fig:AllOrfMomProf}
\end{figure}

The centreline velocity profile and the change in streamwise velocity are compared for the three orifice geometries at the end of the measurement domain in figure~\ref{fig:AllOrfBLProf}. 
The velocity profiles were obtained by averaging the SPIV data from the last 3~mm of the measurement domain in the streamwise direction up to $x/h \approx 100$. 
The velocity profile of the pitched orifice exceeds the velocity of the wall-normal orifices at all $y$, in some places by over 10\% of $U_e$ because the pitched orifice jet has both a shallower ``wake'' and induces higher flow speeds near the wall. 
Conversely, the maximum velocity deficit in the wakes of the wall-normal jets were roughly the same despite being offset by $y/\delta=0.3$ in height relative to each other (figure~\ref{fig:AllOrfBLProf}b).
Between the wake region and the region of enhanced velocity near the wall, a transitional layer exists in all three cases where $\frac{\partial u}{\partial y}$ is roughly zero (figure~\ref{fig:AllOrfBLProf}a). 

The enhancement of the flow velocity near the wall by the $\alpha=45^\circ$, $\beta=0^\circ$ SJA is not only much larger than it is in the other two cases, but the region of elevated velocity is also much thicker in the $y$-direction (figure~\ref{fig:AllOrfBLProf}). 
In terms of near-wall flow speed on centreline, the $\alpha=90^\circ$, $\beta=90^\circ$ jet is a distant second. 
%However, the $\alpha=90^\circ$, $\beta=90^\circ$ jet is not generally superior to the $\alpha=90^\circ$, $\beta=0^\circ$ jet, it just happened to achieve maximum velocity enhancement on centreline, whereas the $\alpha=90^\circ$, $\beta=0^\circ$ jet has higher flow speeds off centreline. 
But while the $\alpha=90^\circ$, $\beta=90^\circ$ jet produced a larger velocity enhancement on centreline than the $\alpha=90^\circ$, $\beta=0^\circ$ jet, away from the centreline the $\alpha=90^\circ$, $\beta=0^\circ$ jet was more effective at increasing the near-wall flow speed.

To account for the effect of the actuation across the entire plane, the fluid momentum was integrated in the spanwise direction. 
%The resulting momentum profiles are compared in figure~\ref{fig:AllOrfMomProf}. 
The resulting momentum profiles are compared in figure~\ref{fig:AllOrfMomProf}, where $\dot{M}_e$ is the momentum of the flow outside the BL and $\Delta \dot{M}_(y)$ is the difference in the integrated momentum between the actuated and baseline flows.
Accounting for the flow field variability in the spanwise direction reduces the relative differences in performance between the pitched and wall-normal orifices. 
%The wakes of the wall-normal orifices are three-dimensional while the region of reduced momentum in the pitched jet is roughly two-dimensional; thus, integrating the momentum in the spanwise direction leads to a relative reduction in the depth of the wall-normal orifice wakes. 
%Conversely, the accelerated fluid in the pitched jet is concentrated near centreline, while the region of accelerated fluid is more uniform in the spanwise direction for the wall-normal jets. 
%Therefore, when the flow conditions across the full exit plane are considered, the advantage that the pitched orifice jet has over the other two geometries on centreline shrinks. 
Furthermore, while the pitched jet still achieves higher flow momentum than the other two cases do at all $y/\delta$, the slope of the momentum profiles in figure~\ref{fig:AllOrfMomProf}b suggest that closer to the wall the wall-normal jets may match or exceed the momentum of the pitched jet. 
What is clear from the measurement data is that on average the $\alpha=90^\circ$, $\beta=0^\circ$ jet is more effective at enhancing the momentum of the near-wall BL than the $\alpha=90^\circ$, $\beta=90^\circ$ jet is when the full measurement domain is considered.

\subsection{Unsteady velocity field on the jet centreline}
\label{sec:unsteadymix}

\begin{figure}
\centering
\includegraphics[width=5in]{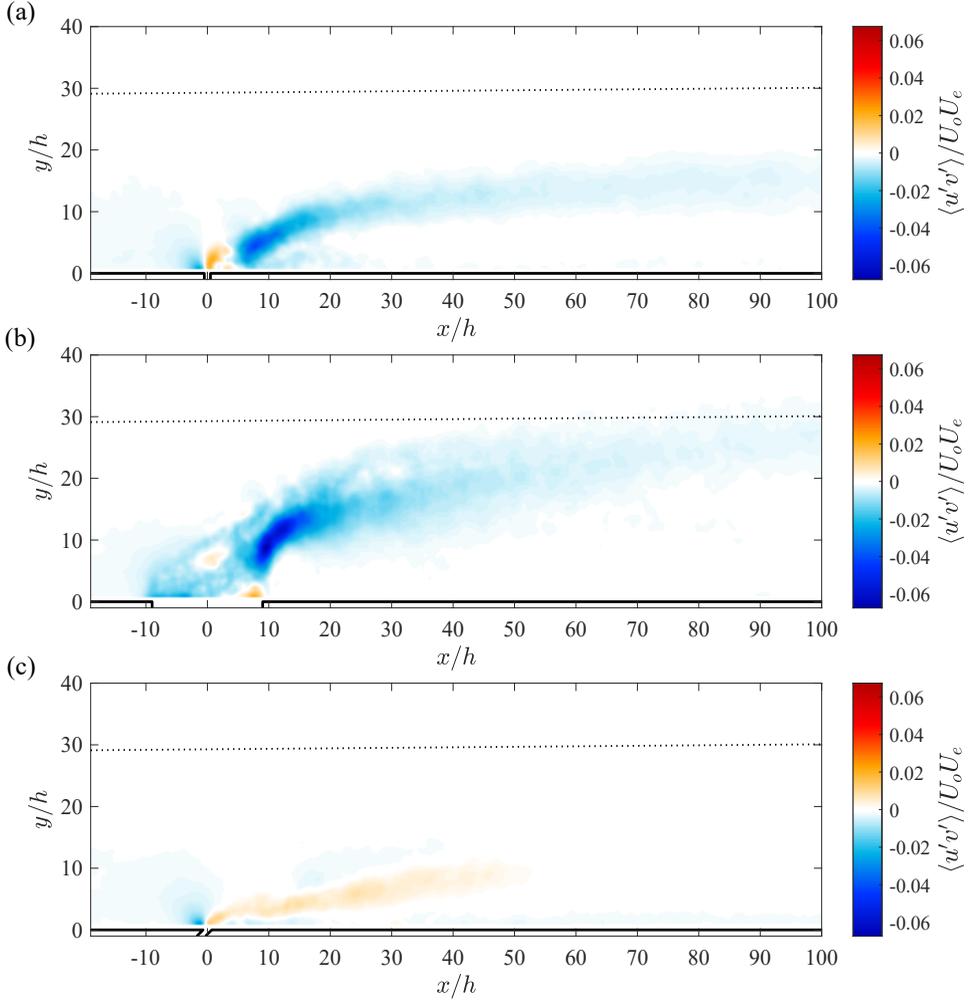}
\caption{Time-averaged Reynolds stress contours on the jet centreline for the (a) $\alpha=90^\circ$, $\beta=0^\circ$, (b) $\alpha=90^\circ$, $\beta=90^\circ$, and (c) $\alpha=45^\circ$, $\beta=0^\circ$ SJAs.}
\label{fig:AllOrfReStress}
\end{figure}

To analyze the unsteady flow field within the SJBLI domain a Reynolds decomposition of the instantaneous velocity field was conducted. 
Per \ref{eq:uprime} the velocity fluctuations ($u'$) were computed by subtracting the time-averaged velocity ($\overline{u}$) from the instantaneous velocity. 
Note that the velocity fluctuation $u'$ is composed of a combination of the periodic and stochastic fluctuations. 
To investigate unsteady mixing, Reynolds shear stress ($\langle u'v' \rangle$) contour plots on the centreline of the three jets are presented in figure~\ref{fig:AllOrfReStress}. 
A particular Reynolds stress value can be assigned to one of four quadrants based on where a specific instantaneous velocity falls on a plot of $u'$ vs $v'$. 
Negative values of $\langle u'v' \rangle$ are associated with enhanced mixing within the BL corresponding to either ``ejections'' --- which move low momentum fluid away from the wall (quadrant two: $u'<0$, $v'>0$) --- or ``sweeps'' --- which entrain high momentum fluid towards the wall (quadrant four: $u'>0$, $v'<0$). 
Conversely, positive values of $\langle u'v' \rangle$ enhance the natural gradient of momentum within a BL by either moving low momentum fluid towards the wall (quadrant three: $u'<0$, $v'<0$) or by moving high momentum fluid away from the wall (quadrant one: $u'>0$, $v'>0$).

\begin{equation} 
\label{eq:uprime}
u(\bm{x},t)=\overline{u}(\bm{x})-u'(\bm{x},t)
\end{equation}

All three jets exhibit similar features in the flow field around the orifice (figure~\ref{fig:AllOrfReStress}). 
Upstream of the orifice a region of negative Reynolds shear stress is associated with the localized influence of the suction phase of the actuator which draws higher momentum fluid down towards the actuator orifice (Q4). 
Alternatively, the high velocity jet produced by the blowing cycle of the actuator which is bent downstream by the cross-flow is responsible for the region of positive Reynolds shear stress around the orifice's downstream corner (Q1). 
Downstream of the actuator orifice the unsteady flow fields of the pitched and wall-normal jets follow different trends.

In the $\alpha=90^\circ$, $\beta=0^\circ$ and $\alpha=90^\circ$, $\beta=90^\circ$ jets a streak of negative $\langle u'v' \rangle$ traces a trajectory which generally follows the path swept out by the primary vortex structures. 
In this region of the flow field the action of the synthetic jet vortices enhances unsteady mixing within the BL. 
Conversely, in the $\alpha=45^\circ$, $\beta=0^\circ$ jet the region of positive $\langle u'v' \rangle$ originating at the orifice extends into the far field. 
In this jet the periodic ejections of high momentum fluid from the pitched jet present as Q1 events and unsteady mixing is considerably lower than it is in the other two jets.

To quantify the intensity of the velocity fluctuations induced by the different jets, the kinetic energy associated with the fluctuations was computed. 
Here the quantity $K(x)$ was obtained by integrating the local kinetic energy of the velocity fluctuations in the wall-normal ($y$) and spanwise ($z$) directions (\ref{eq:Kfluc}).

\begin{figure}
\centering
\includegraphics[width=4in]{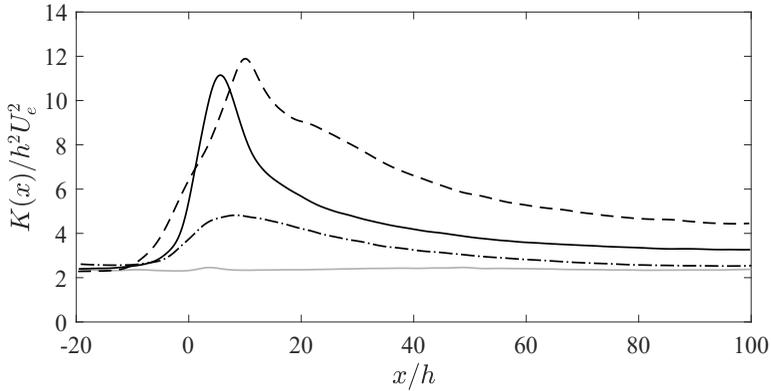}
\caption{Streamwise variation in the time-averaged, integrated fluctuation kinetic energy for the baseline flow (\protect\greyline), $\alpha=90^\circ$, $\beta=0^\circ$ jet (\sampleline{}), $\alpha=90^\circ$, $\beta=90^\circ$ jet (\sampleline{dash pattern=on .475em off .3em on .475em off .3em}), and $\alpha=45^\circ$, $\beta=0^\circ$ jet (\sampleline{dash pattern=on .55em off .2em on .05em off .2em}) cases.}
\label{fig:AllOrfKfluc}
\end{figure}

\begin{equation} 
\label{eq:Kfluc}
K(x)=\iint \frac{1}{2} \Big( \big \langle u'^2 \big \rangle + \big \langle v'^2 \big \rangle + \big \langle w'^2 \big \rangle \Big) dydz
\end{equation}

The results for the baseline flow field and the three jets are presented in figure~\ref{fig:AllOrfKfluc}. 
In all three jets the fluctuating kinetic energy peaks a short distance downstream from the orifice and then decreases downstream as the vortices dissipate.
The kinetic energy is markedly higher in the jets which issue normal to the wall than it is in the pitched jet. 
The $\alpha=90^\circ$, $\beta=90^\circ$ jet in particular maintains a higher magnitude of $K(x)$ in the far field because, after the vortex ring axis-switches, the majority of the ring length is aligned in the spanwise direction. 
While streamwise vortices induce mean mixing in the flow field, as shown in \S~\ref{sec:meanmix}, the alternating sign of the wall-normal velocity induced on the upstream versus the downstream side of spanwise-oriented vortices leads to high unsteadiness as the vortices advect downstream. 
Finally, fluctuating kinetic energy remains relatively low in the $\alpha=45^\circ$, $\beta=0^\circ$ jet, which is to be expected since pitching the orifice aligns the jet closer to the direction of the cross-flow which in turn decreases the magnitude of the mean shear stress associated with the SJBLI.

\section{Influence of actuation conditions}
\label{sec:results3}

The results presented in \S~\ref{sec:results1} and \ref{sec:results2} were all obtained at a single condition, $C_b=1$, $St=1.33$ ($U_o$ = 12~ms$^{-1}$, $U_e$ = 12~ms$^{-1}$). 
In the following section the influence of blowing ratio and Strouhal number on the SJBLI in the centreline plane, for all three orifice geometries, is explored.

\subsection{Trajectory and propagation speed of periodic structures}
\label{sec:vorttraj2}

\begin{figure}
\centering
\includegraphics[width=5in]{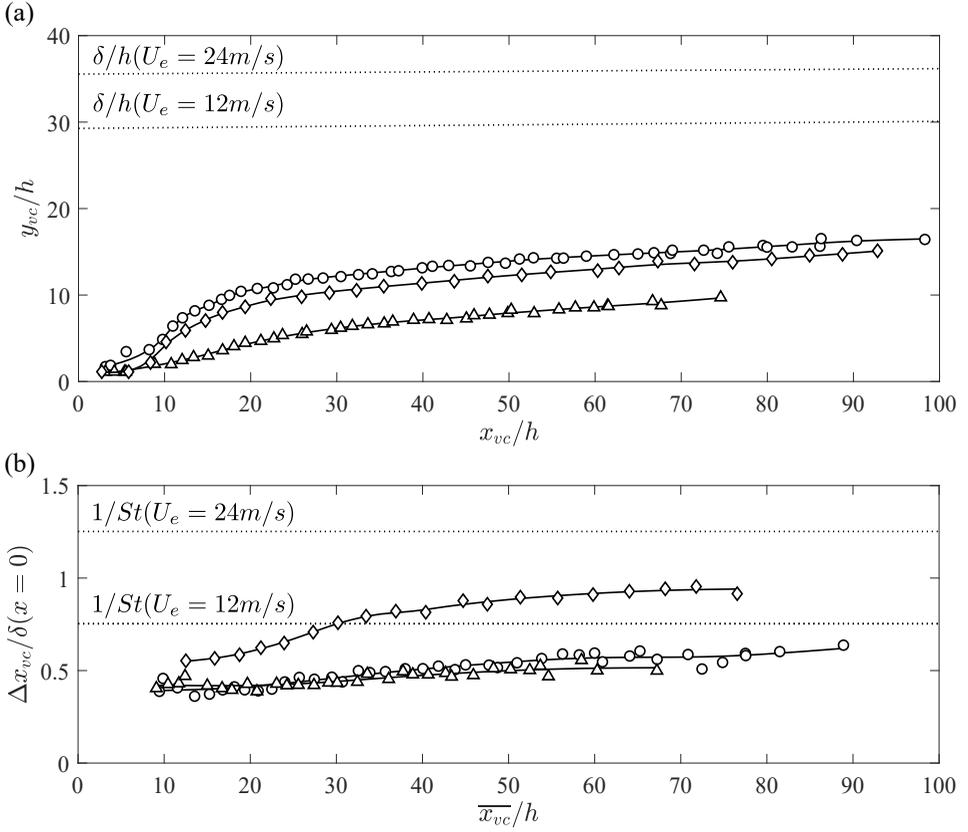}
\caption{$\alpha=90^\circ$, $\beta=0^\circ$ jet centreline vortex centre (a) locations and (b) streamwise spacing from phase-locked data for $C_b=1.0$, $St=1.33$ ($\circ$), $C_b=0.5$, $St=1.33$ ($\vartriangle$), and $C_b=0.5$, $St=0.8$ ($\lozenge$).}
\label{fig:AllCondVortTraj}
\end{figure}

\begin{figure}
\centering
\includegraphics[width=5in]{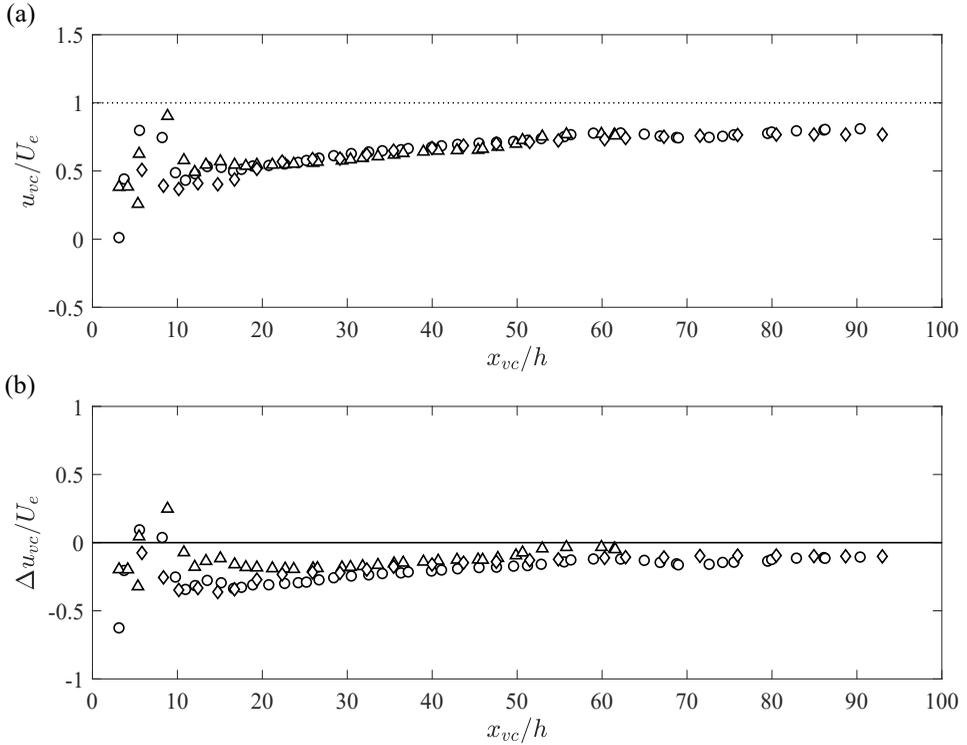}
\caption{$\alpha=90^\circ$, $\beta=0^\circ$ jet centreline vortex (a) absolute streamwise velocity and (b) change in streamwise velocity for $C_b=1.0$, $St=1.33$ ($\circ$), $C_b=0.5$, $St=1.33$ ($\vartriangle$), and $C_b=0.5$, $St=0.8$ ($\lozenge$).}
\label{fig:AllCondVortVel}
\end{figure}

The influence of the actuation and cross-flow conditions on the dominant vortex trajectory and streamwise spacing is depicted in figure~\ref{fig:AllCondVortTraj}a and figure~\ref{fig:AllCondVortTraj}b, respectively, for the $\alpha=90^\circ$, $\beta=0^\circ$ jet. 
Additionally, the absolute and relative streamwise velocity of the dominant vortices is presented in figure~\ref{fig:AllCondVortVel}a and figure~\ref{fig:AllCondVortVel}b, respectively. 
The data presentation format and parameter definitions associated with figure~\ref{fig:AllCondVortTraj} and \ref{fig:AllCondVortVel} follow the descriptions given for figure~\ref{fig:AllOrfVortTraj} and \ref{fig:AllOrfVortVel} in \S~\ref{sec:results2}. 
Furthermore, the results at the $C_b=1$, $St=1.33$ condition, which were previously presented in figure~\ref{fig:AllOrfVortTraj} and \ref{fig:AllOrfVortVel}, are repeated here for comparison.

Comparing the $C_b=0.5 $, $St=1.33$ and $C_b=1.0$, $St=1.33$ SJBLI conditions provides insight into the effect of reducing $C_b$ while holding $St$ constant. 
Reducing the synthetic jet average blowing velocity significantly decreases the distance the jet penetrates into the cross-flow (figure~\ref{fig:AllCondVortTraj}a) which is an expected outcome since the ratio of the jet wall-normal momentum to the cross-flow streamwise momentum has dropped. 

The reduction in $C_b$ between $C_b=1.0$, $St=1.33$ and $C_b=0.5 $, $St=1.33$ was achieved by reducing the average blowing velocity of the actuator. 
In quiescent conditions the circulation strength of the vortex ring scales linearly with $U_o^2/f$, all else being equal \citep{Straccia_Farnsworth_JFM_2020}. 
%Thus a reduction in $U_o$ from $12ms^{-1}$ to $6ms^{-1}$ would imply that circulation strength drops by a factor of four. 
%However, how a reduction of $U_o$ influences the circulation strength of the upstream (CCW) or downstream (CW) side of the vortex ring formed in a cross-flow is more complicated. 
Thus it would be reasonable to predict that a reduction in $U_o$ from 12~ms$^{-1}$ to 6~ms$^{-1}$ should result in a drop in circulation strength by a factor of four. 
However, predicting how a reduction of $U_o$ would influence the circulation strength of the upstream (CCW) or downstream (CW) side of the vortex ring formed in a cross-flow is more complicated.
Specifically, as $C_b$ drops the outstroke velocity profile at the orifice exit becomes more skewed, such that CW vorticity production is favored over CCW vorticity \citep{Sau_Mahesh_2008}. 
The net result of reducing $C_b$ from 1 to 0.5 while holding $St$ constant is that the CW-rotating hairpin vortices produced by the $\alpha=90^\circ$, $\beta=0^\circ$ SJA have lower circulation strength, albeit by less than a factor of four. 
Circulation strength calculations conducted by integrating the velocity field along circuits that enclosed the vortex in the centreline phased-locked data indicated that the CW vortices at $C_b=0.5 $, $St=1.33$ were between two to three times weaker than at $C_b=1.0$, $St=1.33$. 
The lower CW vortex strength affected not just the jet penetration but also the rate of the self-induced deformations which is evident from the fact that the S-curve trajectory of the CW vortex is much shallower at the lower $C_b$ (figure~\ref{fig:AllCondVortTraj}a).

The vortex spacing for the two $St=1.33$ cases are fairly similar despite moderate differences in penetration height and circulation strength (figure~\ref{fig:AllCondVortTraj}b). 
While the $C_b=0.5$, $St=1.33$ hairpin vortices remain closer to the wall in slower moving fluid (figure~\ref{fig:AllCondVortTraj}a) than the $C_b=1.0$, $St=1.33$ vortices, they also induce a smaller upstream velocity on themselves due to their lower circulation strength, i.e., their $\Delta u_{vc}/U_e$ is closer to zero (figure~\ref{fig:AllCondVortVel}b).
The net result is that the $C_b=0.5$, $St=1.33$ vortices travel somewhat slower (figure~\ref{fig:AllCondVortVel}a) and are spaced slightly closer together in the streamwise direction (figure~\ref{fig:AllCondVortTraj}b) than the $C_b=1.0$, $St=1.33$ vortices, but the difference is small.

The third SJBLI condition tested was achieved by doubling the cross-flow speed to $U_e$ = 24~ms$^{-1}$, while holding $U_o$ = 12~ms$^{-1}$, which reduced $St$ to 0.8 and $C_b$ to 0.5. 
Although the $C_b=0.5$, $St=0.8$ jet did not penetrate as deeply into the cross-flow as the $C_b=1$, $St=1.33$ jet, the difference is fairly small compared with the low penetration of the $C_b=0.5$, $St=1.33$ jet (figure~\ref{fig:AllCondVortTraj}a). 
This finding may come as something of a surprise given that the actuator driving conditions ($U_o$ and $f$) are the same between the $C_b=0.5$, $St=0.8$ and $C_b=1$, $St=1.33$ cases, but the vortices in the $C_b=0.5$, $St=0.8$ case encounter a much faster moving cross-flow. 
While these two jets would produce vortex rings with identical circulation strength in quiescent conditions the difference in $C_b$ alters how circulation is distributed around the ring. 
The CW side of the vortex ring, which eventually develops into the hairpin vortices, is stronger in the $C_b=0.5$, $St=0.8$ jet than in the $C_b=1$, $St=1.33$ jet due to enhanced CW vorticity production in the more highly skewed orifice exit velocity profile. 
The fact that the $C_b=0.5$, $St=0.8$ jet penetration depth and spatial rate of self-induced deformations (i.e., the slope of vortex S-curve trajectory) is only slightly lower than that of the $C_b=1$, $St=1.33$ jet implies that the CW vortices have around twice the circulation strength (figure~\ref{fig:AllCondVortTraj}a), a supposition supported by vortex circulation strength calculations on centreline.

The lower $St$ of the $C_b=0.5$, $St=0.8$ jet leads to a larger streamwise spacing between CW vortices as compared with the $St=1.33$ jets (figure~\ref{fig:AllCondVortTraj}b). 
In fact, the ratio of $\Delta x_{vc}/\delta(x=0)$ between the $C_b=0.5$, $St=0.8$ and $C_b=1.0$, $St=1.33$ jets is roughly 1.62, only slightly lower than the ratio of their $St$, which is 1.66. 
The normalized relative streamwise velocity of the vortices in the $C_b=0.5$, $St=0.8$ jet is also similar to the $C_b=1.0$, $St=1.33$ jet, albeit slightly closer to zero indicating a slightly weaker self-induced velocity relative to the cross-flow speed (figure~\ref{fig:AllCondVortVel}b).

It is interesting that the hairpin vortices at all three jet conditions travel downstream at similar rates relative to $U_e$, after the initial self-induced deformations subside (figure~\ref{fig:AllCondVortVel}a). 
This convergence of propagation speeds is a testament to the competing effects that the vortex trajectory and self-induction have on the vortex velocity when the vortex circulation strength is varied in the $\alpha=90^\circ$, $\beta=0^\circ$ jet.
%
%While it is not practical to present the vortex trajectory, spacing, and velocity data for all conditions and orifice geometries tested, we will outline the largest differences. 

\begin{figure}
\centering
\includegraphics[width=5in]{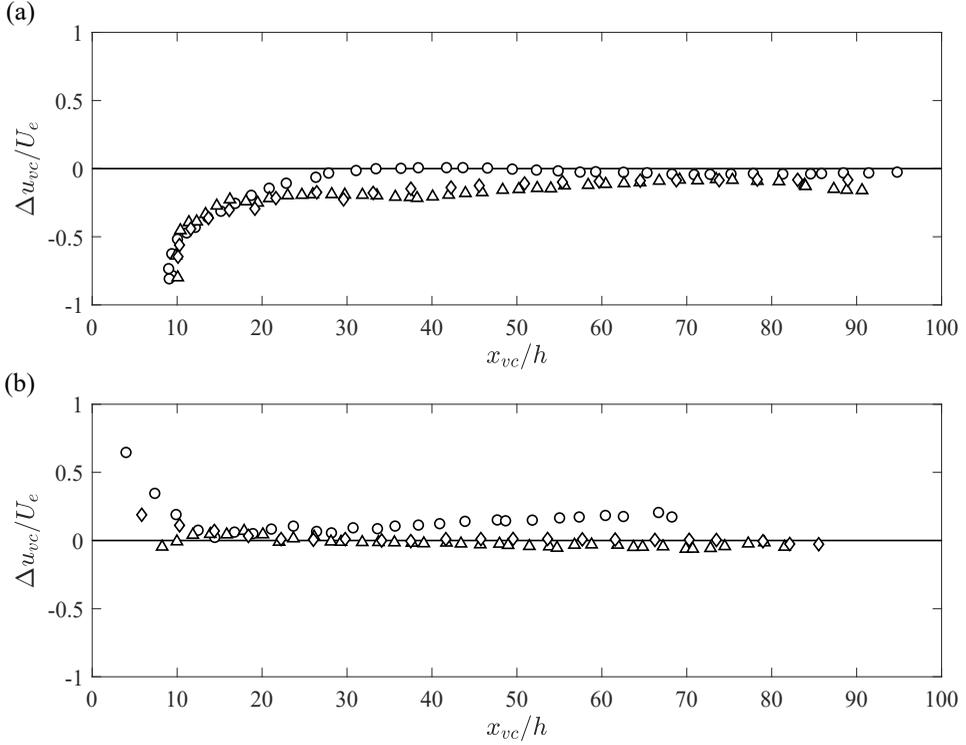}
\caption{Change in vortex streamwise velocity in the (a) $\alpha=90^\circ$, $\beta=90^\circ$ and (b) $\alpha=45^\circ$, $\beta=0^\circ$ jets for $C_b=1.0$, $St=1.33$ ($\circ$), $C_b=0.5$, $St=1.33$ ($\vartriangle$), and $C_b=0.5$, $St=0.8$ ($\lozenge$).}
\label{fig:AllCondVortVel2}
\end{figure}

The jet trajectory from the $\alpha=90^\circ$, $\beta=90^\circ$ SJA follows similar trends with changing $C_b$ and $St$ as the $\alpha=90^\circ$, $\beta=0^\circ$ SJA case does. 
Namely, penetration is highest at $C_b=1$, $St=1.33$, somewhat lower at $C_b=0.5$, $St=0.8$, and much lower at $C_b=0.5$, $St=1.33$. 
The most notable differences between the jets formed by the wall-normal orifice geometries occur in the vortex spacing and velocity. 
While the $\alpha=90^\circ$, $\beta=90^\circ$ jet vortex rings at a $C_b$ of 0.5 initially undergo similar self-induced deformations as they did at $C_b=1$ early in the jet development, the now weaker CCW side of the vortex ring has a diminished influence farther downstream. 
%Instead of acting similarly to a vortex ring with constant circulation strength about its circumference the $C_b=0.5$ vortex rings are dominated by the relatively stronger CW rotating side. 
The behaviour of the vortex diverges further from that of an ideal vortex ring with constant circulation strength about its circumference and is increasingly dominated by its relatively stronger CW rotating side. 
The resulting $C_b=0.5$ periodic structures behave more like the hairpin vortices described in \S~\ref{sec:AR18Beta0}, inducing an upstream pointing velocity on themselves and thus travel at a slower streamwise velocity than the baseline flow (i.e., $\Delta u_{vc}/U_e<0$)(figure~\ref{fig:AllCondVortVel2}a). 
Therefore, the vortices have a tighter streamwise spacing at $C_b=0.5$, $St=1.33$ than they did at $C_b=1$, $St=1.33$.

The influence changing $C_b$ and $St$ has on the $\alpha=45^\circ$, $\beta=0^\circ$ jet is quite different from the situation with the $\alpha=90^\circ$, $\beta=0^\circ$ jet because their dominant vortices rotate in opposite directions. 
Reducing $C_b$ to 0.5 in the $\alpha=45^\circ$, $\beta=0^\circ$ jet caused its BL penetration to drop significantly, for both of the low $C_b$ cases. 
However, for this orifice, the $C_b=0.5$, $St=0.8$ jet travelled the shortest distance from the wall.
As described earlier, reducing $C_b$ causes the CW side of the vortex ring to become stronger and the CCW side to become weaker. 
Unlike with the other orifice geometries this change further weakens the dominant vortex instead of mitigating some of the reduction in the strength of the vortex relative to the cross-flow. 
Therefore, instead of traveling faster than the baseline flow, as the arch-shaped vortices did at the $C_b=1$, $St=1.33$ conditions, the vortices at $C_b=0.5$, $St=0.8$ traveled at roughly the same speed as the baseline flow ($\Delta u_{vc}/U_e=0$), and the vortices at $C_b=0.5$, $St=1.33$ traveled somewhat slower ($\Delta u_{vc}/U_e<0$)(figure~\ref{fig:AllCondVortVel2}b).
The result is a tighter streamwise spacing between the periodic structures at $C_b=0.5$, $St=1.33$ than at $C_b=1.0$, $St=1.33$, despite the actuation frequency being the same at the two conditions.

\subsection{Velocity profile on centreline at \texorpdfstring{$x/h=100$}{TEXT}}
\label{sec:exituprof}

\begin{figure}
\centering
\includegraphics[width=4in]{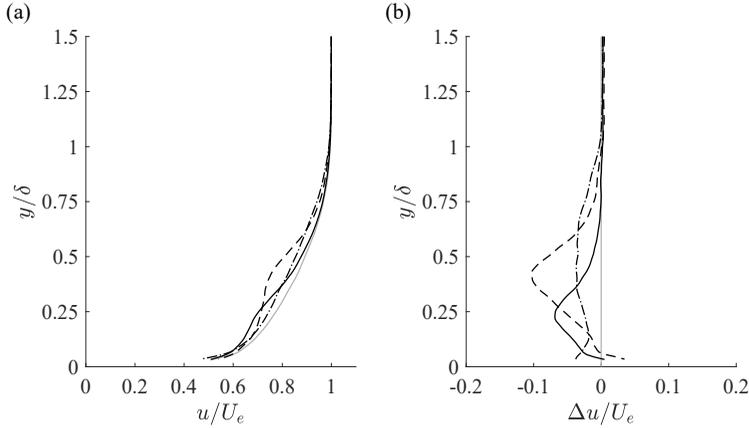}
\caption{Centerline time-averaged (a) streamwise velocity and (b) change in streamwise velocity profiles at $x/h \approx 100$ for the baseline flow (\protect\greyline), $\alpha=90^\circ$, $\beta=0^\circ$ jet (\sampleline{}), $\alpha=90^\circ$, $\beta=90^\circ$ jet (\sampleline{dash pattern=on .475em off .3em on .475em off .3em}), and $\alpha=45^\circ$, $\beta=0^\circ$ jet (\sampleline{dash pattern=on .55em off .2em on .05em off .2em}) at $C_b=0.5$, $St=1.33$.}
\label{fig:AllOrfBLProf2}
\end{figure}

\begin{figure}
\centering
\includegraphics[width=4in]{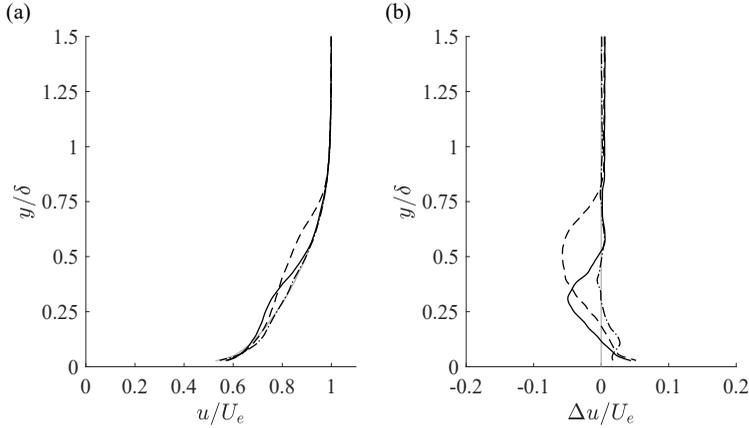}
\caption{Centerline time-averaged (a) streamwise velocity and (b) change in streamwise velocity profiles at $x/h \approx 100$ for the baseline flow (\protect\greyline), $\alpha=90^\circ$, $\beta=0^\circ$ jet (\sampleline{}), $\alpha=90^\circ$, $\beta=90^\circ$ jet (\sampleline{dash pattern=on .475em off .3em on .475em off .3em}), and $\alpha=45^\circ$, $\beta=0^\circ$ jet (\sampleline{dash pattern=on .55em off .2em on .05em off .2em}) at $C_b=0.5$, $St=0.8$.}
\label{fig:AllOrfBLProf3}
\end{figure}

The velocity profiles at $C_b=0.5$, $St=1.33$ and $C_b=0.5$, $St=0.8$ for all three orifices are presented in figures \ref{fig:AllOrfBLProf2} and figure~\ref{fig:AllOrfBLProf3}, respectively.
To compare the effect of actuation at all three SJBLI conditions tested figure~\ref{fig:AllOrfBLProf} should be referenced in conjunction with figures \ref{fig:AllOrfBLProf2} and figure~\ref{fig:AllOrfBLProf3}.

One striking observation when comparing figures \ref{fig:AllOrfBLProf}, \ref{fig:AllOrfBLProf2}, and \ref{fig:AllOrfBLProf3} is that the superiority of the $\alpha=45^\circ$, $\beta=0^\circ$ SJA at $C_b=1.0$ does not extend to either of the $C_b=0.5$ conditions tested. 
In fact, at $C_b=0.5$, $St=1.33$ the $\alpha=45^\circ$, $\beta=0^\circ$ SJA actually decreased the centreline flow velocity everywhere within the BL (figure~\ref{fig:AllOrfBLProf2}b). 
The results are somewhat more favorable at $C_b=0.5$, $St=0.8$ where the $\alpha=45^\circ $, $\beta=0^\circ $ jet did enhance near-wall flow velocity, but the effect is much weaker than at $C_b=1.0$, $St=1.33$ and more localized (figure~\ref{fig:AllOrfBLProf3}b). 

At both conditions plotted in figures \ref{fig:AllOrfBLProf2} and figure~\ref{fig:AllOrfBLProf3} the $\alpha=45^\circ$, $\beta=0^\circ$ SJA was the least effective of the three actuators at enhancing the near-wall flow speed, highlighting an important difference between the flow control mechanisms of the pitched versus wall-normal orifices. 
While the wall-normal orifices modify the BL shape via mixing and flow redistribution within the BL the pitched orifice acts primarily through direct momentum addition. 
Thus the effectiveness of a pitched orifice is highly dependent on the $C_b$. 
If $C_b$ is set at a level such that the streamwise component of the outstroke is too low relative to the flow speed near the wall then the actuation may act as a net sink on the fluid momentum (see for example figure~\ref{fig:AllOrfBLProf2}b). 

At $C_b=0.5$, $St=1.33$ the $\alpha=90^\circ$, $\beta=0^\circ$ and $\alpha=90^\circ$, $\beta=90^\circ$ SJAs had a similar influence on the centreline BL profile as they did at $C_b=1.0$, $St=1.33$, but the modifications are merely more muted (figure~\ref{fig:AllOrfBLProf2}b). 
Specifically, the wakes are shallower in depth and located closer to the wall while the velocity enhancement at the wall is weaker. 
These changes are to be expected for a reduction in the momentum output by the actuator.

At $C_b=0.5$, $St=0.8$ the $\alpha=90^\circ$, $\beta=0^\circ$ and $\alpha=90^\circ$, $\beta=90^\circ$ jet velocity profiles also contain shallow wakes which remain relatively close to the wall (figure~\ref{fig:AllOrfBLProf3}b). 
What is interesting is that the near-wall flow velocity enhancement is still very robust in this case (figure~\ref{fig:AllOrfBLProf3}b). 
In fact, the enhancement of $\Delta u/U_e$ near the wall by the $\alpha=90^\circ$ SJAs at $C_b=0.5$, $St=1.33$ is similar in magnitude to what was observed at $C_b=1.0$, $St=1.33$, despite operating at a lower blowing ratio (compare figure~\ref{fig:AllOrfBLProf}b to figure~\ref{fig:AllOrfBLProf3}b). 
In the following sections the analysis of the mean and fluctuating velocity fields on the jet centreline reveals some interesting differences between the influence of the actuation at $C_b=1.0$, $St=1.33$ and $C_b=0.5$, $St=0.8$.

\subsection{Average vertical mixing on the jet centreline}
\label{sec:deltavfield}

\begin{figure}
\centering
\includegraphics[width=5in]{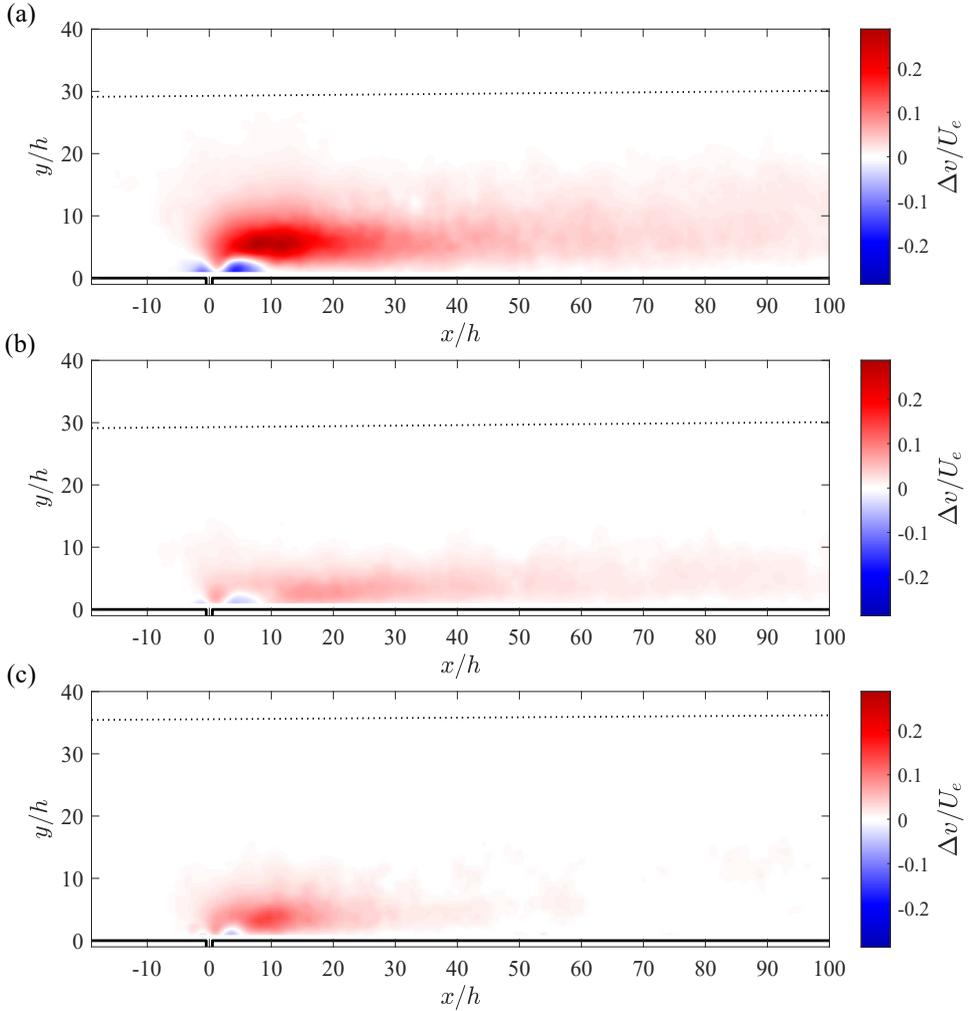}
\caption{Contours of time-averaged wall-normal velocity change on the $\alpha=90^\circ$, $\beta=0^\circ$ jet centreline for (a) $C_b=1.0$, $St=1.33$, (b) $C_b=0.5$, $St=1.33$, and (c) $C_b=0.5$, $St=0.8$.}
\label{fig:AllCondvfield}
\end{figure}

To compare the mean flow redistribution induced by the $\alpha=90^\circ$, $\beta=0^\circ$ SJA at the three conditions tested, contours of $\Delta v$ were plotted on the centreline plane in figure~\ref{fig:AllCondvfield}. 
Although the magnitude of the induced wall-normal velocity on centreline varies between the cases the prominent flow field features are the same. 
%Just upstream and downstream of the orifice a region of downward moving fluid (blue contours) is the manifestation of the localized influence of the suction cycle. 
Just upstream and downstream of the orifice a region of downward moving fluid (blue contours) is the manifestation of the localized influence of the suction cycle along with the downwards velocity induction around the early vortex ring.
The path of the jet out of the orifice during the blowing cycle introduces a streak of upward moving fluid through the middle of the blue contours. 
This region of upward moving fluid in the mean velocity field extends downstream from the orifice, with peak upwash occurring between $x/h=10$ and 15 before decreasing downstream. 

%The relative strength of the upwash at $C_b=0.5$, $St=1.33$ is much lower than at $C_b=1.0$, $St=1.33$, which is the expected outcome of reducing the blowing ratio (figure~\ref{fig:AllCondvfield}). 
%It is interesting however that halving the blowing ratio reduced the strength of the maximum upwash by a factor of about 3.6. 
%By comparison, the ratio of maximum upwash to cross-flow speed at $C_b=0.5$, $St=0.8$ dropped by half relative to the $C_b=1.0$, $St=1.33$ condition, in line with the halving of blowing ratio. 
%While the low relative strength of the upwash at $C_b=0.5$, $St=0.8$ is consistent with the expected effect of reducing blowing ratio, the implied reduction in mean flow redistribution within the BL is at odds with the noteworthy effectiveness of the jet at this condition in enhancing the near-wall flow velocity, as seen in figure~\ref{fig:AllOrfBLProf3}b.
%Therefore, another factor must be at play.
The relative strength of the upwash at the two $C_b=0.5$ conditions was lower than at $C_b=1.0$, which is the expected outcome of reducing the blowing ratio (figure~\ref{fig:AllCondvfield}). 
Of the two $C_b=0.5$ cases, the $C_b=0.5$, $St=0.8$ jet achieved a higher peak $\Delta v/U_e$ value than the $C_b=1.0$, $St=1.33$ jet did, although, it still had only half the peak $\Delta v/U_e$ value measured in the $C_b=1.0$, $St=1.33$ jet.
While it is unsurprising that changing the actuation conditions from $C_b=1.0$, $St=1.33$ to $C_b=0.5$, $St=0.8$ led to a decrease in the relative strength of upwash on centreline, the implied reduction in mean flow redistribution within the BL is at odds with the noteworthy effectiveness of the $C_b=0.5$, $St=0.8$ jet in enhancing the near-wall flow velocity, as seen in figure~\ref{fig:AllOrfBLProf3}b.
Therefore, another factor must be at play.

\subsection{Reynolds stress field on the jet centreline}
\label{sec:Restressfield}

\begin{figure}
\centering
\includegraphics[width=5in]{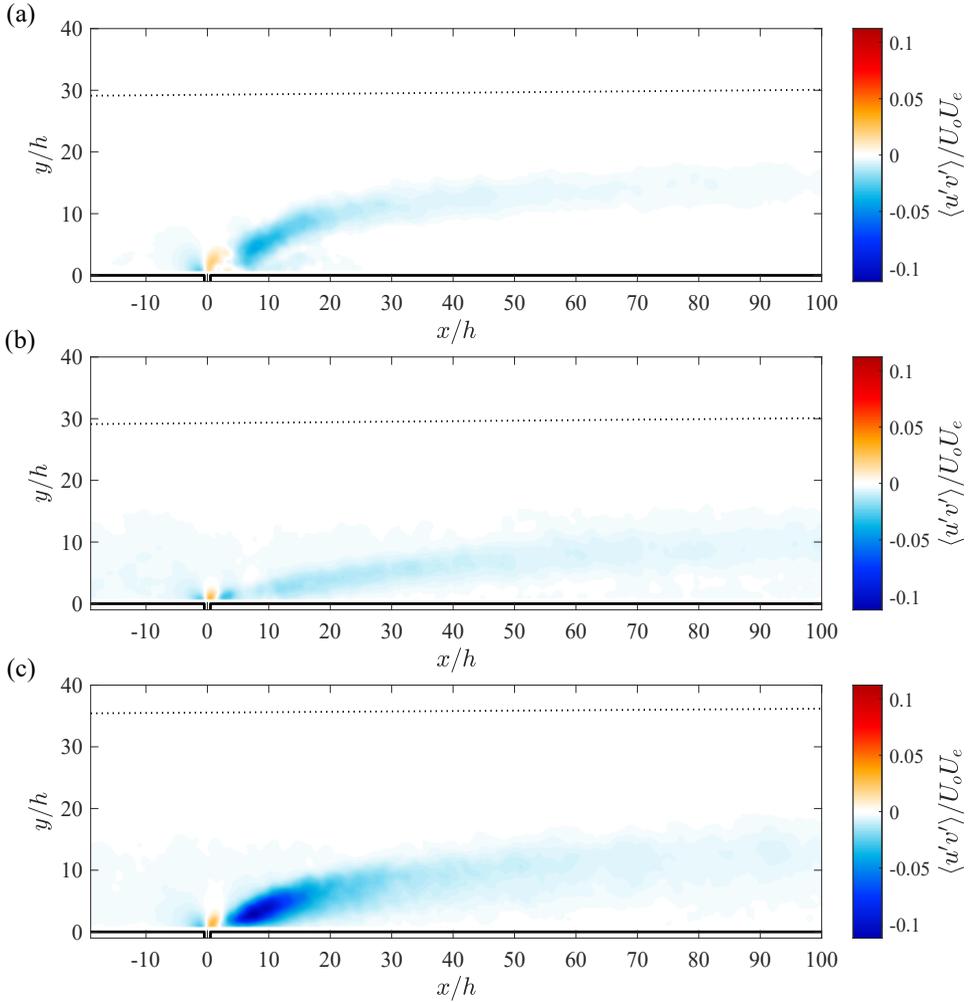}
\caption{Time-averaged Reynolds stress contours on the $\alpha=90^\circ$, $\beta=0^\circ$ jet centreline for (a) $C_b=1.0$, $St=1.33$, (b) $C_b=0.5$, $St=1.33$, and (c) $C_b=0.5$, $St=0.8$.}
\label{fig:AllCondReStress}
\end{figure}

To help understand the surprisingly good performance of the $\alpha=90^\circ$, $\beta=0^\circ$ SJA at $C_b=0.5$, $St=0.8$ a Reynolds decomposition, per \ref{eq:uprime}, of the instantaneous velocity fields was conducted at the three driving conditions. 
The resulting Reynolds stress fields on centreline are presented in figure~\ref{fig:AllCondReStress}. 
Note that the data plotted in figure~\ref{fig:AllCondReStress}a are the same as in figure~\ref{fig:AllOrfReStress}a but with the contour levels now rescaled for easier comparison with the $C_b=0.5$ condition results. 
What is conspicuous in these results is the remarkably high Reynolds shear stress in the $C_b=0.5$, $St=0.8$ jet as compared with the $C_b=1.0$, $St=1.33$ condition, despite the blowing ratio being much lower. 
In fact the peak value of $\langle u'v' \rangle$ at $C_b=0.5$, $St=0.8$ is 2.7 times higher in magnitude than measured at $C_b=1.0$, $St=1.33$. 

Differences in vortex spacing may be one factor that contributes to the observed disparity in unsteady mixing.
At high $St$ subsequent generations of periodic structures are spaced more closely together, and the velocity induction from the spanwise portion of one vortex destructively interferes with the velocity induction of its neighboring structures. 
Reducing $St$ increases the spacing between the structures which in turn reduces the mutual interaction between different generations of vortices and allows each vortex to induce stronger unsteady vertical mixing. 
Furthermore, the formation of relatively stronger CW rotating vortices at the lower blowing ratio --- as described in \S~\ref{sec:vorttraj2} --- also enables strong velocity induction by the hairpin vortices. 
The phase-locked velocity fields (not presented here) support this interpretation. 
Thus while the relative strength of the mean mixing was lower at $C_b=0.5$, $St=0.8$ as compared with $C_b=1.0$, $St=1.33$, the unsteady mixing was much higher, resulting in an unexpectedly strong enhancement of the near-wall flow speed at the lower $C_b$ and $St$. 
\section{Summary and conclusions}
\label{sec:conclusions}

The vortex dynamics resulting from the interaction of an $AR=18$ rectangular orifice synthetic jet with a turbulent boundary layer was experimentally investigated by measuring the three dimensional three-component velocity field within the jet using stereoscopic particle image velocimetry. 
Three common orifice geometries and three practical actuation conditions were tested to explore the impact of factors of actuator design and operation that govern the jet flow physics. 
The measurements revealed a variety of different vortex structures and enabled valuable connections to be made between the vortex dynamics and the jet's influence on the boundary layer.

%The vortices produced by the synthetic jet actuator while operating in a cross-flow looked and behaved differently from the vortex rings which form in quiescent conditions. 
One prominent effect of the cross-flow interaction was to diminish the circulation strength of the upstream side of the initial vortex ring while strengthening the downstream side. 
When the synthetic jet interacted strongly with the cross-flow, as was the case in the wall-normal spanwise-oriented orifice, the weak upstream side of the ring quickly decayed, and the downstream flow field was dominated by a train of vortices which rotated clockwise in the spanwise direction. 
While the initial vortex structure was connected to the wall through legs which branched off from the primary vortex, downstream a new topology developed in the form of hairpin vortices which were connected to each other via their legs. 
Rotating the orifice $90^\circ$ to align the long axis with the streamwise direction produced a more intact vortex ring which propagated rapidly away from the wall and axis-switched once before the self-induced deformations subsided. 
Although the circulation strength still varied around the circumference of the vortex rings from the streamwise-oriented orifice, evidenced by the presence of leg and bridge vortices, the differences were smaller than those seen in the spanwise-oriented jet at the same conditions. 

It should be noted that hairpin vortices are not exclusive to spanwise-oriented jets, and that distorted vortex rings are not restricted to actuators with streamwise-oriented orifices. 
It is possible to generate both types of structures from either orientation of the orifice. 
The difference between these two orifices is the conditions at which the jet transitions from containing one type of vortex structure to the other. 
Because jets from spanwise-oriented orifices interact strongly with the boundary layer, due to the large cross-section they present to the cross-flow, a larger blowing ratio is required to produce a structure which behaves like a vortex ring than the blowing ratio required for a jet issuing from a streamwise-oriented orifice. 
This blowing ratio threshold likely depends also on the aspect ratio of the orifice, the thickness of the boundary layer, and whether the flow is laminar or turbulent. 
For example, \citet{Wang_Feng_JFM_2020} found that their $AR=3$ spanwise-oriented rectangular orifice synthetic jet produced axis-switching vortex rings in a laminar cross-flow at a blowing ratio of one. 
Additionally, in the present study the centreline velocity field of the streamwise-oriented synthetic jet suggests that the vortices produced at a blowing ratio of 0.5 began to behave more like hairpin vortices and less like the modified vortex rings observed at a blowing ratio of one.

Another factor influencing the type of vortex structures produced by the synthetic jet was the pitch angle of the orifice. 
Pitching a spanwise-oriented orifice downstream had the opposite effect on the circulation strength of the vortex ring as the cross-flow did. 
Namely, the upstream side of the vortex ring became stronger and the downstream side weaker. 
Therefore, at a blowing ratio of one the pitched orifice produced a train of counterclockwise-rotating arch-shaped vortices which remained connected to the wall as they advected downstream. 

The effect that the jets had at a blowing ratio of one on the boundary layer varied significantly with the type of vortex structure formed. 
The wall-normal jets induced mixing which transported high momentum fluid towards the wall, accelerating the fluid there, while low momentum fluid transported off the wall was deposited into a roughly circular wake higher in the cross-flow.
Alternatively, the pitched jet accelerated a thick layer of fluid near the wall through direct momentum injection, and, although a layer of fluid higher in the cross-flow was decelerated slightly by the jet, there was no localized wake.

From the perspective of delaying separation, the pitched jet is believed to be more likely to offer effective control at a blowing ratio of one than the wall-normal jets did because the layer of accelerated fluid it produced was thicker and the increase in streamwise velocity produced was larger than those in the other jets. 
Interestingly, when the actuator average blowing velocity was decreased, reducing blowing ratio to 0.5, the pitched jet became a net detriment on or sink for the cross-flow momentum while the wall-normal jets were still able to enhance the near-wall flow speed, albeit to a lesser degree than they did at the higher blowing ratio. 
The differences in the relative performance of the different jets at a lower blowing ratio relates to the mechanism by which they alter the flow field. 
Whereas the wall-normal jets were still able to increase mixing in the turbulent boundary layer at the lower blowing ratio the pitched jet no longer had sufficient excess momentum relative to the cross-flow to be able to accelerate the near-wall fluid. 

%\hl{The performance of the wall-normal synthetic jets with the two orifice orientations was subject to a trade-off between the near-wall persistence of the periodic structures and the strength of their induced velocity.
The performance of the wall-normal synthetic jets in the two orifice orientations tested was subject to a trade-off between the near-wall persistence of the periodic structures and the strength of their induced velocity.
The more complete vortex rings produced by the streamwise-oriented orifice induced a strong near-wall vertical velocity when the vortex ring first formed, but they also propagated away from the wall quickly, which diminished their influence downstream.
Conversely, the wall-normal velocity induction and rate of mixing was more modest in the spanwise-oriented synthetic jet, but the periodic structure which developed into the hairpin vortices remained relatively close to the wall, extending the domain of their near-wall influence.
The spanwise-oriented synthetic jet also enhanced the near-wall flow speed in the areas to the sides of the jet centreline more than the streamwise-oriented jet did.
The net result of the trade-offs between the two wall-normal orifice orientations was that the downstream difference in the near-wall flow speed was small, but a slight advantage was measured for a spanwise orifice orientation when the total footprint of the jet was considered instead of just the jet centreline.

An interesting discovery was made at the third condition tested where a blowing ratio of 0.5 was achieved by increasing the cross-flow speed, instead of by reducing the actuator average blowing velocity. 
Increasing the cross-flow speed reduced the Strouhal number of the jet, i.e., it increased the spacing between the vortices in the boundary layer. 
While the relative rate of wall-normal fluid transport within the mean field dropped significantly from what it had been at a blowing ratio of one, the wall-normal jets exhibited particularly high unsteady mixing at the lower Strouhal number. 
The unsteady mixing was driven by the alternating direction of vertical velocity induction from the spanwise-oriented portion of the vortices and increased when the vortices were spaced further apart, i.e., when they interacted with each other more weakly. 
Furthermore, when the cross-flow speed was doubled the clockwise-rotating vortices produced by the wall-normal jets had roughly double the circulation strength that they had at a blowing ratio of one, despite the fact that the driving conditions of the actuator were unchanged. 
This phenomenon can be attributed to the fact that, at the reduced blowing ratio, the imbalance in the jet vorticity production due to the cross-flow interaction was enhanced such that it skewed further in favor of clockwise vorticity. 
Thus the detrimental impact that reducing the blowing ratio had on the relative strength of induced velocity from the vortices in the jets was partially mitigated in the wall-normal jets --- which were dominated by clockwise-rotating structures --- and amplified in the pitched jet --- which was dominated by counterclockwise-rotating vortices.

The present experimental study sought to investigate the fundamental vortex dynamics of synthetic jet-boundary layer interactions, but the conditions and geometries tested were selected intentionally for their relevance to the application of active flow control. 
The insight provided by this work can help connect particular vortex structures to their characteristic influence on the surrounding flow field, whether that is the height and footprint of the jet, the intensity of mean and unsteady mixing, or the location within the boundary layer where streamwise momentum of the flow is increased or diminished. 
Furthermore, the insight gained on how orifice pitch angle, skew angle and jet blowing ratio can be used to alter the types of vortices produced by the actuator can be a helpful guide for selectively generating synthetic jets that are composed of specific vortex structures. 
The ability to create a jet dominated by clockwise-rotating or counterclockwise-rotating vortices versus vortex pairs with equal strength has significant practical implications. 
Finally, the results have underscored the importance of understanding the underlying physics because the dynamics within a synthetic jet may change considerably when critical parameters are varied.

\bigskip

\textbf{Acknowledgements}

This material is based upon work supported by the National Science Foundation Graduate Research Fellowship Programme under Grant No. (DGE 1144083).

\textbf{Declaration of Interests}

The authors report no conflict of interest.

\bibliographystyle{jfm}
\bibliography{CU_jabref_database}

\end{document}